{}%disable hyper at arxiv
{}%use pdflatex at arxiv

\documentclass[10pt,a4paper]{article}

\newif\ifpublic\publictrue

\ifpublic\else\usepackage{showkeys}\fi
\usepackage[bookmarks=true,hyperfigures=true,hidelinks,bookmarksnumbered,unicode]{hyperref}
\usepackage[nosort]{cite}
\usepackage[parsep]{collref}
\usepackage[english]{babel}
\usepackage[T1]{fontenc}
\usepackage[utf8]{inputenc}
\usepackage{lmodern}
\usepackage{amsfonts,amsbsy,dsfont,euscript,mathrsfs,fixmath}
\usepackage{amsmath,amssymb}
\usepackage{scalerel}
\usepackage{enumerate}
\usepackage[x11names]{xcolor}
\usepackage{graphicx}
\usepackage{tikz}
\def\showkeysrefformat#1{{\normalfont\tiny\ttfamily#1}}
\makeatletter
\def\SK@@ref#1>#2\SK@{{\@inlabelfalse\leavevmode\vbox to\z@{\vss\SK@refcolor\rlap{\vrule\raise .75em \hbox{\showkeysrefformat{#2}}}}}}
\makeatother

\usepackage[a4paper,text={160mm,247mm},centering]{geometry}
\allowdisplaybreaks[3]
\numberwithin{equation}{section}
\def\[{\begin{equation}\begin{aligned}}
\def\]{\end{aligned}\end{equation}}

\usepackage[font=small,labelfont=bf,width=0.85\textwidth]{caption}

\expandafter\def\expandafter\bfseries\expandafter{\bfseries\ifmmode\else\boldmath\fi}
\expandafter\def\expandafter\mdseries\expandafter{\mdseries\ifmmode\else\unboldmath\fi}
\expandafter\def\expandafter\normalfont\expandafter{\normalfont\ifmmode\else\unboldmath\fi}

\RequirePackage{verbatim}

\makeatletter
\newwrite\bibinl@out
\newenvironment{bibtex}[1][\jobname]{%
\immediate\openout\bibinl@out #1.bib%
\immediate\write\bibinl@out{\@percentchar generated from `\jobname' starting line \the\inputlineno^^J}%
\def\verbatim@processline{\immediate\write\bibinl@out{\the\verbatim@line}}%
\@bsphack\let\do\@makeother\dospecials\catcode`\^^M\active\verbatim@start%
}
{\immediate\closeout\bibinl@out\@esphack}
\makeatother

\let\barefrac=\frac
\renewcommand{\frac}[2]{\mathinner{\barefrac{#1}{#2}}}

\let\baresqrt=\sqrt
\makeatletter
\renewcommand{\sqrt}{\@ifnextchar[\@sqrt@space@a\@sqrt@space@b}
\def\@sqrt@space@a[#1]#2{\mathinner{\mathchoice{\mkern-3mu}{\mkern-3mu}{}{}\baresqrt[#1]{#2}}}
\def\@sqrt@space@b#1{\mathinner{\mathchoice{\mkern-3mu}{\mkern-3mu}{}{}\baresqrt{#1}}}
\makeatother

\makeatletter
\let\per@dot@old=\.
\def\.{\ifmmode\def\per@dot@sel{\mkern3mu}\else\def\per@dot@sel{\per@dot@old}\fi\per@dot@sel}
\makeatother

\let\barefootnote=\footnote
\renewcommand{\footnote}[1]{\barefootnote{#1\vspace{3pt}}}

\newcommand{\vfrac}[2]{\ifmmode\mathinner{\textstyle^{#1}\!/\!_{#2}}\else$^{#1}\!/\!_{#2}$\fi}

\DeclareMathOperator{\diag}{diag}

\DeclareMathOperator{\Tr}{Tr}

\DeclareMathOperator{\im}{im}

\newcommand{\Real}{\mathds{R}}

\newcommand{\Integer}{\mathds{Z}}

\let\Re\relax\DeclareMathOperator{\Re}{Re}
\let\Im\relax\DeclareMathOperator{\Im}{Im}

\DeclareMathOperator{\sech}{sech}

\newcommand{\ind}[1]{{\scriptscriptstyle{#1}}}

\makeatletter
\newcommand*\bigcdot{\mathpalette\bigcdot@{.5}}
\newcommand*\bigcdot@[2]{\mathbin{\vcenter{\hbox{\scalebox{#2}{$\m@th#1\bullet$}}}}}
\makeatother

\newcommand{\al}{\alpha}

\usepackage{mathtools}
\newcommand{\alg}[1]{\mathfrak{#1}}
\newcommand{\grp}[1]{\mathrm{#1}}
\DeclareMathOperator{\rank}{rank}

\newcommand{\com}[2]{[#1,#2]}
\newcommand{\anticom}[2]{\{#1,#2\}}

\def\<{\big\langle}
\def\>{\big\rangle}

\DeclareMathOperator{\kernel}{ker}
\DeclareMathOperator{\dimension}{dim}

\newcommand{\geom}[1]{\mathrm{#1}}

\newcommand{\AdS}{\geom{AdS}}

\newcommand{\Sp}{\geom{S}}

\newcommand{\To}{\geom{T}}

\DeclareMathOperator{\Vol}{Vol}

\DeclareFontEncoding{LS1}{}{}
\DeclareFontSubstitution{LS1}{stix}{m}{n}
\DeclareSymbolFont{stixsymbols}{LS1}{stixscr}{m}{n}
\SetSymbolFont{stixsymbols}{bold}{LS1}{stixscr}{b}{n}
\DeclareMathSymbol{\kay}{\mathalpha}{stixsymbols}{"6B}
\DeclareMathSymbol{\hay}{\mathalpha}{stixsymbols}{"68}
\DeclareMathAlphabet{\mathdsl}{U}{bbm}{m}{sl}

\newcommand{\field}{\Psi}
\newcommand{\fields}{\Psi}
\newcommand{\hc}{\mathrm{h.c.}}

\def\ys{||\mathbf{y}_+||^2}
\def\zs{||\mathbf{z}_+||^2}
\def\yz{\mathbf{y}_+\cdot\mathbf{z}_+}
\def\tys{||\mathbf{y}_-||^2}
\def\tzs{||\mathbf{z}_-||^2}
\def\tytz{\mathbf{y}_-\cdot\mathbf{z}_-}
\def\yyt{\mathbf{y}_+\cdot\mathbf{y}_-}
\def\zzt{\mathbf{z}_+\cdot\mathbf{z}_-}
\def\yzt{\mathbf{y}_+\cdot\mathbf{z}_-}
\def\ytz{\mathbf{y}_-\cdot\mathbf{z}_+}
\def\Glc{\mathrm{\Gamma}}
\def\Olc{\mathrm{\Omega}}

\newcommand{\interval}[1]{(#1]}

\usepackage{slashed}

\providecommand{\href}[2]{#2}

\makeatletter
\def\mr@ignsp#1 {\ifx\:#1\@empty\else #1\expandafter\mr@ignsp\fi}
\newcommand{\multiref}[1]{\begingroup%
\xdef\mr@no@sparg{\expandafter\mr@ignsp#1 \: }%
\def\mr@comma{}\def\mr@dash{-}%
\@for\mr@refs:=\mr@no@sparg\do{%
\ifx\mr@refs\mr@dash\def\mr@comma{}--\else%
\mr@comma\def\mr@comma{,}\ref{\mr@refs}\fi}%
\endgroup}
\renewcommand{\eqref}[1]{(\multiref{#1})}
\makeatother

\makeatletter
\newcommand{\namedref}[2]{\hyperref[#2]{#1~\ref*{#2}}}
\newcommand{\secref}{\@ifstar{\namedref{Section}}{\namedref{sec.}}}
\newcommand{\appref}{\@ifstar{\namedref{Appendix}}{\namedref{app.}}}
\newcommand{\tabref}{\@ifstar{\namedref{Table}}{\namedref{tab.}}}
\newcommand{\figref}{\@ifstar{\namedref{Figure}}{\namedref{fig.}}}
\newcommand{\foottref}{\@ifstar{\namedref{Footnote}}{\namedref{foot.}}}
\makeatother

\let\oldbib=\thebibliography
\def\thebibliography{\phantomsection\addcontentsline{toc}{section}{\refname}\oldbib}

\let\oldtoc=\tableofcontents
\def\tableofcontents{\phantomsection\addcontentsline{toc}{section}{\contentsname}\oldtoc}

\providecommand{\hypersetup}[1]{}
\providecommand{\texorpdfstring}[2]{#1}
\hypersetup{plainpages=false}
\hypersetup{pdfpagemode=UseNone}
\hypersetup{bookmarksnumbered=true}
\hypersetup{pdfstartview=FitH}
\hypersetup{colorlinks=false}
\hypersetup{citebordercolor={1 1 1}}
\hypersetup{urlbordercolor={1 1 1}}
\hypersetup{linkbordercolor={1 1 1}}
\makeatletter
\let\@keywords\@empty
\let\@subject\@empty
\providecommand{\keywords}[1]{\gdef\@keywords{#1}}
\providecommand{\subject}[1]{\gdef\@subject{#1}}
\def\thetitle{\@title}
\def\theauthor{\@author}
\def\thesubject{\@subject}
\def\thedate{\@date}
\def\thekeywords{\@keywords}
\makeatother
\AtBeginDocument{
\hypersetup{pdftitle={\thetitle}}
\hypersetup{pdfauthor={\theauthor}}
\hypersetup{pdfsubject={\thesubject}}
\hypersetup{pdfkeywords={\thekeywords}}}

\newif\ifshownote
\shownotetrue

\ifpublic\shownotefalse\fi

\ifshownote

\ifpdf\else\RequirePackage[active]{srcltx}\fi
\RequirePackage{xcolor}
\RequirePackage{ulem}

\newcommand{\remark}[2][]{{\normalfont\sffamily\hspace{1ex}
\def\emph{\textsl}\def\textbullet{$\bullet$}
\def\tmparga{#1}
\def\tmpargb{BH}\ifx\tmparga\tmpargb\color[rgb]{0.7,0,0}\fi
\def\tmpargb{AR}\ifx\tmparga\tmpargb\color[rgb]{0,0.7,0}\fi
\def\tmpargb{FS}\ifx\tmparga\tmpargb\color[rgb]{0,0,0.7}\fi
\def\tmpargb{}\ifx\tmparga\tmpargb\normalfont\color{red}\fi
\def\tmpargb{}\ifx\tmparga\tmpargb\else \textbf{#1:}\fi
#2\hspace{1ex}}}

\else

\newcommand{\remark}[2][]{\ignorespaces}

\fi

\title{Supersymmetry and integrability \texorpdfstring{\\}{} of the elliptic \texorpdfstring{$\AdS_3 \times \Sp^3 \times \To^4$}{AdS3 × S3 × T4} superstring}
\author{Ben Hoare\texorpdfstring{$^{a,}$\footnote{\texttt{ben.hoare@durham.ac.uk}}}{} and Fiona K.~Seibold\texorpdfstring{$^{b,c,}$\footnote{\texttt{fiona.seibold@epfl.ch}}}{}}

\begin{document}

\pdfbookmark[1]{Title Page}{title}
\thispagestyle{empty}

\vspace*{2cm}
\begin{center}
\begingroup\Large\bfseries\thetitle\par\endgroup
\vspace{1cm}

\renewcommand{\thefootnote}{\roman{footnote}}
\begingroup\theauthor\par\endgroup
\vspace{1cm}

\textit{$^a $Department of Mathematical Sciences, Durham University, Durham DH1 3LE, UK}\\
\vspace{1mm}
\textit{$^b $Theoretical Particle Physics Laboratory, EPFL, 1015 Lausanne, Switzerland}\\
\vspace{1mm}
\textit{$^c $Deutsches Elektronen-Synchrotron DESY, Notkestr.~85, 22607 Hamburg, Germany}
\vspace{5mm}

\vfill

\textbf{Abstract}\vspace{5mm}

\begin{minipage}{12cm}
We construct a 1-parameter family of Ramond-Ramond fluxes supporting the elliptic $\AdS_3 \times \Sp^3 \times \To^4$ metric with constant dilaton and preserving 8 of the 16 supercharges of the undeformed background.
On the supersymmetric locus, we compute the tree-level worldsheet S-matrix in uniform light-cone gauge up to quadratic order in fermions and find that it non-trivially satisfies the classical Yang-Baxter equation.
Moreover, imposing classical integrability and symmetries, we conjecture compatible processes quartic in fermions.
We also investigate different limits of interest, including trigonometric deformations and the limit to the $\AdS_2 \times \Sp^2 \times \To^6$ superstring.
Our results provide strong evidence for a supersymmetric and integrable elliptic deformation of the $\AdS_3 \times \Sp^3 \times \To^4$ superstring supported by Ramond-Ramond flux and a constant dilaton.

\end{minipage}

\vspace*{4cm}

\end{center}

\setcounter{footnote}{0}
\renewcommand{\thefootnote}{\arabic{footnote}}

\newpage

%%%%%%%%%%%%%%%%%%%%%%%%%%%%%%%%%%%%%%%%%%%%%%%%%%%%%%%%%%%%%%%%%%%%%%%%%%%%%%%%
\section{Introduction}\label{sec:introduction}

The type IIB $\AdS_3 \times \Sp^3 \times \To^4$ supergravity background preserves 16 supercharges, half of the maximal 32 supercharges for superstrings in 10 dimensions.
It features a constant dilaton and can be supported by a mix of Neveu-Schwarz-Neveu-Schwarz (NS-NS) and Ramond-Ramond (R-R) 3-form fluxes.
Free strings propagating in such backgrounds are classically integrable~\cite{Babichenko:2009dk,Cagnazzo:2012se,Wulff:2014kja}.
That is, the equations of motion for the superstring worldsheet theory can be recast as a flatness condition for a Lax connection, from which it is possible to construct an infinite tower of conserved charges in involution.
The integrability of the theory is closely tied to the formulation of the 2-dimensional worldsheet theory as a sigma model on a semi-symmetric space, in close analogy to the $\AdS_5 \times \Sp^5$ superstring~\cite{Metsaev:1998it,Berkovits:1999zq,Bena:2003wd}.
Both $\AdS_3$ and $\Sp^3$ are symmetric spaces and the curved part of the geometry possesses a $\grp{SU}(1,1)_L \times \grp{SU}(1,1)_R \times \grp{SU}(2)_L \times \grp{SU}(2)_R$ isometry group (we use the labels $L$ and $R$ to distinguish two copies of the same group, corresponding to left- and right-acting symmetries respectively).
In the supersymmetric theory, these bosonic symmetries gather into the $\grp{PSU}(1,1|2)_L \times \grp{PSU}(1,1|2)_R$ supergroup.
This factorised structure makes the theory particularly rich and leads to a large landscape of integrable deformations.

\medskip

Over the past twenty years, many integrable deformations of sigma models and superstrings have been discovered, for a review see~\cite{Hoare:2021dix}.
These modify the space-time geometry in which the strings propagate, thereby breaking some or all of the symmetries while preserving the (classical and/or quantum) integrability of the theory.
A particularly interesting setup is the principal chiral model (PCM) with underlying Lie group $\grp{G}=\grp{SU}(2)$ and its integrable deformations, as considered by Cherednik in~\cite{Cherednik:1981df}.
The undeformed, or rational, theory admits a global $\grp{SU}(2)_L \times \grp{SU}(2)_R$ symmetry and describes the classical motion of a free string on a three-sphere $\Sp^3$.
The trigonometric deformation produces a squashed-sphere geometry and deforms the right-acting $\grp{SU}(2)_R$ symmetry, so that only $\grp{SU}(2)_L \times \grp{U}(1)_R$ remains manifest.
\unskip\footnote{It is also possible to break both the left-acting and right-acting symmetries so that only $\grp{U}(1)_L \times \grp{U}(1)_R$ remains manifest while preserving integrability~\cite{Fateev:1996ea}.}
The elliptic deformation further breaks this to just $\grp{SU}(2)_L$ and all the right-acting symmetries are broken.
While being less symmetric, the trigonometric and elliptic deformations still preserve the classical integrability of the model and an explicit Lax connection can be written down.
By analytic continuation, it is also possible to construct an elliptic deformation of the PCM on $\grp{G}=\grp{SL}(2;\Real)$.
Bringing these building blocks together gives an integrable elliptic deformation of the $\AdS_3 \times \Sp^3$ string, or the $\AdS_3 \times \Sp^3 \times \To^4$ string if the four-torus directions are also included~\cite{Hoare:2023zti}.
An interesting question is then if this integrable deformation can be extended to superstrings.

\medskip
The trigonometric deformation is closely related (up to a closed B-field) to the inhomogeneous Yang-Baxter (also called $\eta$) deformation~\cite{Klimcik:2002zj,Klimcik:2008eq}.
More precisely, it corresponds to a twisted version of the unilateral deformation when only the right-acting symmetry is broken~\cite{Fukushima:2020kta,Hamidi:2025sgg}.
The inhomogeneous Yang-Baxter deformation can be generalised to the PCM on an arbitrary simple Lie group $\grp{G}$, as well as to symmetric~\cite{Delduc:2013fga} and semi-symmetric spaces~\cite{Delduc:2013qra,Hoare:2014oua}.
The broken symmetries are promoted to Drinfel'd-Jimbo type quantum groups~\cite{Delduc:2013fga,Delduc:2014kha}.
The supergravity background supporting the deformed $\AdS_3 \times \Sp^3 \times \To^4$ geometry was obtained from the semi-symmetric space sigma model in~\cite{Seibold:2019dvf}.
For the unilateral deformation, when only the right-acting $\grp{PSU}(1,1|2)_R$ superisometry is broken, the background preserves 8 supercharges~\cite{Hoare:2022asa}.
The elliptic deformation of the $\grp{SU}(2)$ PCM has recently been extended to higher-rank groups $\grp{G}=\grp{SL}(N)$~\cite{Lacroix:2023qlz}, but integrable elliptic deformations of symmetric and semi-symmetric space sigma models are not known.
To make progress towards such a construction, in~\cite{Hoare:2023zti} an embedding of the elliptic $\AdS_3 \times \Sp^3 \times \To^4$ metric into type IIB supergravity was proposed.
The background has the following properties: a vanishing NS-NS 3-form $H_3 = dB_2 =0$, a constant dilaton, and a vanishing R-R 1-form $F_1$.
Moreover, the R-R 3-form and 5-form can be written as a linear combination of four auxiliary 3-forms $f_{3}^{(i)}$ with $i=1,2,3,4$.
Schematically, $F_3 = \sum_{i=1}^4 \mathbf{x}_4^{(i)} f_{3}^{(i)}$ and $F_5 = \sum_{k=1}^3 \sum_{i=1}^4 \mathbf{x}_k^{(i)} f_3^{(i)} \wedge J_{2,k}$, where $J_{2,k}$ are three orthogonal 2-forms on $\To^4$.
The supergravity equations impose constraints on the norms and scalar products of the vectors of parameters $\mathbf{x}^{(i)}$.
In this paper we investigate the supersymmetry and integrability of this theory.

\medskip

Without a (deformed) semi-symmetric space sigma model description, there are two main ways to test if the elliptic $\AdS_3 \times \Sp^3 \times \To^4$ superstring is classically integrable.
The first is through the construction of a Lax connection encoding the equations of motion of the Green-Schwarz action.
The second is through the computation of the worldsheet S-matrix, which is the path we follow in this paper.
For an integrable theory, the presence of a tower of (higher-spin) conserved charges drastically constrains the worldsheet S-matrix.
For massive excitations, there cannot be particle production, and the S-matrix factorises: a $n \rightarrow n$ scattering event can be decomposed into a sequence of $2 \rightarrow 2$ scattering processes~\cite{Zamolodchikov:1978xm}.
Consistency then requires the two-particle S-matrix to satisfy the quantum Yang-Baxter equation.

\medskip

Computing the S-matrix requires fixing a gauge to isolate the physical degrees of freedom.
In other words, one expands the Green-Schwarz action around a specific classical solution and studies interactions above this ``vacuum.''
Integrable string sigma models are customarily analysed in a uniform light-cone gauge~\cite{Arutyunov:2004yx,Arutyunov:2005hd}.
This picks out two commuting isometries $T$ and $\Phi$ of the supergravity background, and fixes the light-cone combinations $X^+ = \tau$ and $P_{X^-}=\sigma$, where $\tau$ and $\sigma$ are time-like and space-like coordinates respectively on the string worldsheet and $P_{X^-}$ denotes the momentum conjugate to the coordinate $X^-$.
For closed strings, the worldsheet has the topology of a cylinder.
The next step is to take the decompactification limit, sending the radius of the cylinder to infinity.
This results in a 2d theory on a plane with well-defined asymptotic states.
For the theory at hand, the curved part of the geometry ($\AdS_3 \times \Sp^3$) leads to four bosonic and four fermionic massive excitations.
The presence of the flat directions from the $\To^4$ leads to four massless bosons and four massless fermions.
The worldsheet S-matrix is a function of the string tension $\mathrm{T}$ and it can be expanded in inverse powers of $\mathrm{T}$.
In particular, the tree-level S-matrix $\mathbb{T}$ is defined by $\mathbb{S}= 1+ i \mathbb{T} /\mathrm{T} + \dots$ and in an integrable theory it should satisfy the classical Yang-Baxter equation.
\unskip\footnote{See~\cite{Demulder:2023bux} for a review on the S-matrix approach to studying integrable strings in $\AdS_3$ backgrounds.}
The tree-level S-matrix in the pure R-R theory was computed in~\cite{Rughoonauth:2012qd,Sundin:2013ypa} and the mixed-flux theory was considered in~\cite{Hoare:2013pma}.
Moreover, the tree-level S-matrix for the two-parameter Yang-Baxter deformation of $\AdS_3 \times \Sp^3 \times \To^4$ with pure R-R flux has also been computed in~\cite{Seibold:2021lju}.
The S-matrix of the $\AdS_3 \times \Sp^3 \times \To^4$ superstring has mainly been investigated in the BMN light-cone gauge with the isometries corresponding to shifts in $T$ and $\Phi$ lying in the diagonal of $\grp{SU}(1,1)_L \times\grp{SU}(1,1)_R$ and $\grp{SU}(2)_L \times \grp{SU}(2)_R$ respectively.
However, these are no longer symmetries of the elliptic deformation.
We are therefore forced to consider an alternative gauge for which the isometries completely lie in $\grp{SU}(1,1)_L$ and $\grp{SU}(2)_L$ respectively.
Such different choices of gauge gives rise to worldsheet Hamiltonians and S-matrices related by $JT$ deformations~\cite{Frolov:2019xzi,Borsato:2023oru}.
\unskip\footnote{Similar alternative light-cone gauge-fixings need to be considered to analyse other types of integrable deformations, for instance homogeneous Yang-Baxter deformations of $\AdS_5 \times \Sp^5$~\cite{Borsato:2024sru}.}

\medskip

In integrable theories it is possible to go beyond perturbation theory.
The exact S-matrix, to all order in string tension, can often be bootstrapped from symmetries up to overall phases.
This was successfully achieved for the $\AdS_3 \times \Sp^3 \times \To^4$ superstring in~\cite{Borsato:2013qpa,Borsato:2014exa,Lloyd:2014bsa} and also for its two-parameter Yang-Baxter deformation~\cite{Hoare:2014oua} by harnessing its $q$-deformed symmetries.
In these two cases, the S-matrix factorises: it is possible to write the $256 \times 256$ S-matrix as $\mathbb{S} = \mathcal S \otimes \mathcal S$, where $\mathcal S$ is of dimension $16 \times 16$ and itself also satisfies the quantum Yang-Baxter equation.
The massive sector further decomposes into four $4 \times 4$ blocks, encoding left-left ($\mathcal S_{++}$), right-right ($\mathcal S_{--}$), left-right ($\mathcal S_{+-}$) and right-left ($\mathcal S_{-+}$) scattering.
These $4 \times 4$ blocks have also been obtained by directly bootstrapping the quantum Yang-Baxter equation~\cite{deLeeuw:2021ufg} using a method that also led to a new 8-vertex elliptic S-matrix.
We will show that for the elliptic $\AdS_3 \times \Sp^3 \times \To^4$ superstring with 8 supersymmetries, while the worldsheet S-matrix does solve the Yang-Baxter equation, it is not straightforwardly compatible with such a factorised structure, and the relation, if any, to the elliptic S-matrix of~\cite{deLeeuw:2021ufg} remains to be understood.

\medskip

The paper is organised as follows.
In \secref{sec:sugra-susy} we present the elliptic $\AdS_3 \times \Sp^3 \times \To^4$ supergravity background we are interested in and analyse its supersymmetries.
We show that for generic deformation parameters, the background preserves 8 supersymmetries if and only if the vector of parameters controlling the R-R fluxes span a 2-plane.
Moreover, these supersymmetries combine into one copy of the $\alg{psu}(1,1|2)$ superalgebra, which is left untouched by the deformation.
In the trigonometric limit, we construct two branches of supersymmetric solutions and identify the special point corresponding to the unilateral inhomogeneous Yang-Baxter deformation.
In \secref{sec:lcgf} we analyse the theory in the uniform light-cone gauge and we compute the tree-level worldsheet S-matrix for massive excitations in~\secref{sec:Smatrix}.
We find that it satisfies the classical Yang-Baxter equation, providing strong evidence that the theory is classically integrable.
We check that we reproduce known results in the rational and trigonometric limits, and also investigate another limit where the S-matrix interpolates between the $\AdS_3 \times \Sp^3$ S-matrix (in an alternative light-cone gauge) and the $\AdS_2 \times \Sp^2$ S-matrix in the usual light-cone gauge.
We conclude and discuss future directions in~\secref{sec:conclusions}.
Our convention for Dirac matrices and spinors is summarised in \appref{app:gammamatrix}. The equivalence between the supersymmetry constraints and the restriction of the R-R parameters to a 2-plane is shown in \appref{app:Gram}. Finally, we present the Killing vectors and Killing spinors of the elliptic background in \appref{app:c1} and \appref{app:c2} respectively.

%%%%%%%%%%%%%%%%%%%%%%%%%%%%%%%%%%%%%%%%%%%%%%%%%%%%%%%%%%%%%%%%%%%%%%%%%%%%%%%%
\section{Supergravity and supersymmetry}\label{sec:sugra-susy}

In this paper we study type IIB supergravity backgrounds for $\AdS_3 \times \Sp^3 \times \To^4$ and deformations of the general form
\begin{equation}\begin{gathered}\label{eq:backclass}
G = g_{\mu\nu}(\fields^\rho) d\fields^\mu d\fields^\nu + \mathrm{T} \sum_{m=6}^9 d\fields^m d\fields^m ~, \qquad
H_3 = 0~, \qquad \Phi = 0 ~,
\\
F_1 = 0 ~, \qquad
F_3 = F_{3,4}(\fields^\rho) ~, \qquad
F_5 = \sum_{i=1}^3 F_{3,i}(\fields^\rho) \wedge J_{2,i} ~,
\end{gathered}\end{equation}
where $H_3$ is the NS-NS flux, $F_{1,3,5}$ are the R-R fluxes, $\Phi$ is the dilaton and $\mathrm{T}$ is the effective string tension.
\unskip\footnote{It is straightforward to restore the dependence on the constant dilaton $\Phi$, hence we have set it to zero.}
The index $\mu,\nu,\rho,\ldots = 0,\ldots,5$ runs over the $\AdS_3$ and $\Sp^3$ directions, i.e.~$\{\field^\rho\} = \{T,U,V,\Phi,X,Y\}$, while $m,n,p,\ldots = 6,\ldots,9$ labels the $\To^4$ directions.
We also introduce the index $M,N,P,\ldots = 0,\ldots,9$ running over all 10 directions.
The 2-forms
\begin{equation}
J_{2,1} = \mathrm{T} (d\fields^6 \wedge d\fields^7 - d\fields^8 \wedge d\fields^9) ~, \qquad \!\!
J_{2,2} = \mathrm{T} (d\fields^6 \wedge d\fields^8 + d\fields^7 \wedge d\fields^9) ~, \qquad \!\!
J_{2,3} = \mathrm{T} (d\fields^6 \wedge d\fields^9 - d\fields^7 \wedge d\fields^8) ~,
\end{equation}
are three orthogonal self-dual 2-forms on the four-torus.
Finally, the 3-form and 5-form R-R fluxes are parametrised in terms of four closed 3-forms, $F_{3,i}$, $i = 1,\ldots, 4$, which we take to only have legs in and only depend on the $\AdS_3$ and $\Sp^3$ directions.
We furthermore assume that they are self-dual $\star_6 F_{3,i} = F_{3,i}$, implying that $(d \star_6 F_{3,i} = 0)=|F_{3,i}|^2 = 0$, and orthogonal, i.e.~$F_{3,i} \cdot F_{3,j} = 0$ for $i \neq j$.
Under these assumptions the type IIB supergravity equations simplify to
\begin{equation}\label{eq:typeiib}
R_{\mu\nu} - \frac{1}{4}\big(\sum_{i=1}^4 (F_{3,i})_{\mu\rho\sigma}(F_{3,i})_{\nu}{}^{\rho\sigma}\big) = 0 ~.
\end{equation}

There are two type IIB supergravity Killing spinor equations, which come from the invariance of the dilatino $\lambda$ and gravitino $\psi_\ind{M}$ fields under supersymmetry variations.
Introducing tangent-space indices $A,B,C\ldots = 0,\ldots,9$ lowered and raised with $\eta_{\ind{AB}} = \diag(-1,1,1,1,1,1,1,1,1,1)_{\ind{AB}}$ and its inverse $\eta^{\ind{AB}}$, the Killing spinor equations are
\begin{align}\label{eq:dilatino}
\delta\lambda = 0 \qquad \Leftrightarrow \qquad &\big(\Gamma^\ind{M} \partial_\ind{M} \Phi + \frac{1}{12} \sigma_3 \slashed{H} + e^{\Phi} (i \sigma_2 \slashed{F}_1 + \frac{1}{12} \sigma_1 \slashed{F}_3 )\big) \epsilon = 0 ~,
\\\label{eq:gravitino}
\delta \psi_\ind{M} = 0 \qquad \Leftrightarrow \qquad & D_\ind{M} \epsilon = \big( \partial_\ind{M} - \frac{1}{4} \slashed{\omega}_\ind{M} + \frac{1}{8} \sigma_3 \slashed{H}_\ind{M} + \frac18 \mathcal S \Gamma_\ind{M} \big) \epsilon = 0 ~.
\end{align}
Here $\epsilon=(\epsilon^1, \epsilon^2)$ is a doublet of 32-component Majorana-Weyl spinors and $\sigma_{1,2,3}$ are Pauli matrices acting on the index $I=1,2$, which we have suppressed.
$\Gamma^\ind{A}$ are $32 \times 32$ 10d Dirac matrices satisfying the Clifford algebra $\anticom{\Gamma^\ind{A}}{\Gamma^\ind{B}} = 2 \eta^{\ind{AB}}$ and we define their curved-space counterparts $\Gamma_\ind{M} = e_\ind{M}^\ind{A} \Gamma_\ind{A}$, where $e_\ind{M}^\ind{A}$ denotes the vielbein associated to the space-time metric, $G_\ind{MN} = e_\ind{M}^\ind{A}e_\ind{N}^\ind{B} \eta_{\ind{AB}}$.
In the following we use Dirac matrices adapted to $\AdS_3 \times \Sp^3 \times \To^4$ backgrounds and their deformations of the form~\eqref{eq:backclass}, which are defined in \appref{app:gammamatrix}.
Slashed quantities are contracted with the Dirac matrices as $\slashed{F}_r \equiv F_{\ind{A}_1 \dots \ind{A}_r} \Gamma^{\ind{A}_1 \dots \ind{A}_r}$, with $\Gamma^{\ind{A}_1 \dots \ind{A}_r} = \Gamma^{\ind{A}_1} \dots \Gamma^{\ind{A}_r}$.
Furthermore, $\omega$ denotes the spin connection
\begin{equation}
\label{eq:spin-connection}
\omega_\ind{M}{}^{\ind{AB}} = - e^{\ind{A}\ind{N}} \partial^{\vphantom{\ind{A}}}_{[\ind{M}} e^{\ind{B}}_{\ind{N}]} + e^{\ind{B}\ind{N}} \partial^{\vphantom{\ind{A}}}_{[\ind{M}} e^{\ind{A}}_{\ind{N}]} + e^{\ind{A}\ind{N}}e^{\ind{B}\ind{P}}\partial^{\vphantom{\ind{B}}}_{[\ind{N}} e^\ind{C}_{\ind{P}]} e_{\ind{C}\ind{M}} ~,
\end{equation}
and the R-R bispinor is given by
\begin{equation}
\label{eq:R-R-bispinor}
\mathcal S = - e^\Phi\bigg( i \sigma_2 \slashed{F}_1 + \frac{1}{3!} \sigma_1 \slashed{F}_3 + \frac{1}{2 \cdot 5!} i \sigma_2 \slashed{F}_5 \bigg)~.
\end{equation}
The round and square brackets denote symmetrisation and antisymmetrisation of the enclosed indices respectively, with an overall $1/n!$ normalisation.

The spinors $\epsilon^I$ are Majorana-Weyl and since we are considering type IIB supergravity solutions they have the same chirality.
In our conventions this means that
\begin{equation}\label{eq:chiral}
(1+\Gamma^{11})\epsilon^I = 0 ~,
\end{equation}
where $\Gamma^{11} = \Gamma^0 \Gamma^1 ... \Gamma^9$, so that each spinor has 16 non-vanishing components.
We denote the 32-dimensional space spanned by the two chiral spinors by $\mathfrak{s}$.
Unless otherwise specified, we read all equations acting on $\epsilon = (\epsilon^1,\epsilon^2)$ as acting on $\epsilon \in \mathfrak{s}$.

Since we will primarily be interested in the number of 6d supersymmetries, it is useful to consider the subspace determined by
\begin{equation}\label{eq:6dspin}
(1-\Gamma^6\Gamma^7\Gamma^8\Gamma^9)\epsilon^I = 0 ~.
\end{equation}
This leaves a total of 16 non-vanishing components spanning a 16-dimensional space that we denote $\mathfrak{s}_6$.
We denote the complement, i.e., the subspace determined by $(1+\Gamma^6\Gamma^7\Gamma^8\Gamma^9)\epsilon^I = 0$, by $\bar{\mathfrak{s}}_6$.
To study the pp-wave limit it will also be helpful to introduce the light-cone Dirac matrices
\begin{equation}
\Glc^\pm = \frac{1}{2} ( \Gamma^0 \pm \Gamma^3 )~,
\end{equation}
which satisfy
\begin{equation}
\Glc^\pm \Glc^\pm =0~, \qquad \Glc^+ \Glc^- +\Glc^- \Glc^+ = -1~.
\end{equation}
We have that $\Glc^\pm : \mathfrak{s}_6 \to \mathfrak{s}_6$ and they can be used to write the direct sum decomposition $\mathfrak{s}_6 = \mathfrak{s}_{6+} \oplus \mathfrak{s}_{6-}$, where $\mathfrak{s}_{6\pm} = \kernel\Glc^\pm\vert_{\mathfrak{s}_6}$ are both 8-dimensional.
\unskip\footnote{The subspace $\bar{\mathfrak{s}}_6$ similarly admits a direct sum decomposition $\bar{\mathfrak{s}}_6 = \bar{\mathfrak{s}}_{6+} \oplus \bar{\mathfrak{s}}_{6-}$ where $\mathfrak{s}_{6\pm} = \kernel\Glc^\pm\vert_{\bar{\mathfrak{s}}_6}$ are also both 8-dimensional.}

For the class of backgrounds~\eqref{eq:backclass}, the dilatino equation~\eqref{eq:dilatino} becomes
\begin{equation}\label{eq:dilatino3}
\sigma_1 \slashed{F}_3 \epsilon = 0 ~,
\end{equation}
while the gravitino equation~\eqref{eq:gravitino} can be written in the form
\begin{equation}\label{eq:gravitino3}
D_\ind{M}\epsilon \equiv (\partial_\ind{M} - \Omega_\ind{M})\epsilon = 0 ~, \qquad
\Omega_\ind{M} = \frac{1}{4}\slashed{\omega}_{\ind{M}} + \frac{1}{8 \cdot 3!}\sigma_1 \slashed{F}_3 \Gamma_\ind{M} + \frac{1}{16\cdot 5!} i\sigma_2 \slashed{F}_5 \Gamma_\ind{M} ~.
\end{equation}
This can be locally solved for any $\epsilon\in\mathfrak{s}$ that satisfies the compatibility condition $[\partial_\ind{M},\partial_\ind{N}] \epsilon = F_\ind{MN} \epsilon = 0$ where
\begin{equation}\label{eq:curvom}
F_\ind{MN} = \partial_\ind{M} \Omega_\ind{N} - \partial_\ind{N} \Omega_\ind{M} - [\Omega_\ind{M},\Omega_\ind{N}] ~.
\end{equation}
Therefore, the total number of supersymmetries of a background is given by
\begin{equation}\label{eq:susycount}
\dimension(\kernel \sigma_1 \slashed{F}_3 \cap \bigcap_{M,N} \kernel F_\ind{MN}) ~.
\end{equation}

Since $\ker \sigma_1 \slashed{F}_3 \supseteq \mathfrak{s}_6$,
\unskip\footnote{This follows from the identity $\sigma_1 \slashed{F}_3 (1+\Gamma^6\Gamma^7\Gamma^8\Gamma^9) = 0$, which is satisfied by the self-duality of the 3-form $F_{3,4}$.}
the dilatino equation~\eqref{eq:dilatino3} is automatically satisfied for $\epsilon \in \mathfrak{s}_6$.
It is also the case that
\unskip\footnote{If we split $\Omega_\ind{M} = \Omega^{\omega}_\ind{M} + \Omega^F_\ind{M}$, where the first part contains the contribution from the spin connection and the second contains the contribution from the R-R fluxes, then $\Omega^{\omega}_\mu : \mathfrak{s}_6 \to \mathfrak{s}_6$ and $\Omega^{\omega}_\mu : \bar{\mathfrak{s}}_6 \to \bar{\mathfrak{s}}_6$, while $\Omega^{\omega}_m = 0$.
Furthermore, we have $\ker \Omega_\mu^F \supseteq \bar{\mathfrak{s}}_6$, $\ker \Omega_m^F \supseteq \mathfrak{s}_6$ and $\im\Omega_\ind{M}^F \subseteq \mathfrak{s}_6$ since $(\sigma_1 \slashed{F}_3 ,i\sigma_2 \slashed{F}^5)\Gamma_\mu (1-\Gamma^6\Gamma^7\Gamma^8\Gamma^9)= (\sigma_1 \slashed{F}_3 ,i\sigma_2 \slashed{F}^5)\Gamma_m (1+\Gamma^6\Gamma^7\Gamma^8\Gamma^9) =
(1-\Gamma^6\Gamma^7\Gamma^8\Gamma^9)(\sigma_1 \slashed{F}_3 ,i\sigma_2 \slashed{F}^5)\Gamma_\ind{M} = 0$
by the self-duality of the 3-forms $F_{3,i}$.
Recall that these equations should be read as acting on $\epsilon \in \mathfrak{s}$ with $\epsilon^I$ satisfying \eqref{eq:chiral}.}
\begin{equation}\begin{aligned}
&\Omega_\mu : \mathfrak{s}_6 \to \mathfrak{s}_6 ~, \qquad
&&\Omega_\mu : \bar{\mathfrak{s}}_6 \to \bar{\mathfrak{s}}_6 ~,
\\
&\Omega_m : \mathfrak{s}_6 \to 0 ~, \qquad
&&\Omega_m : \bar{\mathfrak{s}}_6 \to \mathfrak{s}_6 ~.
\end{aligned}\end{equation}
Since no background fields depend on the coordinates parametrising the $\To^4$, it follows that
\begin{equation}\begin{aligned}
& F_{\mu\nu}:\mathfrak{s}_6 \to \mathfrak{s}_6 ~, \qquad &&
F_{\mu\nu}: \bar{\mathfrak{s}}_6 \to \bar{\mathfrak{s}}_6 ~,
\\
& F_{\mu n}:\mathfrak{s}_6 \to 0 ~, \qquad &&
F_{\mu n}: \bar{\mathfrak{s}}_6 \to \mathfrak{s}_6 ~,
\\
& F_{m n}:\mathfrak{s}_6 \to 0 ~, \qquad &&
F_{m n}: \bar{\mathfrak{s}}_6 \to 0 ~.
\end{aligned}\end{equation}

As we will describe in more detail shortly, the 16 Killing spinors of the undeformed $\AdS_3 \times \Sp^3 \times \To^4$ supergravity background are all valued in $\mathfrak{s}_6$.
We refer to these as 6d supersymmetries since they survive the dimensional reduction to 6 dimensions.
Our goal is to determine which of these 6d supersymmetries persist after we deform the background.
This does not exclude the possibility that there may be certain special values of the deformation parameters, or simplifying limits such as the pp-wave limit, for which the number of supersymmetries is enhanced, i.e., spinors valued in $\bar{\mathfrak{s}}_6$ become Killing.

Given that the dilatino equation is satisfied for $\epsilon \in \mathfrak{s}_6$ and $F_{Mn}:\mathfrak{s}_6 \to 0$, the number of 6d supersymmetries is equal to the number of independent $\epsilon \in \mathfrak{s}_6$ such that $F_{\mu\nu} \epsilon = 0$ for all $\mu$ and $\nu$.
It follows that the number of 6d supersymmetries of the background is given by
\begin{equation}\label{eq:6dsusycount}
\dimension(\bigcap_{\mu,\nu} \kernel F_{\mu\nu}) ~,
\end{equation}
where $F_{\mu\nu}$ are understood as linear operators acting on $\mathfrak{s}_6$.

\subsection{The elliptic background}

The elliptic $\AdS_3 \times \Sp^3 \times \To^4$ supergravity background we are interested in is the one proposed in \cite{Hoare:2023zti}. Its metric is given by
\begin{equation}\begin{aligned}\label{eq:ellipticmetricAdS3xS3}
G &= G_{AdS_3} + G_{\Sp^3} + G_{\To^4}~,
\\
G_{AdS_3} &= \frac{\alpha_1}{4} \Tr\big(g_\ind{A}^{-1}d g_\ind{A}\sigma_1\big)^2
-\frac{\alpha_2}{4} \Tr\big(g_\ind{A}^{-1}d g_\ind{A}(i\sigma_2)\big)^2
+\frac{\alpha_3}{4} \Tr\big(g_\ind{A}^{-1}d g_\ind{A}\sigma_3\big)^2 ~,
\\
G_{\Sp^3} &= \frac{\alpha_1}{4} \Tr\big(g_\ind{S}^{-1}d g_\ind{S}(i\sigma_1)\big)^2
+\frac{\alpha_2}{4} \Tr\big(g_\ind{S}^{-1}d g_\ind{S}(i\sigma_2)\big)^2
+\frac{\alpha_3}{4} \Tr\big(g_\ind{S}^{-1}d g_\ind{S}(i\sigma_3)\big)^2 ~,
\\
G_{\To^4} &= \mathrm{T} \big((d \field^6)^2 + (d\Psi^7)^2 + (d\Psi^8)^2 + (d\Psi^9)^2\big) ~,
\end{aligned}\end{equation}
where
\begin{equation}\label{eq:coordinates}
g_\ind{A} = e^{i T \sigma_2}e^{U \sigma_3} e^{V\sigma_1} \in \grp{SL}(2;\Real) ~,
\qquad
g_\ind{S} = e^{i \Phi \sigma_2}e^{i X \sigma_3} e^{iY\sigma_1} \in \grp{SU}(2) ~,
\end{equation}
$\sigma_{1,2,3}$ are the Pauli matrices, and $\alpha_1, \alpha_2, \alpha_3 > 0$ are deformation parameters
\unskip\footnote{Here we have indicated strict inequalities, however we will later consider a limit where one of the deformation parameters $\alpha_1$ or $\alpha_3$ vanishes.}
that scale with the effective string tension $\mathrm{T}$.
The coordinates $\{T,U,V\}$ parametrise $\AdS_3$, the coordinates $\{\Phi,X,Y\}$ parametrise $\Sp^3$ and the coordinates $\{\field^6,\Psi^7,\Psi^8,\Psi^9\}$ parametrise $\To^4$.
The metric on $\Sp^3$ is that of the elliptic $\grp{SU}(2)$ PCM first introduced in \cite{Cherednik:1981df}, and the metric on $\AdS_3$ is its $\grp{SL}(2;\Real)$ counterpart.
In the deformed metric on $\Sp^3$ the three deformation parameters are on the same footing, however, in the deformed metric on $\AdS_3$ $\alpha_2$ is distinguished as controlling the time-like direction, while $\alpha_1$ and $\alpha_3$ are associated with space-like directions.

A convenient choice for the vielbein ($e^\ind{A}= e^\ind{A}_\ind{M} d\fields^\ind{M}$) is
\begin{equation}
\label{eq:vielbein}
\begin{aligned}
e^0 &= -\frac{\sqrt{\alpha_2}}2\Tr\big(g_\ind{A}^{-1}d g_\ind{A}(i\sigma_2)\big) = \sqrt{\alpha_2} (\cosh 2U \cosh 2V dT + \sinh 2V dU ) ~, \\
e^1 &= -\frac{\sqrt{\alpha_1}}2\Tr\big(g_\ind{A}^{-1}d g_\ind{A}\sigma_1\big) = \sqrt{\alpha_1} (\sinh 2 U dT - dV)~, \\
e^2 &= -\frac{\sqrt{\alpha_3}}2\Tr\big(g_\ind{A}^{-1}d g_\ind{A}\sigma_3\big) = \sqrt{\alpha_3} (-\cosh 2 U \sinh 2 V dT - \cosh 2 V dU)~, \\
e^3 &= -\frac{\sqrt{\alpha_2}}2\Tr\big(g_\ind{S}^{-1}d g_\ind{S}(i\sigma_2)\big) = \sqrt{\alpha_2} (\cos 2 X \cos 2 Y d\Phi - \sin 2 Y dX)~, \\
e^4 &= -\frac{\sqrt{\alpha_1}}2\Tr\big(g_\ind{S}^{-1}d g_\ind{S}(i\sigma_1)\big) = \sqrt{\alpha_1} (-\sin 2 X d\Phi + dY)~, \\
e^5 &= -\frac{\sqrt{\alpha_3}}2\Tr\big(g_\ind{S}^{-1}d g_\ind{S}(i\sigma_3)\big) = \sqrt{\alpha_3} (\cos 2 X \sin 2Y d\Phi + \cos 2 Y dX)~, \\
e^6 &= \sqrt{\mathrm{T}} d\field^6~, \quad e^7 = \sqrt{\mathrm{T}} d\Psi^7~, \quad e^8 = \sqrt{\mathrm{T}} d\Psi^8~,\quad e^9 = \sqrt{\mathrm{T}} d\Psi^9~.
\end{aligned}
\end{equation}
This metric is supplemented by the following dilaton and fluxes
\begin{equation}\begin{gathered}
\Phi = 0 ~, \qquad H_3 = 0~, \qquad
F_1 = 0 ~, \qquad
F_3 = F_{3,4}(\fields^\rho) ~, \qquad
F_5 = \sum_{i=1}^3 F_{3,i}(\fields^\rho) \wedge J_{2,i} ~.
\end{gathered}\end{equation}
The index $\mu,\nu,\rho,\dots = 0,\ldots,5$ runs over the $\AdS_3$ and $\Sp^3$ directions, i.e.~$\{\field^\rho\} = \{T,U,V,\Phi,X,Y\}$, while $r,\ldots = 6,\ldots,9$ labels the $\To^4$ directions.
The 2-forms
\begin{equation}
J_{2,1} = e^6 \wedge e^7 - e^8 \wedge e^9 ~, \qquad
J_{2,2} = e^6 \wedge e^8 + e^7 \wedge e^9 ~, \qquad
J_{2,3} = e^6 \wedge e^9 - e^7 \wedge e^8 ~,
\end{equation}
are three orthogonal self-dual 2-forms on the four-torus.
The 3-form fluxes are given by
\begin{equation}\label{eq:fluxes}
F_{3,i} = \mathbf{x}^{(1)}_i f_3^{(1)}
+ \mathbf{x}^{(2)}_i f_3^{(2)}
+ \mathbf{x}^{(3)}_i f_3^{(3)}
+ \mathbf{x}^{(4)}_i f_3^{(4)} ~, \qquad i = 1,2,3,4 ~,
\end{equation}
with the auxiliary fluxes
\begin{equation}\begin{gathered}\label{eq:fluxbasis}
f_3^{(1)} = \mathrm{T}^{-\frac12} d(e^1 \wedge e^4) = 2 \mathrm{T}^{-\frac12}(
e^0 \wedge e^2 \wedge e^4
- e^1 \wedge e^3 \wedge e^5) ~,
\\
f_3^{(2)} = \mathrm{T}^{-\frac12}d(e^0 \wedge e^3) = 2 \mathrm{T}^{-\frac12}(
e^0 \wedge e^4 \wedge e^5 + e^1 \wedge e^2 \wedge e^3) ~,
\\
f_3^{(3)} = -\mathrm{T}^{-\frac12}d (e^2 \wedge e^5) = 2\mathrm{T}^{-\frac12} (
e^0 \wedge e^1 \wedge e^5
- e^2 \wedge e^3 \wedge e^4) ~,
\\
f_3^{(4)} = 2\mathrm{T}^{-\frac12} (e^0 \wedge e^1 \wedge e^2 + e^3 \wedge e^4 \wedge e^5) ~.
\end{gathered}\end{equation}
At this point let us emphasise that since this ansatz for the elliptic background is built out of left-invariant $\grp{SL}(2;\Real)$ and $\grp{SU}(2)$ Maurer-Cartan forms, it will have a left-acting $\grp{SL}(2;\Real)_L \times \grp{SU}(2)_L$ symmetry.
The corresponding Killing vectors are presented in \appref{app:c1}.

For this background to solve the supergravity equations of motion the coefficients of the fluxes are required to satisfy
\begin{equation}\begin{aligned}\label{eq:solution}
||\mathbf{x}^{(1)}||^2 & = \frac{\mathrm{T}(\alpha_2 + \alpha_3 - \alpha_1)}{\alpha_2 \alpha_3} - ||\mathbf{x}^{(4)}||^2 ~,
& \qquad
\mathbf{x}^{(1)} \cdot \mathbf{x}^{(4)} & = \mathbf{x}^{(2)} \cdot \mathbf{x}^{(3)} ~,
\\
||\mathbf{x}^{(2)}||^2 & = \frac{\mathrm{T}(\alpha_2 - \alpha_1 - \alpha_3)}{\alpha_1 \alpha_3} + ||\mathbf{x}^{(4)}||^2 ~,
& \qquad
\mathbf{x}^{(2)} \cdot \mathbf{x}^{(4)} & = -\mathbf{x}^{(1)} \cdot \mathbf{x}^{(3)} ~,
\\
||\mathbf{x}^{(3)}||^2 & = \frac{\mathrm{T}(\alpha_2 + \alpha_1 - \alpha_3)}{\alpha_1 \alpha_2} - ||\mathbf{x}^{(4)}||^2 ~,
& \qquad
\mathbf{x}^{(3)} \cdot \mathbf{x}^{(4)} & = \mathbf{x}^{(1)} \cdot \mathbf{x}^{(2)} ~.
\end{aligned}
\end{equation}
The $\grp{O}(4)_{\mathrm{T-d}}$ invariance of these equations, with the index $i=1,2,3,4$ transforming in the vector representation, follows from the T-duality group of the four-torus.
This is different to the formal $\grp{O}(4)_{\To^4}$ symmetry of the four-torus acting on $\field^m$, $m=6,7,8,9$, also in the vector representation.
However, they have a common $\grp{SO}(3)$ subgroup.
The $\grp{SO}(3)$ subgroup of $\grp{O}(4)_{\mathrm{T-d}}$ that rotates $F_{3,i}$, $i=1,2,3$, can also be understood as rotating the three self-dual 2-forms $J_{2,i}$.
On the other hand, the three self-dual 2-forms transform in a representation of $\grp{SO}(4)_{\To^4}$.
More precisely, using the isomorphism $\grp{SO}(4)_{\To^4} \cong (\grp{SU}(2) \times \grp{SU}(2))/\Integer_2$, they transform in the $\mathbf{3}$ of one of the two $\grp{SU}(2)$ factors.

\medskip

The rational limit of the supergravity background~\eqref{eq:ellipticmetricAdS3xS3,-,eq:fluxbasis} is given by setting
\begin{equation}
\alpha_1 = \alpha_2 = \alpha_3 = \lambda^2 \mathrm{T} ~,
\end{equation}
where $\lambda$ is a real parameter, which we take to be strictly positive, controlling the relative size of $\AdS_3 \times \Sp^3$ and $\To^4$.
From the supergravity conditions~\eqref{eq:solution} we find $||\mathbf{x}^{(1)}||^2 = ||\mathbf{x}^{(3)}||^2 = - ||\mathbf{x}^{(2)}||^2$, hence for a real solution we must have
\begin{equation}
||\mathbf{x}^{(4)}||^2 = \frac{1}{\lambda^2} ~, \qquad \mathbf{x}^{(1)} = \mathbf{x}^{(2)} = \mathbf{x}^{(3)} =0~.
\end{equation}
This background only depends on $f_3^{(4)}$, which is proportional to the volume forms of $\AdS_3$ and $\Sp^3$, and is well-known to have 16 supersymmetries~\cite{Maldacena:1997re}.
These supersymmetries are precisely given by the 6d supersymmetries with Killing spinors $\epsilon \in \mathfrak{s}_6$.
If $\mathbf{x}^{(4)}_4 = \lambda^{-1}$ then the background only has a 3-form flux and we find $\bigcap_{M,N}\ker F_\ind{MN} = \mathfrak{s}$ and $\ker \sigma_1\slashed{F}_3 = \mathfrak{s}_6$.
On the other hand if $\mathbf{x}^{(4)}_4 = 0$ then the background only has a 5-form flux and $\bigcap_{M,N}\ker F_\ind{MN} = \mathfrak{s}_6$ and $\ker \sigma_1\slashed{F}_3 = \mathfrak{s}$.
With non-vanishing 3-form and 5-form fluxes both kernels are $\mathfrak{s}_6$.
As explained above, from this point forward, we would like to ask which of these 6d supersymmetries persist in the trigonometric and elliptic backgrounds.
Therefore we are interested in the space of those $\epsilon \in \mathfrak{s}_6$ that satisfy $F_{\mu\nu}\epsilon = 0$ for all $\mu$ and $\nu$, with the number of 6d supersymmetries then given by eq.~\eqref{eq:6dsusycount}.

\subsection{The pp-wave limit}

To gain insight into the supersymmetries of the full elliptic $\AdS_3 \times \Sp^3 \times \To^4$ supergravity background, we consider the pp-wave limit.
In particular, we will focus on the 6d supersymmetries, i.e. restricting to $\epsilon \in \mathfrak{s}_6$, and investigate for what choice of R-R fluxes this is enhanced.

The pp-wave limit is given by introducing
\begin{equation}
\label{eq:lc-ppwave}
X^+ = \frac{1}{2}(T+\Phi) ~, \qquad
X^- = - T + \Phi ~,
\end{equation}
rescaling
\begin{equation}
\begin{gathered}
X^+ \to X^+ ~, \qquad X^- \to \varepsilon^2 X^- ~,
\\
(U,V,X,Y,\field_{6,7,8,9}) \to \varepsilon (U,V,X,Y,\Psi_{6,7,8,9}) ~, \qquad (\alpha_{1,2,3},\mathrm{T}) \to \varepsilon^{-2} (\alpha_{1,2,3},\mathrm{T}) ~,
\end{gathered}
\end{equation}
and taking $\varepsilon \to 0$.
We also define the Lorentz-rotated vielbein
\begin{equation}
\tilde{e}^0 = \tfrac12 \varepsilon (e^0 + e^3) + \tfrac12 \varepsilon^{-1} (e^0 - e^3) ~,
\qquad
\tilde{e}^3 = \tfrac12 \varepsilon (e^0 + e^3) - \tfrac12 \varepsilon^{-1} (e^0 - e^3) ~,
\end{equation}
which satisfies $- (e^0)^2 + (e^3)^2 = -(\tilde{e}^0)^2 + (\tilde{e}^3)^2$.
Using that $T = X^+ - \frac{1}{2} X^-$ and $\Phi = X^+ + \frac{1}{2}X^-$, we find that the $\varepsilon \to 0$ limit of the vielbein is
\begin{equation}
\begin{split}
\tilde{e}^0 & = \sqrt{\alpha_2} (dX^+ - \tfrac12 dX^- + (U^2 +V^2 + X^2 + Y^2) dX^+ + V dU + Y dX) ~,
\\
e^1 &= \sqrt{\alpha_1} (2 U dX^+ - dV)~, \\
e^2 &= \sqrt{\alpha_3}(- 2 V dX^+ - dU)~, \\
\tilde{e}^3 & = \sqrt{\alpha_2} (dX^+ + \tfrac12 dX^- - (U^2 +V^2 + X^2 + Y^2) dX^+ - V dU - Y dX) ~,
\\
e^4 &= \sqrt{\alpha_1} (- 2 X dX^+ + dY)~, \\
e^5 &= \sqrt{\alpha_3} (2Y dX^+ + dX)~, \\
e^6 &= \sqrt{\mathrm{T}} d\field_6~, \quad e^7 = \sqrt{\mathrm{T}} d\Psi_7~, \quad e^8 = \sqrt{\mathrm{T}} d\Psi_8~,\quad e^9 = \sqrt{\mathrm{T}} d\Psi_9~.
\end{split}
\end{equation}
To determine the limit of the auxiliary fluxes, we use that
\begin{equation}
\lim_{\varepsilon\to0} \varepsilon e^0 = \lim_{\varepsilon\to0} \varepsilon e^3 = \tfrac12 (\tilde{e}^0 + \tilde{e}^3) = \sqrt{\alpha_2}dX^+ \equiv e^+ ~,
\end{equation}
which immediately gives us that
\begin{equation}\begin{gathered}\label{eq:fluxbasispp}
f_3^{(1)} = f_3^{(3)} = 2 \mathrm{T}^{-\frac12}
e^+ \wedge (e^1 \wedge e^5
+ e^2 \wedge e^4) ~,
\\
f_3^{(2)} = f_3^{(4)} = 2 \mathrm{T}^{-\frac12}
e^+ \wedge ( e^1 \wedge e^2
+ e^4 \wedge e^5) ~.
\end{gathered}\end{equation}
Since we have $f_3^{(1)} = f_3^{(3)}$ and $f_3^{(2)} = f_3^{(4)}$, in the pp-wave limit the background only depends on
\begin{equation}
\mathbf{y}_+ = \mathbf{x}^{(2)} + \mathbf{x}^{(4)} ~, \qquad
\mathbf{z}_+ = \mathbf{x}^{(1)} + \mathbf{x}^{(3)} ~.
\end{equation}
For a supergravity solution these vectors are constrained as
\begin{equation}\label{eq:sugra}
\ys+\zs = \mathrm{T}\frac{\alpha_2^2-(\alpha_1-\alpha_3)^2}{\alpha_1\alpha_2\alpha_3} ~,
\end{equation}
which, as expected, is implied by the supergravity conditions~\eqref{eq:solution}.

This pp-wave background has at least 16 supersymmetries, 8 of which are 6d supersymmetries.
\unskip\footnote{These supersymmetries are always present due to the pp-wave limit~\cite{Cvetic:2002hi,Cvetic:2002nh} and we do not expect them to generically survive away from the limit.}
In particular, $\sigma_1 \slashed{F}_3\epsilon = F_\ind{MN} \epsilon = 0$ for all $\epsilon \in \mathfrak{s}_{+}=\mathfrak{s}_{6+} \oplus \mathfrak{\bar{s}}_{6+}$.
Therefore, the non-trivial supersymmetries correspond to Killing spinors $\epsilon \in\mathfrak{s}_-$.
Computing $\det F_{\mu\nu}$, we find that the number of 6d supersymmetries is enhanced from 8 to 12 (or 16 for special choices of parameters) when the vectors $\mathbf{y}_+$ and $\mathbf{z}_+$ additionally satisfy
\begin{equation}
\label{eq:susy1}
\Big(\ys - \zs - \frac{\mathrm{T}\big((\alpha_2-\alpha_3)^2 - 3\alpha_1^2 + 2\alpha_1(\alpha_2+\alpha_3)\big)}{\alpha_1\alpha_2\alpha_3}\Big)^2 + 4 (\yz)^2 + \frac{16 \mathrm{T} (\alpha_1-\alpha_2)}{\alpha_2\alpha_3}\zs = 0 ~,
\end{equation}
\begin{equation}
\label{eq:susy2}
\Big(\ys - \zs - \frac{\mathrm{T}\big((\alpha_2-\alpha_1)^2 - 3\alpha_3^2 + 2\alpha_3(\alpha_2+\alpha_1)\big)}{\alpha_1\alpha_2\alpha_3}\Big)^2 + 4 (\yz)^2 + \frac{16 \mathrm{T} (\alpha_3-\alpha_2)}{\alpha_1\alpha_2}\zs = 0 ~.
\end{equation}
Here we restrict ourselves to real R-R fluxes, hence the vectors $\mathbf{y}_+$ and $\mathbf{z}_+$ are valued in $\Real^4$.
This implies that the first two terms in both the equations~\eqref{eq:susy1} and~\eqref{eq:susy2} are non-negative, and the third term must be non-positive for a solution to exist.
Recalling that $\alpha_1,\alpha_2,\alpha_3> 0$, this imposes the following restriction on parameter space
\begin{equation}\label{eq:reality}
0<\alpha_1\leq\alpha_2 ~, \qquad
0<\alpha_3\leq\alpha_2 ~.
\end{equation}
Note that these inequalities also imply that the right-hand side of eq.~\eqref{eq:sugra} is non-negative, ensuring it can admit a solution as well.
The same restriction on parameter space also appears from demanding the quadratic dispersion relation for the bosonic excitations in light-cone gauge leads to a branch with positive energy~\cite{Hoare:2023zti}.

Note that the three equations~\eqref{eq:sugra},~\eqref{eq:susy1} and \eqref{eq:susy2} are not independent.
The difference of eqs.~\eqref{eq:susy1} and~\eqref{eq:susy2} is proportional to eq.~\eqref{eq:sugra}.
Taking a linear combination of eqs.~\eqref{eq:susy1},~\eqref{eq:susy2} and~\eqref{eq:sugra} we find that we can take the other independent equation to be
\begin{equation}\label{eq:susy}
\Big(\ys-\zs - \frac{\mathrm{T}(\alpha_1-\alpha_2+\alpha_3)^2}{\alpha_1\alpha_2\alpha_3}\Big)^2 + 4(\yz)^2 = \frac{16\mathrm{T}^2(\alpha_2-\alpha_1)(\alpha_2-\alpha_3)}{\alpha_1\alpha_2^2\alpha_3} ~.
\end{equation}
Introducing the dimensionless parameters
\begin{equation}\label{eq:gamalp}
\gamma_1 = \frac{\sqrt{\mathrm{T}}\alpha_2}{\sqrt{\alpha_1\alpha_2\alpha_3}} ~,
\qquad
\gamma_2 = \frac{\sqrt{\mathrm{T}}(\alpha_1 - \alpha_3)}{\sqrt{\alpha_1\alpha_2\alpha_3}} ~,
\qquad
\gamma_3 = \frac{\sqrt{\mathrm{T}}(\alpha_1-\alpha_2+\alpha_3)}{\sqrt{\alpha_1\alpha_2\alpha_3}} ~,
\qquad
\gamma_\pm = \gamma_1 \pm \gamma_2 ~,
\end{equation}
the allowed region of parameter space~\eqref{eq:reality} becomes
\begin{equation}\begin{aligned}\label{eq:realitygamma}
\gamma_\pm > - \gamma_3 ~, \qquad \gamma_3 \leq 0 ~.
\\
\gamma_\pm \geq \gamma_3 ~,\qquad \gamma_3 > 0 ~,
\end{aligned}\end{equation}
and the two equations~\eqref{eq:sugra} and~\eqref{eq:susy} simplify to
\begin{align}
\label{eq:sugraup}
\ys+\zs & = \gamma_+ \gamma_- ~,
\\
\label{eq:susyup}
\big(\ys - \zs - \gamma_3^2\big)^2 + 4 (\yz)^2 & = (\gamma_+^2 - \gamma_3^2)(\gamma_-^2 - \gamma_3^2) ~.
\end{align}
As anticipated, the right-hand sides of both are positive in the region~\eqref{eq:realitygamma}.
Therefore, eqs.~\eqref{eq:sugraup} and~\eqref{eq:susyup} can be solved parametrically in terms of an angle $\phi \in \interval{-\pi,\pi}$ as
\begin{equation}\begin{split}\label{eq:eqsol}
\ys & = \frac{1}{2}(\gamma_+\gamma_- + \gamma_3^2 + \sqrt{(\gamma_+^2 -\gamma_3^2)(\gamma_-^2 - \gamma_3^2)}\cos\phi) ~,
\\
\zs & = \frac{1}{2}(\gamma_+\gamma_- - \gamma_3^2 - \sqrt{(\gamma_+^2 -\gamma_3^2)(\gamma_-^2 - \gamma_3^2)} \cos\phi) ~,
\\
\yz & = \frac{1}{2}\sqrt{(\gamma_+^2 -\gamma_3^2)(\gamma_-^2 - \gamma_3^2)} \sin \phi ~,
\end{split}\end{equation}
where, without loss of generality, we take the positive branch of the square root.

In the rational limit, $\alpha_1 = \alpha_2 = \alpha_3 = \lambda^2\mathrm{T}$, we have $\gamma_1 = \gamma_3 = \gamma_\pm = \lambda^{-1}$ and $\gamma_2 = 0$, and the solution~\eqref{eq:eqsol} simplifies to
\begin{equation}
\ys = \lambda^{-2} ~, \qquad \zs = \yz = 0 ~,
\end{equation}
hence $\mathbf{z}_+ = 0$ and we recover the familiar $\AdS_3 \times \Sp^3$ pp-wave background with 16 6d supersymmetries.
In a particular trigonometric limit $\alpha_1 = \alpha_3 = \lambda^2\mathrm{T}(1+\kappa^2)^{-1}$, $\alpha_2 = \lambda^2\mathrm{T}$, where $\kappa$ is a real parameter, we have $\gamma_1 = \gamma_\pm = \lambda^{-1}(1+\kappa^2)$, $\gamma_2 = 0$ and $\gamma_3 = \lambda^{-1}( 1-\kappa^2)$.
The solution~\eqref{eq:eqsol} becomes
\begin{equation}\begin{split}\label{eq:eqsoltrig}
\ys & = \frac{1+\kappa^4 + 2 \kappa^2\cos\phi}{\lambda^2} ~, \qquad
\zs = \frac{2\kappa^2(1 - \cos\phi)}{\lambda^2} ~, \qquad
\yz = \frac{2\lambda^{-2}\kappa^2\sin\phi}{\lambda^2} ~.
\end{split}\end{equation}
For $\phi = 0$ we again have $\mathbf{z}_+ = 0$ and we recover the pp-wave background discussed in~\cite{Hoare:2022asa}, which has 16 6d supersymmetries.
For $\phi \neq 0$ the background has 12 6d supersymmetries.
In the elliptic case with $\alpha_{1,2,3}$ unconstrained apart from the conditions~\eqref{eq:reality}, we also find that the pp-wave background has 12 6d supersymmetries, but now for all values of $\phi$ including $\phi = 0$.
\unskip\footnote{
Note that away from the aforementioned trigonometric limit it is not generically possible to choose $\phi$ such that $\mathbf{z}_+ = 0$.
However, if we set $\gamma_3 = 0$, i.e. $\alpha_2 = \alpha_1 + \alpha_3$, we again find that for $\phi = 0$ we have $\mathbf{z}_+ = 0$ and the background has 16 6d supersymmetries.
For both these cases, the trigonometric limit and $\gamma_3 = 0$, the enhancement to 16 6d supersymmetries in the pp-wave limit does not survive in the full deformed background.}

We can alternatively take the pp-wave limit by setting
\begin{equation}
X^+ = \frac{1}{2}(T-\Phi) ~, \qquad X^- = - T - \Phi ~.
\end{equation}
Doing so we find the same set of equations except now with $\mathbf{y}_+ \to \mathbf{y}_-$ and $\mathbf{z}_+ \to \mathbf{z}_-$, where
\begin{equation}
\mathbf{y}_- = \mathbf{x}^{(2)} - \mathbf{x}^{(4)} ~, \qquad
\mathbf{z}_- = \mathbf{x}^{(1)} - \mathbf{x}^{(3)} ~.
\end{equation}
It follows that to have enhanced, i.e.~more than 8, 6d supersymmetries, we require these vectors to be constrained as
\begin{align}
\label{eq:sugraupt}
\tys+\tzs & = \gamma_+ \gamma_- ~,
\\
\label{eq:susyupt}
\big(\tys - \tzs - \gamma_3^2\big)^2 + 4 (\tytz)^2 & = (\gamma_+^2 - \gamma_3^2)(\gamma_-^2 - \gamma_3^2) ~,
\end{align}
which we can again solve parametrically in terms of an angle $\psi \in \interval{-\pi,\pi}$ as
\begin{equation}\begin{split}\label{eq:eqsolt}
\tys & = \frac{1}{2}(\gamma_+\gamma_- + \gamma_3^2 - \sqrt{(\gamma_+^2 -\gamma_3^2)(\gamma_-^2 - \gamma_3^2)}\cos\psi) ~,
\\
\tzs & = \frac{1}{2}(\gamma_+\gamma_- - \gamma_3^2 + \sqrt{(\gamma_+^2 -\gamma_3^2)(\gamma_-^2 - \gamma_3^2)} \cos\psi) ~,
\\
\tytz & = - \frac{1}{2}\sqrt{(\gamma_+^2 -\gamma_3^2)(\gamma_-^2 - \gamma_3^2)} \sin \psi ~.
\end{split}\end{equation}
The analysis of the number of supersymmetries is then identical to the original pp-wave limit.

\subsection{Supersymmetries of the deformed background}

Thus far we have discussed the 6d supersymmetries of the undeformed background and the pp-wave limit of the deformed background.
We now turn to the 6d supersymmetries of the full elliptic background, as well as its trigonometric limits.
To count the number of supersymmetries it is more straightforward to work with $F^{\ind{AB}} = e^{\ind{A}\ind{M}}e^{\ind{B}\ind{N}}F_\ind{MN}$.
These matrices, 12 of which are non-vanishing, are constant and turn out to be block diagonal, with two $8 \times 8$ blocks.
Even so, the explicit expressions for $\det F^{\ind{AB}}$ are long, hence we will not reproduce them here.
Importantly, upon imposing the supergravity conditions~\eqref{eq:solution} only three of these determinants are independent.

To find a solution to the equations $\det F^{\ind{AB}} = 0$, we start by noting that the pp-wave limits are limits of the background, not of the parameters.
Therefore, the pp-wave supersymmetry conditions~\eqref{eq:susyup} and \eqref{eq:susyupt} will still be required for supersymmetry of the full deformed background.
Together with the supergravity conditions~\eqref{eq:solution}, which we recall imply the pp-wave supergravity conditions~\eqref{eq:sugraup} and \eqref{eq:sugraupt}, the 10 inner products of the vectors $\mathbf{y}_\pm$ and $\mathbf{z}_\pm$ are parametrised in terms of the two angles $\phi$ and $\psi$ as
\begin{equation}\begin{aligned}\label{eq:solngam}
&
\begin{aligned}
\ys & = \frac{1}{2}(\gamma_+\gamma_- + \gamma_3^2 + \sqrt{(\gamma_+^2 -\gamma_3^2)(\gamma_-^2 - \gamma_3^2)}\cos\phi) ~,
\\
\zs & = \frac{1}{2}(\gamma_+\gamma_- - \gamma_3^2 - \sqrt{(\gamma_+^2 -\gamma_3^2)(\gamma_-^2 - \gamma_3^2)} \cos\phi) ~,
\end{aligned}
& \quad
& \yz = \frac{1}{2}\sqrt{(\gamma_+^2 -\gamma_3^2)(\gamma_-^2 - \gamma_3^2)} \sin \phi ~,
\\
&\begin{aligned}
\tys & = \frac{1}{2}(\gamma_+\gamma_- + \gamma_3^2 - \sqrt{(\gamma_+^2 -\gamma_3^2)(\gamma_-^2 - \gamma_3^2)}\cos\psi) ~,
\\
\tzs & =\frac{1}{2}(\gamma_+\gamma_- - \gamma_3^2 + \sqrt{(\gamma_+^2 -\gamma_3^2)(\gamma_-^2 - \gamma_3^2)} \cos\psi) ~,
\end{aligned}
& \quad
& \tytz = - \frac{1}{2}\sqrt{(\gamma_+^2 -\gamma_3^2)(\gamma_-^2 - \gamma_3^2)} \sin \psi ~,
\\
&\yyt = - \gamma_1\gamma_3 ~,
&
&\zzt = - \gamma_2 \gamma_3 ~,
\\
&\yzt = 0 ~,
&
&\ytz = 0 ~.
\end{aligned}\end{equation}
Substituting this ansatz into $\det F^\ind{AB}$ we find that the determinants all vanish when the two angles $\phi$ and $\psi$ are related as
\begin{equation}\begin{split}\label{eq:anglegamma}
\cos\psi = \frac{4\gamma_1^2\gamma_2^2\cos\phi - (\gamma_1^2+\gamma_2^2-\gamma_3^2)\sqrt{(\gamma_+^2 -\gamma_3^2)(\gamma_-^2 - \gamma_3^2)} \sin^2\phi}{4\gamma_1^2\gamma_2^2 +(\gamma_+^2 -\gamma_3^2)(\gamma_-^2 - \gamma_3^2)\sin^2\phi} ~,
\\
\sin\psi = \frac{2\gamma_1\gamma_2(\gamma_1^2+\gamma_2^2-\gamma_3^2 +\sqrt{(\gamma_+^2 -\gamma_3^2)(\gamma_-^2 - \gamma_3^2)}\cos\phi)\sin\phi}{4\gamma_1^2\gamma_2^2 + (\gamma_+^2 -\gamma_3^2)(\gamma_-^2 - \gamma_3^2)\sin^2\phi} ~.
\end{split}\end{equation}
Since we have fixed both $\phi\in \interval{-\pi,\pi}$ and $\psi\in \interval{-\pi,\pi}$, it follows that the cosine and sine of the half angle are
\begin{equation}\begin{split}\label{eq:halfanglegamma}
\cos\frac{\psi}{2} & = \frac{\sqrt{1-\frac{\sqrt{(\gamma_+^2 -\gamma_3^2)(\gamma_-^2 - \gamma_3^2)}}{\gamma_1^2+\gamma_2^2-\gamma_3^2}}\cos\frac\phi2}{\sqrt{1-\frac{\sqrt{(\gamma_+^2 -\gamma_3^2)(\gamma_-^2 - \gamma_3^2)}}{\gamma_1^2+\gamma_2^2-\gamma_3^2}\cos\phi}} ~,
\qquad
\sin\frac{\psi}{2} =\operatorname{sgn}\big(\frac{\gamma_1\gamma_2}{\gamma_1^2+\gamma_2^2-\gamma_3^2}\big) \frac{\sqrt{1+\frac{\sqrt{(\gamma_+^2 -\gamma_3^2)(\gamma_-^2 - \gamma_3^2)}}{\gamma_1^2+\gamma_2^2-\gamma_3^2}}\sin\frac\phi2}{\sqrt{1-\frac{\sqrt{(\gamma_+^2 -\gamma_3^2)(\gamma_-^2 - \gamma_3^2)}}{\gamma_1^2+\gamma_2^2-\gamma_3^2}\cos\phi}} ~.
\end{split}\end{equation}
As shown in \appref{app:Gram}, the conditions~\eqref{eq:solngam} and the relation between angles~\eqref{eq:anglegamma} imply that the vectors $\mathbf{y}_\pm,\mathbf{z}_\pm\in\Real^4$ lie in a 2-plane.
As a consequence, on the supersymmetric locus~\eqref{eq:solngam,eq:anglegamma} there is a residual $\grp{O}(2)_{\mathrm{T-d}} \subset \grp{O}(4)_{\mathrm{T-d}}$ symmetry that leaves the background R-R-fluxes invariant.
This rotation will reappear as a symmetry of the tree-level S-matrix in \secref{sec:lcgf}.

\begin{figure}
\centering
\begin{tikzpicture}[scale=0.75,baseline=0pt]
\fill [lightgray] (0,0)--(0,5)--(5,5)--(5,0);
\draw[->, very thick] (5,0)--(7,0) node[right] {$\alpha_1$};
\draw[very thick] (-1,0)--(0,0);
\draw[very thick,dashed] (0,0)--(5,0);
\draw[->, very thick] (0,5)--(0,7) node[above] {$\alpha_3$};
\draw[very thick] (0,-1)--(0,0);
\draw[very thick,dashed] (0,0)--(0,5);
\draw[-,very thick] (5,5)--(0,5) node[left] {$\alpha_2> 0$};
\draw[-,very thick] (5,5)--(5,0) node[below] {$\alpha_2> 0$};
\draw[-,dashed] (0,5)--(5,0);
\node at (3.5,3.5) {$\gamma_3 > 0$};
\node at (1.5,1.5) {$\gamma_3 \leq 0$};
\draw[-,white] (2,-8)--(0,-8) node[left] {\color{black}$-\gamma_3\geq0$};
\draw[-,white] (2,-7)--(2,-10) node[below] {\color{black}$-\gamma_3\geq0$};
\fill [lightgray] (2,-8)--(7,-8)--(7,-3)--(2,-3);
\draw[->, very thick] (-1,-10)--(7,-10) node[right] {$\gamma_+$};
\draw[->, very thick] (0,-11)--(0,-3) node[above] {$\gamma_-$};
\draw[-,very thick,dashed] (7,-8)--(2,-8);
\draw[-,very thick,dashed] (2,-3)--(2,-8);
\end{tikzpicture}
\hspace{20pt}
\begin{tikzpicture}[scale=0.75,baseline=0pt]
\path [fill=lightgray,draw=black,very thick,dashed] (3,0)--(0,3)--(3,6);
\path [fill=lightgray,draw=black,very thick] (3,6)--(3,0);
\path [fill=lightgray,draw=black,very thick,dashed] (0,3)--(-3,6)--(-3,0)--(0,3);
\draw[->,very thick] (-4,3)--(4,3) node[right] {$s$};
\draw[->,very thick] (0,-1)--(0,7) node[above] {$t$};
\node at (3.2,2.7) {$1$};
\node at (-3.4,2.7) {$-1$};
\node at (-0.4,6) {$\hphantom{-}1$};
\node at (-0.4,0) {$-1$};
\node at (2,3.5) {$\gamma_3 > 0$};
\node at (-2,3.5) {$\gamma_3 < 0$};
\draw[-,white] (-1,-8)--(-3,-8) node[left] {\color{black}$\gamma_3>0$};
\draw[-,white] (-1,-7)--(-1,-10) node[below] {\color{black}$\gamma_3>0$};
\fill [lightgray] (-1,-8)--(4,-8)--(4,-3)--(-1,-3);
\draw[->, very thick] (-4,-10)--(4,-10) node[right] {$\gamma_+$};
\draw[->, very thick] (-3,-11)--(-3,-3) node[above] {$\gamma_-$};
\draw[-,very thick] (4,-8)--(-1,-8);
\draw[-,very thick] (-1,-3)--(-1,-8);
\end{tikzpicture}
\caption{The regions of parameter space, defined in eqs.~(\ref{eq:reality}),~(\ref{eq:realitygamma}) and~(\ref{eq:realityst}) respectively, following from the requirement that the supersymmetric background of the elliptic deformation of $\AdS_3\times\Sp^3\times\To^4$ has real R-R fluxes.}\label{fig:regions}
\end{figure}
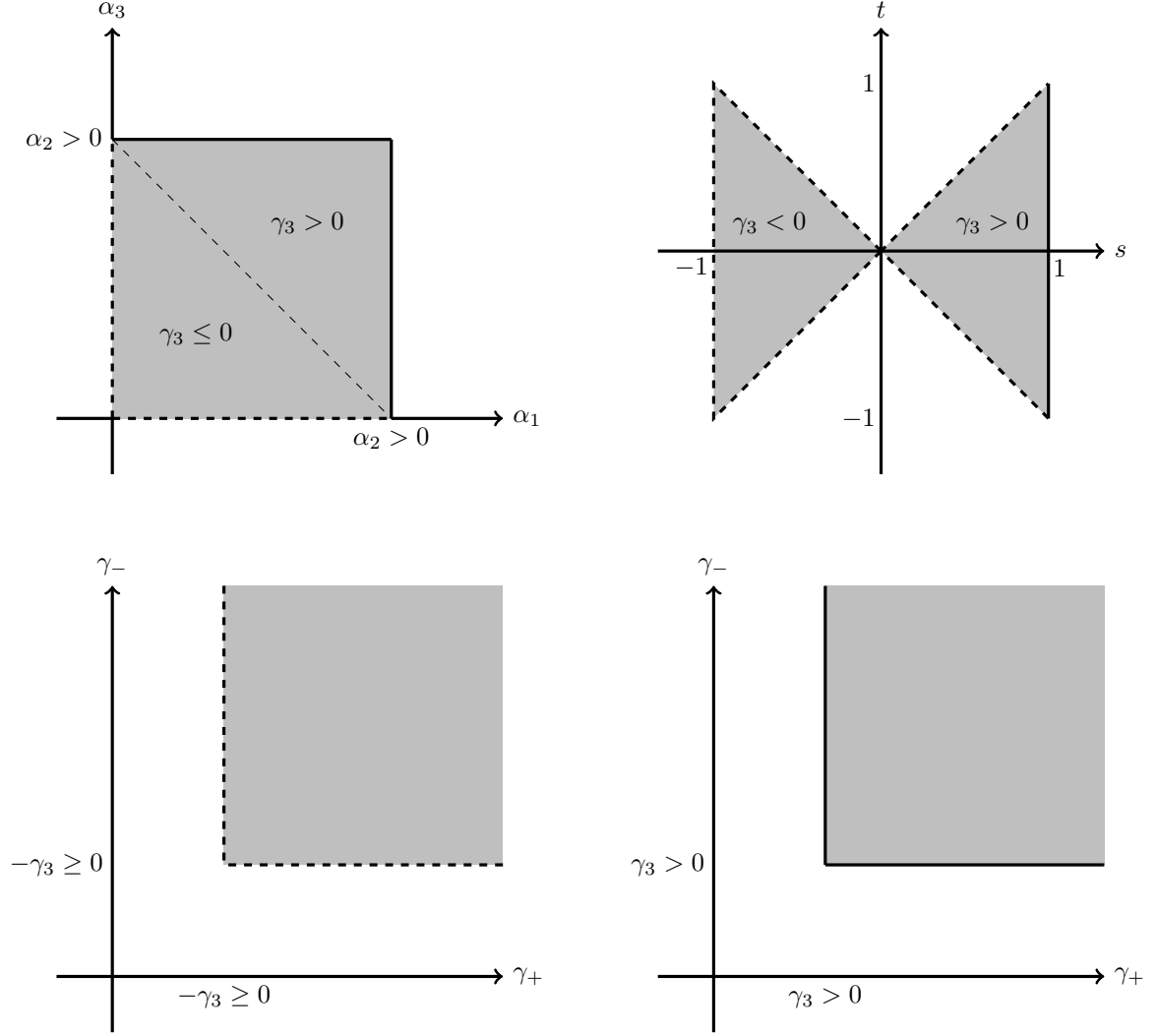

To count the number of supersymmetries and compute the Killing spinors, it is convenient to introduce the parameters $s$ and $t$ defined as
\begin{equation}
\gamma_1 = \frac{s(1-t^2)\gamma_3}{s^2-t^2} ~,
\qquad
\gamma_2 = \frac{t(1-s^2)\gamma_3}{s^2-t^2} ~,
\qquad
\gamma_\pm = \frac{1\mp st}{s\mp t} \gamma_3 ~.
\end{equation}
In terms of these parameters the allowed region of parameter space, defined in eq.~\eqref{eq:realitygamma}, is covered by
\unskip\footnote{Note that the map from $(s,t)$ to $(\gamma_+,\gamma_-)$ is two-to-one.
Without loss of generality, we have picked one of the two regions of the $(s,t)$ plane that covers the region of interest in the $(\gamma_+,\gamma_-)$ plane.
The choice we have made corresponds to the branch
\begin{equation*}
s=\frac{\gamma_3^2 + \gamma_+\gamma_- - \sqrt{(\gamma_+^2 -\gamma_3^2)(\gamma_-^2 - \gamma_3^2)}}{\gamma_3(\gamma_++\gamma_-)} ~,
\qquad
t=-\frac{\gamma_3^2 - \gamma_+\gamma_- + \sqrt{(\gamma_+^2 -\gamma_3^2)(\gamma_-^2 - \gamma_3^2)}}{\gamma_3(\gamma_+-\gamma_-)} ~.
\end{equation*}}
\begin{equation}\begin{aligned}\label{eq:realityst}
s^2 < 1 ~, \qquad t^2 < 1 ~, \qquad s^2 > t^2 ~, \qquad s < 0 ~, \qquad \gamma_3 < 0 ~,
\\
s^2 \leq 1 ~, \qquad t^2 \leq 1 ~, \qquad s^2 > t^2 ~, \qquad s > 0 ~, \qquad \gamma_3 > 0 ~,
\end{aligned}\end{equation}
with the $\gamma_3 \to 0$ limit requiring $s$ and $t$ to be simultaneously sent to zero.
This region is depicted in \figref{fig:regions}.
In this region we have
\begin{equation}
\frac{\sqrt{(\gamma_+^2 -\gamma_3^2)(\gamma_-^2 - \gamma_3^2)}}{\gamma_1^2+\gamma_2^2-\gamma_3^2} = \frac{s^2-t^2}{s^2+t^2} ~,
\end{equation}
hence, substituting into eq.~\eqref{eq:halfanglegamma}, we find
\begin{equation}
\cos\frac\psi2 = \operatorname{sgn} t \frac{t \cos\frac\phi2}{\sqrt{s^2 \sin^2\frac\phi2 + t^2 \cos^2\frac\phi2}} ~, \qquad
\sin\frac\psi2 = \operatorname{sgn} t \frac{s\sin\frac\phi2}{\sqrt{s^2 \sin^2\frac\phi2 + t^2 \cos^2\frac\phi2}} ~,
\end{equation}
from which we find the simple relation between angles
\begin{equation}\label{eq:angleconstraint}
\tan\frac\psi2 = \frac st \tan\frac\phi2 ~.
\end{equation}
Defining
\begin{equation}\label{eq:sigma1234}
\Sigma_1 = \sqrt{1+\tan^2\frac\phi2} ~, \qquad \Sigma_2 = \sqrt{t^2 + s^2\tan^2\frac\phi2} ~, \qquad \Sigma_3 = \sqrt{1+s^2\tan^2\frac\phi2} ~,
\qquad \Sigma_4 = \sqrt{t^2 + \tan^2\frac\phi2} ~,
\end{equation}
the supersymmetric locus~\eqref{eq:solngam,eq:anglegamma} is given by
\begin{equation}\begin{aligned}\label{eq:susylocus}
||\mathbf{y}_+||^2 & = \gamma_3^2\frac{1-t^2}{s^2 - t^2} \frac{\Sigma_3^2}{\Sigma_1^2} ~, \qquad
& ||\mathbf{y}_-||^2 & = \gamma_3^2\frac{s^2(1-t^2)}{s^2 - t^2}\frac{\Sigma_4^2}{\Sigma_2^2} ~,
\\
||\mathbf{z}_+||^2 & = \gamma_3^2\frac{1-s^2}{s^2 - t^2} \frac{\Sigma_4^2}{\Sigma_1^2} ~,
\qquad &
||\mathbf{z}_-||^2 & = \gamma_3^2\frac{t^2(1-s^2)}{s^2 - t^2}\frac{\Sigma_3^2}{\Sigma_2^2} ~,
\\
\mathbf{y}_+\cdot\mathbf{z}_+ & = \gamma_3^2\frac{(1-s^2)(1-t^2)}{s^2 - t^2}\frac{\tan\frac\phi2}{\Sigma_1^2} ~,
\qquad & \mathbf{y}_-\cdot\mathbf{z}_- & = -\gamma_3^2\frac{st(1-s^2)(1-t^2)}{s^2 - t^2} \frac{\tan\frac\phi2}{\Sigma_2^2} ~,
\\
\mathbf{y}_+\cdot\mathbf{y}_- & = - \frac{\gamma_3^2 s (1-t^2)}{s^2-t^2} ~,
\qquad &
\mathbf{z}_+\cdot\mathbf{z}_- & = - \frac{\gamma_3^2 t (1-s^2)}{s^2-t^2} ~,
\\
\mathbf{y}_+\cdot\mathbf{z}_- & = 0 ~,
\qquad &
\mathbf{y}_-\cdot\mathbf{z}_+ & = 0 ~,
\end{aligned}\end{equation}
where $\phi$ is now an additional free parameter on top of the original parameters $\alpha_1,\alpha_2,\alpha_3$ of the elliptic background.
Recalling that these constraints imply that the vectors $\mathbf{y}_\pm$ and $\mathbf{z}_\pm$ lie in a 2-plane, using the identity
\begin{equation}
\Sigma_3^2\Sigma_4^2 - \Sigma_1^2 \Sigma_2^2 = (1-s^2)(1-t^2)\tan^2\frac\phi2 ~,
\end{equation}
it is straightforward to see that the following vectors lie on the supersymmetric locus~\eqref{eq:susylocus}
\begin{equation}\begin{aligned}\label{eq:explicitsolution}
\mathbf{y}_+ & =\gamma_3\frac{\sqrt{1-t^2}}{\sqrt{s^2-t^2}}\frac{\Sigma_3}{\Sigma_1}\Big(0,0,0,1\Big) ~,
\\
\mathbf{z}_+ & = \gamma_3 \frac{\sqrt{1-s^2}}{\sqrt{s^2-t^2}}\frac{\Sigma_4}{\Sigma_1} \Big(-\frac{\Sigma_1\Sigma_2}{\Sigma_3\Sigma_4},0,0,\frac{\sqrt{1-s^2}\sqrt{1-t^2}\tan\frac\phi2}{\Sigma_3\Sigma_4} \Big) ~,
\\
\mathbf{y}_- & = \gamma_3 \frac{s\sqrt{1-t^2}}{\sqrt{s^2-t^2}}\frac{\Sigma_4}{\Sigma_2} \Big(-\frac{\sqrt{1-s^2}\sqrt{1-t^2} \tan\frac\phi2}{\Sigma_3\Sigma_4} , 0 , 0 , - \frac{\Sigma_1\Sigma_2}{\Sigma_3\Sigma_4}\Big) ~,
\\
\mathbf{z}_- & = \gamma_3 \frac{t\sqrt{1-s^2}}{\sqrt{s^2-t^2}}\frac{\Sigma_3}{\Sigma_2} \Big(1 , 0 , 0 , 0 \Big) ~.
\end{aligned}\end{equation}
Other solutions can then be found by acting with an $\grp{SO}(4)_{\mathrm{T-d}}$ transformation on this seed solution.
\unskip\footnote{In principle it is possible to act with an $\grp{O}(4)_{\mathrm{T-d}}$ transformation on this seed solution, however, since on the supersymmetric locus the four vectors $\mathbf{y}_\pm$ and $\mathbf{z}_\pm$ lie in a 2-plane, for any $\grp{O}(4)_{\mathrm{T-d}}$ transformation in the component not connected to the identity, there is an $\grp{SO}(4)_{\mathrm{T-d}}$ transformation that has the same effect.
Therefore, without loss of generality, we will restrict our attention to $\grp{SO}(4)_{\mathrm{T-d}}$ transformations.}
In \appref{app:gammamatrix} we specify the action of $\grp{SO}(4)_{\mathrm{T-d}}$ on $\alg{s}_6$.
This allows us to construct the Killing spinors for different choices of R-R fluxes related by an $\grp{SO}(4)_{\mathrm{T-d}}$ transformation.

Using the explicit seed solution~\eqref{eq:explicitsolution}, we find by construction that the elliptic $\AdS_3 \times \Sp^3 \times \To^4$ background on the supersymmetric locus~\eqref{eq:susylocus} has 8 6d supersymmetries.
Explicit expressions for the Killing spinors solving the dilatino and the gravitino equations \eqref{eq:dilatino} and \eqref{eq:gravitino}, for the supersymmetric choice of R-R fluxes~\eqref{eq:explicitsolution}, are given in \appref{app:c2} . They can be written $\epsilon_{a \alpha A}$, with indices $a=\pm$, $\alpha=\pm$ and $A=\pm$.

\subsection{Killing superalgebra}

Let us now work out the Lie superalgebra generated by the Killing vectors and Killing spinors.
Killing vectors $v_i$ and Killing spinors $\epsilon_j$ satisfy the (anti-)commutation relations~\cite{Figueroa-OFarrill:2004qhu}
\begin{equation}
\begin{aligned}
\com{v_i}{v_j} = \mathcal L_{v_i} v_j~, \qquad \com{v_i}{\epsilon_j} = \mathcal L_{v_i} \epsilon_j~, \qquad \anticom{\epsilon_i}{\epsilon_j} = \bar{\epsilon}_i \Gamma^\ind{M} \epsilon_j \partial_\ind{M}~,
\end{aligned}
\end{equation}
where the Kosmann derivative of a spinor field is given by
\begin{equation}
\label{eq:Kosmann}
\mathcal L_v \epsilon = v^\ind{M} \nabla_\ind{M} \epsilon +\frac{1}{4} \nabla_\ind{M} v_\ind{N} \Gamma^\ind{M} \Gamma^\ind{N} \epsilon~, \qquad \nabla_\ind{M} \epsilon = (\partial_\ind{M} - \frac{1}{4} \omega_\ind{M}^{\ind{AB}} \Gamma_\ind{A} \Gamma_\ind{B})\epsilon~,
\end{equation}
and we define, for a 64-dimensional spinor doublet,
\begin{equation}
\bar{\epsilon} = \epsilon^t (1_2 \otimes \mathcal C)~,
\end{equation}
where $\mathcal{C}$ is the conjugation matrix defined in eq.~\eqref{eq:conjugationmatrix}.
Applying these formulas to the Killing vectors for the undeformed $\grp{SL}(2;\Real) \times \grp{SU}(2)$ presented in \appref{app:c1} and the Killing spinors $\epsilon_{a \alpha A}$ presented in \appref{app:c2}, we find that the Killing spinor bilinears close as
\begin{equation}
\label{eq:comSS}
\begin{aligned}
\anticom{\epsilon_{++\pm}}{\epsilon_{+ - \mp}} &= \pm L_+~, &\quad \anticom{\epsilon_{-+\pm}}{ \epsilon_{-- \mp} } &= \mp L_-~, \\
\anticom{\epsilon_{++\pm}}{\epsilon_{-+\mp}} &= \mp J_+ ~, &\quad \anticom{\epsilon_{+-\pm}}{\epsilon_{--\mp}} & = \pm J_-~, \\
\anticom{\epsilon_{++\pm}}{\epsilon_{--\mp}} & = \mp (L_3-J_3) ~, &\quad
\anticom{\epsilon_{+-\pm}}{ \epsilon_{-+\mp}} & = \pm (L_3+J_3)~,
\end{aligned}
\end{equation}
where we have introduced the ladder Killing vectors
\begin{equation}
L_3 = -i\frac{v_3}{2}~, \qquad L_\pm = -\frac{1}{2}( v_1 \mp i v_2) ~, \qquad J_3 = -i \frac{v_6}{2} ~, \qquad J_\pm = -\frac{i}{2} (v_5 \pm i v_4)~,
\end{equation}
which satisfy the commutation relations
\begin{equation}
\label{eq:comVV}
[L_3, L_\pm] = \pm L_\pm ~, \qquad [L_+, L_-] = 2 L_3~, \qquad [ J_3, J_\pm] = \pm J_\pm~, \qquad \com{J_+}{J_-} = 2 J_3~.
\end{equation}
The Killing vectors $L_3$ and $J_3$ generate shifts in the isometric directions $T$ and $\Phi$ respectively. These coordinates are those used in \secref{sec:lcgf} to fix uniform light-cone gauge and compute the worldsheet S-matrix.
Let us now turn to the action of the ladder Killing vectors on the Killing spinors. Acting with the Kosmann derivative \eqref{eq:Kosmann}, we have that the spinors satisfy
\begin{equation}
\label{eq:comVS}
\begin{aligned}
\com{L_3}{ \epsilon_{\pm \alpha A}} &= \pm \frac{1}{2} \epsilon_{ \pm \alpha A}~, &\qquad \com{L_\pm}{ \epsilon_{\mp \alpha A}} &= \epsilon_{ \pm \alpha A}~, \\ \com{J_3}{ \epsilon_{a \pm A}} &= \pm \frac{1}{2} \epsilon_{a \pm A}~, &\qquad \com{J_\pm}{\epsilon_{a\mp A}} &= \epsilon_{a \pm A}~.
\end{aligned}
\end{equation}
From the relations \eqref{eq:comVV}, \eqref{eq:comVS} and \eqref{eq:comSS}, we recognise an $\alg{sl}(2|2)$ superalgebra. To identify the real form, we note that the ladder Killing vectors satisfy the reality conditions
\unskip\footnote{A vector $v=v^\ind{M} \partial_\ind{M}$ with real components $v^\ind{M} \in \Real$ satisfies the reality condition $v^\dagger = -v - \nabla_\ind{M} v^\ind{M}$, and the contribution proportional to the divergence vanishes for Killing vectors.}
\begin{equation}
L_3^\dagger = L_3~, \qquad L_\pm^\dagger = -L_\mp~, \qquad J_3^\dagger =J_3~, \qquad J_\pm^\dagger = J_\mp~.
\end{equation}
For the Killing spinors, we have that $ (1_2 \otimes \mathcal C (\Gamma^0)^{-1}) : \alg{s}_6 \rightarrow \alg{s}_6$, and a Majorana-Weyl spinor needs to satisfy $\epsilon^\dagger = \epsilon^t (1_2 \otimes \mathcal C (\Gamma^0)^{-1})$. While the Killing spinors $\epsilon_{a \alpha A}$ do not satisfy the Majorana-Weyl condition, the following linear combinations do:
\begin{equation}
\begin{aligned}
&\epsilon_{+++} + i \epsilon_{---}~, &\qquad &\epsilon_{++-} - i \epsilon_{--+}~, &\qquad &\epsilon_{+-+} - i \epsilon_{---}~, &\qquad &\epsilon_{+--} + i\epsilon_{-++}~, \\
&\epsilon_{-++} + i \epsilon_{+--}~, &\qquad &\epsilon_{-+-} - i \epsilon_{+-+}~, &\qquad &\epsilon_{--+} - i \epsilon_{++-}~, &\qquad &\epsilon_{---} +i\epsilon_{+++}~.
\end{aligned}
\end{equation}
Finally, let us note that there is a $\alg{u}(1)$ automorphism $U$ acting on the spinors as
\begin{equation}
\com{U}{\epsilon_{a \alpha \pm}} = \pm \frac{1}{2} \epsilon_{a \alpha \pm}~,
\end{equation}
and commuting with the bosonic generators $L_3, L_\pm, J_3, J_\pm$, which can be used to rescale $\epsilon_{a \alpha +} \rightarrow \rho^{-1} \epsilon_{a \alpha +}$ and $\epsilon_{a \alpha -} \rightarrow \rho \epsilon_{a \alpha -}$, while leaving the commutation relations unaltered.

To conclude, the elliptic $\AdS_3 \times \Sp^3 \times \To^4$ superstring presents a $\alg{psu}(1,1|2)$ superalgebra when the parameters defining the R-R fluxes lie on the supersymmetric locus~\eqref{eq:susylocus}.

\subsection{Trigonometric and rational limits}

Before computing the tree-level S-matrix in \secref{sec:lcgf}, let us briefly analyse various interesting limits and special cases.
Recalling that $\alpha_2$ controls the size of the time-like component of the vielbein~\eqref{eq:vielbein}, while $\alpha_1$ and $\alpha_3$ are associated to space-like components, in the following discussion of limits when $\alpha_2$ is distinguished we call this a time-like deformation, while if $\alpha_1$ or $\alpha_3$ are distinguished we say it is a space-like deformation.
Since $\alpha_1$ and $\alpha_3$ both control the size of space-like components, there is a symmetry if we interchange them, which from~\eqref{eq:solution} also requires interchanging $\mathbf{x}^{(1)}$ and $\mathbf{x}^{(3)}$.
Letting $\lambda > 0$, $\kappa \geq 0$ and $0\leq\tilde{\kappa} < 1$, there are a number of interesting limits that we can consider.

\paragraph{Time-like trigonometric limit.} The time-like trigonometric limit corresponds to taking $\alpha_1 = \alpha_3$.
To explore this limit and make contact with the literature, it is helpful to parametrise it in the following way ($\kappa \geq 0$, $\lambda > 0$):
\begin{equation}\begin{aligned}
& 0 < \alpha_1=\alpha_3 = \frac{\lambda^2 \mathrm{T}}{1+\kappa^2} \leq \alpha_2 = \lambda^2 \mathrm{T} ~,
\\
& \begin{aligned}
&\gamma_1 = \gamma_\pm = \frac{1+\kappa^2}{\lambda} ~, \quad && \gamma_2 = 0 ~,
\\
&s = \frac{1-\kappa^2}{1+\kappa^2} ~, \quad && t = 0 ~,
\end{aligned} \qquad \gamma_3 = \frac{1-\kappa^2}{\lambda} ~.
\end{aligned}\end{equation}
Since this limit involves setting $t=0$, it follows from the relation between angles $\phi$ and $\psi$~\eqref{eq:angleconstraint} that it is somewhat subtle.
In particular, we either need to take $\phi \to 0$ or $\psi \to \pi$ at the same time as taking $t \to 0$.
Correspondingly, the supersymmetric locus splits into two branches in the time-like trigonometric limit.

The first branch is given by taking $\psi \to \pi$ and can be simply found by setting $t = 0$ on the supersymmetric locus~\eqref{eq:susylocus}
\begin{equation}\begin{aligned}
||\mathbf{y}_+||^2 & = \frac{(1+\kappa^2)^2\cos^2\frac{\phi}{2} + (1-\kappa^2)^2 \sin^2\frac{\phi}{2}}{\lambda^2} ~, \qquad
& ||\mathbf{y}_-||^2 & = \frac{(1+\kappa^2)^2}{\lambda^2} ~,
\\
||\mathbf{z}_+||^2 & = \frac{4\kappa^2\sin^2\frac{\phi}{2} }{\lambda^2} ~,
\qquad &
||\mathbf{z}_-||^2 & = 0 ~,
\\
\mathbf{y}_+\cdot\mathbf{z}_+ & = \frac{4\kappa^2\sin\frac{\phi}{2}\cos\frac{\phi}{2}}{\lambda^2} ~,
\qquad & \mathbf{y}_-\cdot\mathbf{z}_- & = 0 ~,
\\
\mathbf{y}_+\cdot\mathbf{y}_- & = - \frac{1-\kappa^4}{\lambda^2} ~,
\qquad &
\mathbf{z}_+\cdot\mathbf{z}_- & = 0 ~,
\\
\mathbf{y}_+\cdot\mathbf{z}_- & = 0 ~,
\qquad &
\mathbf{y}_-\cdot\mathbf{z}_+ & = 0 ~.
\end{aligned}\end{equation}
If we additionally take $\phi \to 0$ we recover the supersymmetric locus of \cite{Hoare:2022asa}.
\unskip\footnote{
Explicitly we have
\begin{equation*}\begin{aligned}
||\mathbf{y}_+||^2 & = ||\mathbf{y}_-||^2 = \frac{(1+\kappa^2)^2}{\lambda^2} ~,\qquad
\mathbf{y}_+\cdot\mathbf{y}_- = - \frac{1-\kappa^4}{\lambda^2} ~,
\\
||\mathbf{z}_+||^2 & = ||\mathbf{z}_-||^2 =
\mathbf{y}_+\cdot\mathbf{z}_+ = \mathbf{y}_-\cdot\mathbf{z}_-
=
\mathbf{z}_+\cdot\mathbf{z}_-
=
\mathbf{y}_+\cdot\mathbf{z}_-
=
\mathbf{y}_-\cdot\mathbf{z}_+ = 0 ~.
\end{aligned}\end{equation*}
Noting that $\mathbf{y}_\pm = \lambda^{-1}(\mathbf{z}_{2}^{\mathrm{there}} \pm \mathbf{z}_{1}^{\mathrm{there}})$ we find that we recover the supersymmetric locus found in \cite{Hoare:2022asa}.
Our explicit seed solution~\eqref{eq:explicitsolution} becomes
\begin{equation*}\begin{aligned}
&\mathbf{y}_+ = \Big(0,0,0,\frac{1+\kappa^2}{\lambda} \Big) ~,
\qquad
&\mathbf{y}_- &= \Big(-\frac{2\kappa}{\lambda},0,0,-\frac{1-\kappa^2}{\lambda}\Big) ~, \qquad \mathbf{z}_+ = \mathbf{z}_- = 0 ~,
\\
&
\mathbf{z}_1^{\mathrm{there}} = (\kappa,0,0,1) ~, \qquad &\mathbf{z}_2^{\mathrm{there}} & = (-\kappa,0,0,\kappa^2) ~.
\end{aligned}\end{equation*}
assuming we take $\phi \to 0^+$ and act with the $\grp{O}(4)_{\mathrm{T-d}}$ transformation $\diag(1,1,1,\operatorname{sgn}(1-\kappa^2))$ to simplify the expressions.
In this way we recover the fluxes found from the unilateral inhomogeneous Yang-Baxter deformation of the $\AdS_3\times \Sp^3$ semi-symmetric space sigma model~\cite{Hoare:2014oua} up to an $\grp{O}(4)_{\mathrm{T-d}}$ transformation \cite{Seibold:2019dvf,Hoare:2022asa}.}

The second branch is given by taking $\phi \to 0$ and can be found by first eliminating $\phi$ in favour of $\psi$ on the supersymmetric locus~\eqref{eq:susylocus} and then setting $t=0$.
Doing so we find
\begin{equation}\begin{aligned}
||\mathbf{y}_+||^2 & = \frac{(1+\kappa^2)^2}{\lambda^2}~, \qquad
& ||\mathbf{y}_-||^2 & =\frac{(1+\kappa^2)^2\sin^2\frac{\psi}{2} + (1-\kappa^2)^2 \cos^2\frac{\psi}{2}}{\lambda^2}~,
\\
||\mathbf{z}_+||^2 & = 0 ~,
\qquad &
||\mathbf{z}_-||^2 & = \frac{4\kappa^2\cos^2\frac{\psi}{2} }{\lambda^2} ~,
\\
\mathbf{y}_+\cdot\mathbf{z}_+ & = 0 ~,
\qquad & \mathbf{y}_-\cdot\mathbf{z}_- & = -\frac{4\kappa^2\sin\frac{\psi}{2}\cos\frac{\psi}{2}}{\lambda^2} ~,
\\
\mathbf{y}_+\cdot\mathbf{y}_- & = - \frac{1-\kappa^4}{\lambda^2} ~,
\qquad &
\mathbf{z}_+\cdot\mathbf{z}_- & = 0 ~,
\\
\mathbf{y}_+\cdot\mathbf{z}_- & = 0 ~,
\qquad &
\mathbf{y}_-\cdot\mathbf{z}_+ & = 0 ~.
\end{aligned}\end{equation}
This branch is related to the first by the formal transformation
\begin{equation}
\mathbf{y}_+ \leftrightarrow \mathbf{y}_- ~, \qquad \mathbf{z}_+ \leftrightarrow \mathbf{z}_- ~, \qquad
\tan\frac{\psi}{2} \to \cot\frac{\phi}{2} ~.
\end{equation}
Taking $\psi \to \pi$, we find that the second branch coincides with the first branch at $\phi = 0$ and recovers the supersymmetric locus of \cite{Hoare:2022asa}.

\paragraph{Space-like trigonometric limit.}
The space-like trigonometric limit corresponds to either taking $\alpha_1 = \alpha_2$ or $\alpha_3 = \alpha_2$.
In the former case we parametrise this limit as ($0\leq\tilde\kappa<1$, $\lambda>0$):
\unskip\footnote{\label{foot:a3a2case}
In the latter case we similarly parametrise the limit as
\begin{equation*}\begin{aligned}
&0 < \alpha_1 = \lambda^2(1-\tilde\kappa^2)\mathrm{T} \leq \alpha_3=\alpha_2 = \lambda^2\mathrm{T} ~, &&
\\
&\begin{aligned}
&\gamma_1 = \frac{1}{\lambda\sqrt{1-\tilde{\kappa}^2}} ~, \qquad && \gamma_2 = -\frac{\tilde\kappa^2}{\lambda\sqrt{1-\tilde{\kappa}^2}} ~,
\\
&s = -t =1 ~, \qquad && \frac{1-s}{-1-t} = -\tilde{\kappa}^2 ~,
\end{aligned} \qquad
& \gamma_3 = \gamma_+ = \frac{\sqrt{1-\tilde{\kappa}^2}}{\lambda} ~.
\end{aligned}\end{equation*}}
\begin{equation}\begin{aligned}
&0 < \alpha_3 = \lambda^2(1-\tilde\kappa^2)\mathrm{T} \leq \alpha_1=\alpha_2 = \lambda^2\mathrm{T} ~, &&
\\
&\begin{aligned}
&\gamma_1 = \frac{1}{\lambda\sqrt{1-\tilde{\kappa}^2}} ~, \qquad && \gamma_2 = \frac{\tilde\kappa^2}{\lambda\sqrt{1-\tilde{\kappa}^2}} ~,
\\
&s = t =1 ~, \qquad && \frac{1-s}{1-t} = \tilde{\kappa}^2 ~,
\end{aligned} \qquad
& \gamma_3 = \gamma_- = \frac{\sqrt{1-\tilde{\kappa}^2}}{\lambda} ~.
\end{aligned}\end{equation}
In terms of $s$ and $t$, the limit corresponds to taking both of these parameters equal to $1$.
The deformation parameter $\tilde{\kappa}$ controls the direction by which we approach this point.

Taking this limit on the supersymmetric locus~\eqref{eq:susylocus} we find that $\Sigma_1 = \Sigma_2 = \Sigma_3 = \Sigma_4$ and the parameter $\phi$ drops out.
We are then left with
\unskip\footnote{
Similarly, in the $\alpha_3 = \alpha_2$ case using the parametrisation in \foottref{foot:a3a2case} we find
\begin{equation*}\begin{gathered}
||\mathbf{y}_+||^2 = ||\mathbf{y}_-||^2 = -\mathbf{y}_+\cdot\mathbf{y}_- = \frac{1}{\lambda^2} ~, \qquad
||\mathbf{z}_+||^2 = ||\mathbf{z}_-||^2 = \mathbf{z}_+\cdot\mathbf{z}_- = \frac{\tilde\kappa^2}{\lambda^2} ~,
\\
\mathbf{y}_+\cdot\mathbf{z}_+ =
\mathbf{y}_-\cdot\mathbf{z}_- =
\mathbf{y}_+\cdot\mathbf{z}_- =
\mathbf{y}_-\cdot\mathbf{z}_+ = 0 ~,
\end{gathered}\end{equation*}
implying $\mathbf{x}^{(3)} = \mathbf{x}^{(2)} = 0$, while $\mathbf{x}^{(1)}$ and $\mathbf{x}^{(4)}$ are orthogonal with norms $\tilde{\kappa}\lambda^{-1}$ and $\lambda^{-1}$ respectively.}
\begin{equation}\begin{gathered}
||\mathbf{y}_+||^2 = ||\mathbf{y}_-||^2 = -\mathbf{y}_+\cdot\mathbf{y}_- = \frac{1}{\lambda^2} ~, \qquad
||\mathbf{z}_+||^2 = ||\mathbf{z}_-||^2 = -\mathbf{z}_+\cdot\mathbf{z}_- = \frac{\tilde\kappa^2}{\lambda^2} ~,
\\
\mathbf{y}_+\cdot\mathbf{z}_+ =
\mathbf{y}_-\cdot\mathbf{z}_- =
\mathbf{y}_+\cdot\mathbf{z}_- =
\mathbf{y}_-\cdot\mathbf{z}_+ = 0 ~.
\end{gathered}\end{equation}
This implies that $\mathbf{x}^{(1)} = \mathbf{x}^{(2)} = 0$, while $\mathbf{x}^{(3)}$ and $\mathbf{x}^{(4)}$ are orthogonal with norms $\tilde{\kappa}\lambda^{-1}$ and $\lambda^{-1}$ respectively.

The space-like trigonometric limit is particularly interesting since the deformed $
\AdS_3 \times \Sp^3$ metric admits a further limit to $\AdS_2\times \Sp^2\times \Real^2$.
This corresponds to taking $\alpha_3 \to 0$ after setting $\alpha_1=\alpha_2$.
The coordinates~\eqref{eq:coordinates} are not well-adapted to taking this limit.
Therefore, we introduce an alternative set of coordinates
\unskip\footnote{This set of coordinates is related to~\eqref{eq:coordinates} by a transformation of the type
\begin{equation*}\begin{gathered}
T = \tilde T + f_T(\tilde U,\tilde V) ~, \qquad \tilde U = U + \dots ~, \qquad \tilde V = V + \dots ~,
\\
\Phi = \tilde \Phi + f_\Phi(\tilde X,\tilde Y) ~, \qquad \tilde X = X + \dots ~, \qquad \tilde Y = Y + \dots ~,
\end{gathered}
\end{equation*}
where the ellipses denote terms higher order in the coordinates $(U,V)$ and $(X,Y)$ respectively.
It follows from the results of~\cite{Borsato:2023oru} that fixing uniform light-cone gauge in either set of coordinates will lead to the same light-cone gauge S-matrix.}
\begin{equation}
g_\ind{A} = e^{i \tilde T \sigma_2}e^{\tilde V \sigma_1} e^{\tilde U\sigma_3} \in \grp{SL}(2;\Real) ~,
\qquad
g_\ind{S} = e^{i \tilde\Phi \sigma_2}e^{i \tilde Y \sigma_1} e^{i\tilde X\sigma_3} \in \grp{SU}(2) ~.
\end{equation}
Setting $\alpha_1 = \alpha_2 = \lambda^2\mathrm{T}$, the vielbein~\eqref{eq:vielbein} satisfies
\begin{equation}\begin{aligned}\label{eq:ads2comb}
-(e^0)^2 + (e^1)^2 & = \lambda^2\mathrm{T} (-\cosh^22\tilde V d\tilde T^2 + d\tilde V^2) ~,
& \qquad (e^3)^2 + (e^4)^2 & = \lambda^2\mathrm{T} (\cos^22\tilde Yd\tilde\Phi^2 + d\tilde Y^2) ~, \qquad
\\
e^0 \wedge e^1 & = -\lambda^2\mathrm{T}\cosh 2\tilde V d\tilde T\wedge d\tilde V ~,
\qquad
& e^3 \wedge e^4 & =\lambda^2\mathrm{T}\cos2\tilde Y d\tilde \Phi\wedge d\tilde Y ~,
\\
e^2 & = \sqrt{\alpha_3} (-d\tilde U - \sinh2\tilde V d\tilde T) ~,
\qquad
& e^5 & = \sqrt{\alpha_3} (d\tilde X + \sin2\tilde Yd\tilde\Phi) ~.
\end{aligned}\end{equation}
As we have already seen, setting $\alpha_1 = \alpha_2$ implies $\mathbf{x}^{(1)} = \mathbf{x}^{(2)} = 0$, hence the auxiliary fluxes $f_3^{(1)}$ and $f_3^{(2)}$ do not contribute and the vielbein only appears in the background in the combinations given in eq.~\eqref{eq:ads2comb}.
Since $e^2$ and $e^5$ vanish if we simply take $\alpha_3 \to 0$, to engineer a non-degenerate limit we first rescale
\begin{equation}
\tilde U \to - \frac{\sqrt{\mathrm{T}}}{\sqrt{\alpha_3}} \tilde U ~,
\qquad
\tilde X \to \frac{\sqrt{\mathrm{T}}}{\sqrt{\alpha_3}} \tilde X ~,
\end{equation}
such that
\begin{equation}
e^2 \to \sqrt{\mathrm{T}} d\tilde U ~, \qquad e^5 \to \sqrt{\mathrm{T}} d\tilde X ~.
\end{equation}
The coordinates $\tilde U$ and $\tilde X$ then parametrise a flat $\Real^2$ decoupled from the $\AdS_2$ and $\Sp^2$ with coordinates $(\tilde T,\tilde V)$ and $(\tilde \Phi, \tilde Y)$ respectively, which we can compactify to a two-torus $\To^2$.
Moreover, starting from the background R-R fluxes in the space-like trigonometric limit
\begin{equation}\begin{aligned}
\mathbf{x}^{(1)} & = \mathbf{x}^{(2)} = 0 ~, \qquad &\mathbf{x}^{(3)} &= (0,0,-\frac{\tilde\kappa}{\lambda},0) ~, \qquad \mathbf{x}^{(4)} = (\frac{1}{\lambda},0,0,0) ~, \\
F_3 & = 0 ~, \qquad & F_5 & = \frac{1}{\lambda} f_3^{(4)} \wedge J_{2,1} - \frac{\tilde{\kappa}}{\lambda} f_3^{(3)} \wedge J_{2,3} ~,
\end{aligned}\end{equation}
when we take $\alpha_3 \to 0$, or equivalently $\tilde\kappa \to 1$, according to the prescription above, we find the metric of $\AdS_2\times \Sp^2 \times \To^6$ supported by the 5-form R-R flux
\begin{equation}\begin{aligned}
F_5 & = \Vol(\AdS_2) \wedge \Re \Olc + \Vol(\Sp^2) \wedge \Im \Olc ~, \qquad &&\Vol(\AdS_2) = e^0 \wedge e^1 ~, \qquad \Vol(\Sp^2) = e^3 \wedge e^4 ~,
\\ &&& \Olc = (e^2 + i e^5) \wedge (e^6+i e^8) \wedge (e^7+i e^9) ~,
\end{aligned}
\end{equation}
where $\Olc$ is a holomorphic $(3,0)$-form on $\To^6$.
In this way we recover the type IIB $\AdS_2 \times \Sp^2 \times \To^6$ background with 8 supersymmetries from~\cite{Sorokin:2011rr}.

Finally, let us briefly note that an analogous construction goes through for the $\alpha_1 \to 0$ limit after setting $\alpha_3 = \alpha_2$.
In this case we use the original set of coordinates~\eqref{eq:coordinates}, now with $e^1$ and $e^4$ and the coordinates $V$ and $Y$ parametrising the flat $\Real^2$, $e^0$ and $e^2$ and the coordinates $T$ and $U$ parametrising $\AdS_2$, and $e^3$ and $e^5$ and the coordinates $\Phi$ and $X$ parametrising $\Sp^2$.

\paragraph{Rational limit.}
To conclude our discussion of supersymmetry we return to the rational limit $\alpha_1 = \alpha_2 = \alpha_3$
\begin{equation}\begin{gathered}
\alpha_1 = \alpha_2 = \alpha_3 = \lambda^2 \mathrm{T} ~,
\\
\gamma_1 = \gamma_\pm = \frac{1}{\lambda} ~, \qquad \gamma_2 = 0 ~, \qquad \gamma_3 = \frac{1}{\lambda} ~, \qquad s = 1 ~.
\end{gathered}\end{equation}
From the time-like and space-like trigonometric limits we recover the rational limit by further taking $\kappa$ or $\tilde\kappa$ to $0$.
In terms of $s$ and $t$, the rational limit corresponds to taking $s \to 1$ for any value of $t \neq \pm1$.
Taking this limit on the supersymmetric locus~\eqref{eq:susylocus}, the parameters $t$ and $\phi$ both drop out and we are left with
\begin{equation}\begin{gathered}
||\mathbf{y}_+||^2 = ||\mathbf{y}_-||^2 = -\mathbf{y}_+\cdot\mathbf{y}_- = \frac{1}{\lambda^2} ~, \qquad
||\mathbf{z}_+||^2 = ||\mathbf{z}_-||^2 = -\mathbf{z}_+\cdot\mathbf{z}_- = 0 ~,
\\
\mathbf{y}_+\cdot\mathbf{z}_+ =
\mathbf{y}_-\cdot\mathbf{z}_- =
\mathbf{y}_+\cdot\mathbf{z}_- =
\mathbf{y}_-\cdot\mathbf{z}_+ = 0 ~.
\end{gathered}\end{equation}
As discussed above, this leaves us with $||\mathbf{x}^{(4)}||^2 = \lambda^{-2}$, $\mathbf{x}^{(1)} = \mathbf{x}^{(2)} = \mathbf{x}^{(3)} = 0$ and the corresponding supergravity background has 16 6d supersymmetries.

%%%%%%%%%%%%%%%%%%%%%%%%%%%%%%%%%%%%%%%%%%%%%%%%%%%%%%%%%%%%%%%%%%%%%%%%%%%%%%%%
\section{Light-cone gauge-fixed theory}\label{sec:lcgf}

We have seen that there is a 1-parameter family of R-R fluxes such that the elliptic $\AdS_3 \times \Sp^3 \times \To^4$ supergravity background presented in \secref{sec:sugra-susy} preserves 8 supersymmetries closing into a $\alg{psu}(1,1|2)$ superalgebra. Given that the bosonic truncation of the theory is classically integrable~\cite{Cherednik:1981df}, the hope is then that also the motion of superstrings propagating in this supersymmetric supergravity background is classically integrable. In this section we analyse the Green-Schwarz action in a uniform light-cone gauge~\cite{Arutyunov:2004yx,Arutyunov:2005hd} and compute the massive worldsheet S-matrix at tree-level (in inverse powers of the string tension). We find that it satisfies the classical Yang-Baxter equation, a necessary condition for classical integrability.

\subsection{Light-cone gauge-fixing}

The Lagrangian for the type IIB Green-Schwarz superstring, to quadratic order in the fermions, reads (we set the string tension $\mathrm{T}=1$; it can be restored by dimensional analysis)
\begin{equation}
\mathcal L = \gamma^{\alpha \beta} \hat{G}_{\ind{MN}} \partial_\alpha X^\ind{M} \partial_\beta X^\ind{N} - \varepsilon^{\alpha \beta} \hat{B}_\ind{MN} \partial_\alpha X^\ind{M} \partial_\beta X^\ind{N} + \mathcal L_{kin}~,
\end{equation}
with
\begin{equation} \label{eq:GBhat}
\begin{aligned}
\hat{G}_\ind{MN} = G_\ind{MN} - \frac{i}{4} \bar{\theta} \Gamma_{(\ind{M}} \slashed{\omega}_{\ind{N})} \theta + \frac{i}{8} \bar{\theta} \sigma_3 \Gamma_{(\ind{M}} H_{\ind{N})\ind{PQ}} \Gamma^{\ind{PQ}} \theta + \frac{i}{8} \bar{\theta} \Gamma_{(\ind{M}} \mathcal S \Gamma_{\ind{N})} \theta~, \\
\hat{B}_\ind{MN} = B_\ind{MN} + \frac{i}{4} \bar{\theta} \sigma_3 \Gamma_{[\ind{M}} \slashed{\omega}_{\ind{N}]}\theta - \frac{i}{8} \bar{\theta} \Gamma_{[\ind{M}} H_{\ind{N}]\ind{PQ}} \Gamma^{\ind{PQ}} \theta - \frac{i}{8} \bar{\theta} \sigma_3 \Gamma_{[\ind{M}} \mathcal S \Gamma_{\ind{N}]} \theta~,
\end{aligned}
\end{equation}
and
\begin{equation}
\mathcal L_{kin} = \gamma^{\alpha \beta} \partial_\alpha X^\ind{M} i \bar{\theta} \Gamma_\ind{M} \partial_\beta \theta + \varepsilon^{\alpha \beta} \partial_\alpha X^\ind{M} i \bar{\theta} \Gamma_\ind{M} \sigma_3 \partial_\beta \theta~, \qquad \bar{\theta} = \theta^\dagger \Gamma^0~.
\end{equation}
The 2d worldsheet of the string is parametrised by the two coordinates $\sigma^\alpha$ with $\alpha = 0,1$, and we will often use the notation $\sigma^0 = \tau$ for the time-like direction and $\sigma^1=\sigma$ for the space-like direction.
For derivatives we use the shorthand notation $\partial_\alpha = \frac{\partial}{\partial \sigma^\alpha}$.
The bosonic fields $X^\ind{M}$ with $M=0,\dots,9$ parametrise the embedding of the string in space-time, and are supplemented by $\theta=(\theta^1, \theta^2)$, a doublet of 10d Majorana-Weyl spinors.
The Weyl-invariant metric on the worldsheet is given by $\gamma^{\alpha \beta}$, while $\varepsilon^{\alpha \beta}$ denotes the antisymmetric tensor with the convention $\varepsilon^{\tau \sigma} = 1$.
The symmetric quantity $G_\ind{MN}$ is the space-time metric, while $B_\ind{MN}$ is the anti-symmetric B-field with associated NS-NS flux $H_3 = dB_2$. The spin connection $\omega$ and the R-R bispinor $\mathcal S$ are as defined in eqs.~\eqref{eq:spin-connection} and \eqref{eq:R-R-bispinor} respectively.

The type IIB action is invariant under worldsheet reparametrisations and fermionic kappa-symmetry. To remove these redundancies it is convenient to impose a uniform light-cone gauge and a compatible kappa-symmetry gauge. For this to be possible, the background must have at least two isometric directions, which we call $X^0 = T$ and $X^3 = \Phi$. We then introduce the light-cone coordinates
\begin{equation}
X^+ = (1-a) T + a \Phi~, \qquad X^- = - T + \Phi~, \qquad T = X^+ - a X^-~, \qquad \Phi = X^+ + (1-a) X^-~,
\end{equation}
where $a \in [0,1]$ is a free parameter,
\unskip\footnote{The light-cone coordinates \eqref{eq:lc-ppwave} used in the pp-wave limit correspond to the symmetric, $a=1/2$, case.}
and the corresponding Dirac matrices with curved-space indices
\begin{equation}
\Gamma_+ = \Gamma_T + \Gamma_\Phi~, \qquad \Gamma_- = -a \Gamma_T + (1-a) \Gamma_\Phi~.
\end{equation}
One way to fix uniform light-cone gauge begins by T-dualising the theory in the coordinate $X^-$.
Denoting the T-dual coordinate by $X^\circ$, up to quadratic order in the fermions the T-dual Lagrangian is given by~\cite{Arutyunov:2014jfa}
\begin{equation}
\label{eq:LTdual}
\mathcal L_{\mathrm{T-dual}} = -\sqrt{-\det \mathring{G}_{\alpha \beta}} - \frac{1}{2} \epsilon^{\alpha \beta} \mathring{E}_{\alpha \beta}~,
\end{equation}
with
\begin{equation} \label{eq:GEcirc}
\begin{aligned}
\mathring{G}_{\alpha \beta} &= \mathring{G}_\ind{MN} \partial_\alpha X^\ind{M} \partial_\beta X^\ind{N} + i \partial_{(\alpha} X^{\bar{\ind{M}}} \bar{\theta} \mathring{\Gamma}_{\bar{\ind{M}}} \partial_{\beta)} \theta+ i \partial_{(\alpha} X^\circ \bar{\theta} \mathring{\Gamma}_{\circ} \sigma_3 \partial_{\beta)} \theta~, \\
\mathring{E}_{\alpha \beta} &= -\mathring{B}_\ind{MN} \partial_\alpha X^\ind{M} \partial_\beta X^\ind{N} + i \partial_{[\alpha} X^{\bar{\ind{M}}} \bar{\theta} \mathring{\Gamma}_{\bar{\ind{M}}} \sigma_3 \partial_{\beta]} \theta + i \partial_{[\alpha} X^\circ \bar{\theta} \mathring{\Gamma}_{\circ} \partial_{\beta]} \theta~,
\end{aligned}
\end{equation}
and
\begin{equation}\label{eq:GBcirc}
\begin{aligned}
\mathring{G}_{\circ \circ} = \frac{1}{\hat{G}_{--}}~, \qquad \mathring{G}_{\circ \bar{\ind{M}}} &= \mathring{G}_{\bar{\ind{M}} \circ} = - \frac{\hat{B}_{-\bar{\ind{M}}}}{\hat{G}_{--}}~, &\, \mathring{G}_{\bar{\ind{M}}\bar{\ind{N}}} &= \hat{G}_{\bar{\ind{M}}\bar{\ind{N}}} - \frac{\hat{G}_{-\bar{\ind{M}}} \hat{G}_{-\bar{\ind{N}}} - \hat{B}_{-\bar{\ind{M}}}\hat{B}_{-\bar{\ind{N}}}}{\hat{G}_{--}}~, \\
\mathring{B}_{\circ \bar{\ind{M}}} &= - \mathring{B}_{\bar{\ind{M}} \circ} = - \frac{\hat{G}_{-\bar{\ind{M}}}}{\hat{G}_{--}}~, &\, \mathring{B}_{\bar{\ind{M}} \bar{\ind{N}}} &= \hat{B}_{\bar{\ind{M}} \bar{\ind{N}}} - \frac{\hat{G}_{-\bar{\ind{M}}} \hat{B}_{- \bar{\ind{N}}} -\hat{B}_{-\bar{\ind{M}}} \hat{G}_{- \bar{\ind{N}}} }{\hat{G}_{--}}~, \\
\mathring{\Gamma}_{\circ} &= \frac{1}{G_{--}} \Gamma_- ~, &\qquad \mathring{\Gamma}_{\bar{\ind{M}}} &= \Gamma_{\bar{\ind{M}}} - \frac{G_{-\bar{\ind{M}}}}{G_{--}} \Gamma_-~.
\end{aligned}
\end{equation}
The indices $\bar{M}$ and $\bar{N}$ run over all coordinates that are not involved in the T-duality, i.e.,~$\bar{M}, \bar{N} \in \{+,\mu \}$. We assume a purely transverse B-field, so that $B_{-+}=B_{- \mu}=0$. In the expressions \eqref{eq:GEcirc} and \eqref{eq:GBcirc}, we only keep terms up to quadratic order in the fermions explaining why in some expressions it is possible to replace hatted quantities by non-hatted ones.

The uniform light-cone gauge-fixing condition on the bosonic degrees of freedom is
\begin{equation} \label{eq:uniform-lcgf}
X^+ = \tau~, \qquad X^\circ = \sigma~.
\end{equation}
The remaining dynamical bosonic degrees of freedom are then the eight transverse coordinates $X^\mu$, and the Lagrangian can be expanded in powers of the transverse fields. We also require that \eqref{eq:uniform-lcgf} is a classical solution to the equations of motion of the T-dual theory so that we can choose a basis for the vielbein that is diagonal in the light-cone directions, with
\begin{equation}
\label{eq:vielbein-diag}
e^0 = dT + O(X^\mu)~, \qquad e^3 = d\Phi + O(X^\mu)~, \qquad e^\nu = O(X^\mu)~.
\end{equation}
This should be complemented with an appropriate kappa-symmetry gauge on the fermions.
Expanding the light-cone gauge-fixed Lagrangian up to quadratic order in the fields, and assuming that the lowest-order contribution to the vielbein is of the form \eqref{eq:vielbein-diag}, we find that the kinetic terms for the fermions are of the form
\begin{equation}
\mathcal L_2 \supset i \bar{\theta} \Glc^- \partial_\tau{\theta}~.
\end{equation}
Therefore, only the fields $\Glc^- \theta$ are dynamical.
A natural kappa-symmetry gauge to impose is then
\begin{equation}
\Glc^+ \theta =0~.
\end{equation}
This halves the number of fermionic degrees of freedom from 16 to 8 complex fermions.

\subsection{Expansion of the Lagrangian}

The above gauge-fixing procedure, in particular \eqref{eq:LTdual}, can now be applied to the elliptic $\AdS_3 \times \Sp^3 \times \To^4$ superstring, characterised by the vielbein \eqref{eq:vielbein} and fluxes \eqref{eq:fluxes}.
To obtain a vielbein that satisfies \eqref{eq:vielbein-diag} for generic values of the deformation parameters, it is convenient to do the rescaling (assuming $\alpha_{1,2,3} > 0$ and setting the string tension $\mathrm{T}=1$ for convenience)
\begin{equation}
T \rightarrow \frac{T}{\sqrt{\alpha_2}}~, \qquad U \rightarrow \frac{U}{\sqrt{\alpha_3}}~, \qquad V \rightarrow \frac{V}{\sqrt{\alpha_1}}~, \qquad \Phi \rightarrow \frac{\Phi}{\sqrt{\alpha_2}}~, \qquad X \rightarrow \frac{X}{\sqrt{\alpha_3}}~, \qquad Y \rightarrow \frac{Y}{\sqrt{\alpha_1}}~.
\end{equation}
The light-cone gauge-fixed Lagrangian is then expanded in powers of the transverse fields,
\begin{equation}
\mathcal L_{g.f.} = \mathcal L_2 + \mathcal L_3 + \mathcal L_4 + \dots~,
\end{equation}
where $\mathcal L_n$ contains terms with $n$ fields, of bosonic (B) or fermionic (F) type.
When restoring the string tension, this corresponds to a large tension expansion around the classical solution \eqref{eq:uniform-lcgf} with $X=Y=U=V=0$ and $\theta=0$.
The quadratic Lagrangian can be written as
\begin{equation}
\mathcal L_2 = \mathcal L_{2,B} + \mathcal L_{2,F}~,
\end{equation}
where $\mathcal L_{2,B}$ describes four massive and four massless bosons, while $\mathcal L_{2,F}$ describes four massive and four massless fermions.
The bosonic Lagrangian $\mathcal L_{2,B}$ was already obtained and analysed in~\cite{Hoare:2023zti}.
It is given by
\begin{equation}
\begin{aligned}
\mathcal{L}_{2,B}&=
\frac{1}{2}\left(\dot{U}^2-U'{}^2+\dot{V}^2-V'{}^2\right)+\frac{2(\al_1-\al_2)}{\al_2\al_3}U^2-\frac{2(\al_2-\al_3)}{\al_1\al_2}V^2-\frac{2\sqrt{\al_1}}{\sqrt{\al_2\al_3}}U\dot{V}-\frac{2(\al_2-\al_3)}{\sqrt{\al_1\al_2\al_3}}\dot{U}V\\
&+\frac{1}{2}\left(\dot{X}^2-X'{}^2+\dot{Y}^2-Y'{}^2\right)+\frac{2(\al_1-\al_2)}{\al_2\al_3}X^2-\frac{2(\al_2-\al_3)}{\al_1\al_2}Y^2-\frac{2\sqrt{\al_1}}{\sqrt{\al_2\al_3}}X\dot{Y}-\frac{2(\al_2-\al_3)}{\sqrt{\al_1\al_2\al_3}}\dot{X}Y\\
&+ \sum_{r=1}^4 \left( \dot{x}_r^2 - x'_r{}^2 \right)~,
\end{aligned}
\label{eq:L2B}
\end{equation}
where we use the shorthand notation $\partial_\tau \Psi = \dot{\Psi}$ and $\partial_\sigma \Psi = \Psi'$.
The equations of motion are $\ddot{x}_r - x_r'' =0$ and
\begin{equation}
\begin{aligned}
0 &= \ddot{U} - U'' -4 \frac{\alpha_1 - \alpha_2}{\alpha_2 \alpha_3} U + 2 \frac{\alpha_1 - \alpha_2 + \alpha_3}{\sqrt{\alpha_1 \alpha_2 \alpha_3}} \dot{V}~, \\
0 &= \ddot{V} - V'' +4 \frac{\alpha_2 - \alpha_3}{\alpha_1 \alpha_2} V - 2 \frac{\alpha_1 - \alpha_2 + \alpha_3}{\sqrt{\alpha_1 \alpha_2 \alpha_3}} \dot{U}~, \\
0 &= \ddot{X} - X'' -4 \frac{\alpha_1 - \alpha_2}{\alpha_2 \alpha_3} X + 2 \frac{\alpha_1 - \alpha_2 + \alpha_3}{\sqrt{\alpha_1 \alpha_2 \alpha_3}} \dot{Y}~, \\
0 &= \ddot{Y} - Y'' +4 \frac{\alpha_2 - \alpha_3}{\alpha_1 \alpha_2} Y - 2 \frac{\alpha_1 - \alpha_2 + \alpha_3}{\sqrt{\alpha_1 \alpha_2 \alpha_3}} \dot{X}~.
\end{aligned}
\end{equation}
The quadratic Lagrangian for the fermions is
\begin{equation}
\begin{aligned}
\mathcal L_{2,F} &= i \sum_{j=1}^2 \zeta_{R,j}^\star (\partial_\tau - \partial_\sigma - i \gamma_3) \zeta_{R,j} +i \sum_{j=1}^2 \zeta_{L,j}^\star (\partial_\tau + \partial_\sigma - i \gamma_3) \zeta_{L,j} \\
&\quad+ \sum_{j,k=1}^2 \left( \zeta_{R,j}^\star (\mathbf{Y}_+)_{jk} \zeta_{L,k} + \zeta_{L,j}^\star (\mathbf{Y}_+^\dagger)_{jk} \zeta_{R,k} - \zeta_{R,j} (\mathbf{Z}_+)_{jk} \zeta_{L,k} - \zeta_{L,j}^\star(\mathbf{Z}_+^\dagger)_{jk} \zeta_{R,k}^\star \right) \\
&\quad +i \sum_{j=3}^4 \zeta_{R,j}^\star (\partial_\tau - \partial_\sigma) \zeta_{R,j} +i \sum_{j=3}^4 \zeta_{L,j}^\star (\partial_\tau + \partial_\sigma) \zeta_{L,j}~.
\label{eq:L2F}
\end{aligned}
\end{equation}
Our convention for the parametrisation of the two Majorana-Weyl spinors $(\theta^1,\theta^2)$ in terms of the right- and left-movers $\zeta_R$ and $\zeta_L$ can be found in~\appref{app:gammamatrix}.
Note that the conjugation rule for the Grassmann variables is $(\zeta_L^\star \zeta_R)^\star = \zeta_R^\star \zeta_L$, and therefore the quadratic Lagrangian $\mathcal L_{2,F}$ is real for real deformation parameters.
It features the shift
\begin{equation}
\gamma_3 = \frac{\al_1-\al_2+\al_3}{\sqrt{\al_1 \al_2 \al_3}}~,
\end{equation}
as well as the matrices
\begin{equation}
\mathbf{Y}_+ = \begin{pmatrix}
(\mathbf{y}_{+})_4 + i (\mathbf{y}_{+})_1 & (\mathbf{y}_{+})_2+i (\mathbf{y}_+)_3 \\
- (\mathbf{y}_{+})_2 + i (\mathbf{y}_{+})_3 & (\mathbf{y}_{+})_4 - i (\mathbf{y}_{+})_1
\end{pmatrix}~, \qquad
\mathbf{Z}_+ = \begin{pmatrix}
(\mathbf{z}_+)_3 + i (\mathbf{z}_+)_2 & -(\mathbf{z}_+)_1-i (\mathbf{z}_+)_4 \\
-(\mathbf{z}_+)_1 + i (\mathbf{z}_+)_4 & -(\mathbf{z}_+)_3 + i (\mathbf{z}_+)_2
\end{pmatrix}~, \qquad
\begin{aligned}
\end{aligned}
\end{equation}
where we recall the relation to the parameters in the R-R fluxes
\begin{equation}
\textbf{y}_{+} = \textbf{x}^{(2)} + \textbf{x}^{(4)}~, \qquad
\textbf{z}_{+} = \textbf{x}^{(1)} + \textbf{x}^{(3)}~.
\end{equation}
The fermions $\zeta_{L,3}, \zeta_{L,4}$ and $\zeta_{R,3}, \zeta_{R,4}$ are the massless fermions, supersymmetric partners of the massless bosons $x_r$ with $r=1,2,3,4$.
On the other hand, $\zeta_{L,1}, \zeta_{L,2}$ and $\zeta_{R,1}, \zeta_{R,2}$ are massive fermions.
Their masses originate from the R-R fluxes in the theory.
For generic deformation parameters it is not possible to diagonalise $\mathbf{Y}_+$ and $\mathbf{Z}_+$ simultaneously.
To simplify the quadratic Lagrangian $\mathcal L_{2,F}$ and write it in terms of $\grp{SO}(4)$-invariant quantities, it is useful to make a unitary rotation of the fermions
\begin{equation}
\label{eq:rotation}
\zeta_R = \mathcal V \tilde{\zeta}_R~, \qquad \zeta_{L} = \mathcal U \mathcal V \tilde{\zeta}_L~,
\end{equation}
with
\begin{align}
\label{eq:rotation-U}
&\det \mathcal U=1~, \qquad \mathcal U \mathcal U^{\dagger} = \mathcal U^\dagger \mathcal U =1~, \qquad \mathcal U = \frac{1}{||\mathbf{y}_+||}\mathbf{Y}_+^\dagger~, \\
&\det \mathcal V = 1~, \qquad \mathcal V \mathcal V^{\dagger} = \mathcal V^\dagger \mathcal V =1~, \qquad \mathcal V = \frac{\sqrt{\beta-\beta_{14}}}{\sqrt{2\beta}}\begin{pmatrix}
1 & -\frac{\beta + \beta_{14}}{\beta_{12}-i \beta_{13}} \\
\frac{\beta+\beta_{14}}{\beta_{12}+ i \beta_{13}} & 1
\end{pmatrix}~,
\label{eq:rotation-V}
\end{align}
and
\begin{equation}
\label{eq:beta}
\begin{aligned}
\beta_{12} &= (\mathbf{y}_+)_1 (\mathbf{z}_+)_2 - (\mathbf{y}_+)_2 (\mathbf{z_+})_1 - (\mathbf{y}_+)_3 (\mathbf{z}_+)_4 + (\mathbf{y}_+)_4 (\mathbf{z}_+)_3~, \\
\beta_{13} &= (\mathbf{y}_+)_1 (\mathbf{z}_+)_3 + (\mathbf{y}_+)_2 (\mathbf{z}_+)_4 - (\mathbf{y}_+)_3 (\mathbf{z}_+)_1 - (\mathbf{y}_+)_4 (\mathbf{z}_+)_2~, \\
\beta_{14} &= (\mathbf{y}_+)_1 (\mathbf{z}_+)_4 - (\mathbf{y}_+)_2 (\mathbf{z}_+)_3 + (\mathbf{y}_+)_3 (\mathbf{z}_+)_2 - (\mathbf{y}_+)_4 (\mathbf{z}_+)_1~, \\
\beta &= \sqrt{\ys \zs - (\yz)^2} = \sqrt{\beta_{12}^2 + \beta_{13}^2 + \beta_{14}^2}~.
\end{aligned}
\end{equation}
The rotated mass matrices are then
\begin{equation}
\label{eq:MNrotated}
\begin{aligned}
\tilde{\mathbf{Y}}_+&\equiv \mathcal V^\dagger \mathbf{Y}_+ \mathcal U \mathcal V = m~, \qquad \tilde{\mathbf{Z}}_+ \equiv\mathcal V^t \mathbf{Z}_+ \mathcal U \mathcal V = \frac{1}{m} (- \beta \sigma_1 + (\yz)\sigma_2)~, \qquad m=||\mathbf{y}_+||~.
\end{aligned}
\end{equation}
This makes it manifest that the quadratic Lagrangian $\mathcal L_{2,F}$ is invariant under the $\grp{U}(1)$ symmetry
\begin{equation}
\label{eq:symL2F}
\tilde{\zeta}_L \rightarrow \mathcal W \tilde{\zeta}_L~, \qquad \tilde{\zeta}_R \rightarrow \mathcal W \tilde{\zeta}_R~, \qquad \mathcal W = \begin{pmatrix} e^{i q} & 0 \\ 0& e^{-i q} \end{pmatrix}~, \qquad \mathcal W^t \sigma_{1,2} \mathcal W = \sigma_{1,2}~.
\end{equation}
When $\mathbf{Z}_+=0$, which is the case in the rational limit and for one of the two supersymmetric time-like trigonometric branches, it is sufficient to rotate $\zeta_L$, and the above $\grp{U}(1)$ is promoted to a manifest $\grp{SU}(2)$ symmetry.
The rotation diagonalises the mass matrix $\mathbf{Y}_+$, and the massive fermions have all the same mass $m$.

In the generic case, the equations of motion for the rotated massive fermionic fields $\tilde{\zeta}_{L,R}$ are
\begin{equation}
\label{eq:eom-F}
\begin{aligned}
0 &=i (\partial_\tau - \partial_\sigma - i \gamma_3) \tilde{\zeta}_{R,j} - m \tilde{\zeta}_{L,j} - (\tilde{\mathbf{Z}}_+^\star)_{jk} \tilde{\zeta}_{L,k}^\star~, \\
0 &=-i (\partial_\tau - \partial_\sigma + i \gamma_3) \tilde{\zeta}_{R,j}^\star - m \tilde{\zeta}_{L,j}^\star - (\tilde{\mathbf{Z}}_+)_{jk} \tilde{\zeta}_{L,k}~, \\
0 &=i (\partial_\tau + \partial_\sigma - i \gamma_3) \tilde{\zeta}_{L,j} - m\tilde{\zeta}_{R,j} + (\tilde{\mathbf{Z}}_+^\dagger)_{jk} \tilde{\zeta}_{R,k}^\star~, \\
0 &=-i (\partial_\tau + \partial_\sigma + i \gamma_3) \tilde{\zeta}_{L,j}^\star - m \tilde{\zeta}_{R,j}^\star + (\tilde{\mathbf{Z}}_+^t)_{jk} \tilde{\zeta}_{R,k}~.
\end{aligned}
\end{equation}
The cubic Lagrangian vanishes, $\mathcal L_3=0$ and the quartic interactions $\mathcal L_4$ are too lengthy to be written explicitly here.

\subsection{Oscillators}

To define the asymptotic states in the scattering processes, we diagonalise the quadratic Hamiltonian by introducing a set of harmonic oscillators.
We focus on the massive sector of the theory, involving the bosonic fields $U,V,X,Y$ and the fermionic fields $\zeta_{L,j},\zeta_{R,j}$ with $j=1,2$.
We recall the mode expansion for the massive bosons, as already found in~\cite{Hoare:2023zti},
\begin{equation}
\begin{aligned}
U &= \int dp \left( \frac{\sqrt{\bar{\omega} + \gamma_3 - \gamma_2}}{2\sqrt{\omega_+} \sqrt{\bar{\omega}} } e^{-i \omega_+ \tau + i p \sigma} a_{+,1} + \frac{\sqrt{\bar{\omega} - \gamma_3 + \gamma_2}}{2\sqrt{\omega_-} \sqrt{\bar{\omega}} } e^{-i \omega_- \tau + i p \sigma} a_{-,1} + \hc \right)~, \\
V &= \int dp \left( i \frac{\sqrt{\bar{\omega} + \gamma_3 + \gamma_2}}{2\sqrt{\omega_+} \sqrt{\bar{\omega}} } e^{-i \omega_+ \tau + i p \sigma} a_{+,1} - i\frac{\sqrt{\bar{\omega} - \gamma_3 - \gamma_2}}{2\sqrt{\omega_-} \sqrt{\bar{\omega}} } e^{-i \omega_- \tau + i p \sigma} a_{-,1} + \hc \right)~, \\
X &= \int dp \left( \frac{\sqrt{\bar{\omega} + \gamma_3 - \gamma_2}}{2\sqrt{\omega_+} \sqrt{\bar{\omega}} } e^{-i \omega_+ \tau + i p \sigma} a_{+,2} + \frac{\sqrt{\bar{\omega} - \gamma_3 + \gamma_2}}{2\sqrt{\omega_-} \sqrt{\bar{\omega}} } e^{-i \omega_- \tau + i p \sigma} a_{-,2} + \hc \right)~, \\
Y &= \int dp \left( i \frac{\sqrt{\bar{\omega} + \gamma_3 + \gamma_2}}{2\sqrt{\omega_+} \sqrt{\bar{\omega}} } e^{-i \omega_+ \tau + i p \sigma} a_{+,2} - i\frac{\sqrt{\bar{\omega} - \gamma_3 - \gamma_2}}{2\sqrt{\omega_-} \sqrt{\bar{\omega}} } e^{-i \omega_- \tau + i p \sigma} a_{-,2} + \hc \right)~,
\end{aligned}
\end{equation}
where $\hc$ denotes the hermitian conjugate, $p$ denotes the momentum of the plane wave, and $\omega_\pm$ is its energy.
The creation and annihilation operators depend on the momentum and obey the canonical commutation relations
\begin{equation}
[a_{\mu,j}(p), a_{\nu,k}(q)] = [a^\dagger_{\mu,j}(p), a^\dagger_{\nu,k}(q)] = 0~, \qquad [a_{\mu,j}(p), a^\dagger_{\nu, k}(q)] = \delta_{\mu \nu}\delta_{jk} \delta(p-q)~,
\end{equation}
where $\mu,\nu =\pm$ and $j,k=1,2$.
The quadratic bosonic Hamiltonian then takes the canonical form
\begin{equation}
H_{2,B} = \int d\sigma \, \mathcal H_{2,B} = \int d\sigma \big(P_U \dot{U} + P_V \dot{V} + P_X \dot{X} + P_Y \dot{Y} - \mathcal L_{2,B} \big) = \int dp \sum_{j=1,2} \sum_{\mu = \pm} \omega_\mu a_{\mu,j}^\dagger a_{\mu,j}~.
\end{equation}
In the above, the conjugate momenta are
\begin{equation}
\begin{aligned}
P_U &= \dot{U} - 2 \frac{\alpha_2 - \alpha_3}{\sqrt{\alpha_1 \alpha_2 \alpha_3}} V~, &\qquad P_V &= \dot{V} - 2 \frac{\alpha_1}{\sqrt{\alpha_1 \alpha_2 \alpha_3}} U~, \\
P_X &= \dot{X} - 2 \frac{\alpha_2 - \alpha_3}{\sqrt{\alpha_1 \alpha_2 \alpha_3}} Y~, &\qquad P_Y &= \dot{Y} - 2 \frac{\alpha_1}{\sqrt{\alpha_1 \alpha_2 \alpha_3}} X~,
\end{aligned}
\end{equation}
and the excitations have dispersion relation
\unskip\footnote{The relation between the parameters $\alpha_{1,2,3}$ and $\gamma_{1,2,3}$ is the same as in eq.~\eqref{eq:gamalp} with $\mathrm{T}=1$.}
\begin{equation}
\label{eq:disp}
\sqrt{\omega_\pm^2 + \gamma_2^2} = \sqrt{p^2 + \gamma_1^2} \pm \gamma_3~, \qquad \gamma_1 = \frac{\alpha_2}{\sqrt{\alpha_1 \alpha_2 \alpha_3}}~, \qquad \gamma_2 = \frac{\alpha_1-\alpha_3}{\sqrt{\alpha_1 \alpha_2 \alpha_3}}~, \qquad \gamma_3 = \frac{\alpha_1 - \alpha_2+\alpha_3}{\sqrt{\alpha_1 \alpha_2 \alpha_3}}~.
\end{equation}
A convenient equivalent way of writing this is
\begin{equation}
\omega_\pm^2= (\bar{\omega} \pm \gamma_3 - \gamma_2)(\bar{\omega}\pm\gamma_3+\gamma_2)~, \qquad \bar{\omega} = \sqrt{p^2 + \gamma_1^2}~.
\end{equation}
Similarly, for the massive fermions we make the ansatz (the relation between tilded (after rotation) and untilded (before rotation) fermions is given in eq.~\eqref{eq:rotation})
\begin{equation}
\begin{aligned}
\label{eq:oscillators-F}
\tilde{\zeta}_{L,j} &= \sum_{k=1,2} \sum_{\mu=\pm} \int dp \left( A_{\mu,jk} e^{-i \omega_\mu \tau + i p \sigma} b_{\mu,k}+ B_{\mu,jk} e^{+i \omega_\mu \tau - i p \sigma} b_{\mu,k}^\dagger \right)~, \qquad j=1,2~, \\
\tilde{\zeta}_{R,j} &= \sum_{k=1,2} \sum_{\mu=\pm} \int dp \left( C_{\mu,jk} e^{-i \omega_\mu \tau + i p \sigma} b_{\mu,k}+ D_{\mu,jk} e^{+i \omega_\mu \tau - i p \sigma} b_{\mu,k}^\dagger \right)~, \qquad j=1,2~.
\end{aligned}
\end{equation}
The fermionic creation and annihilation operators commute with the bosonic ones, and obey anti-commutation relations among themselves,
\begin{equation}
\{b_{\mu,j}(p),b_{\nu,k}(q)\} = \{b^\dagger_{\mu,j}(p),b^\dagger_{\nu,k}(q)\}=0~, \qquad \{b_{\mu,j}(p),b^\dagger_{\nu,k}(q)\} = \delta_{\mu \nu} \delta_{jk} \delta(p-q)~.
\end{equation}
The coefficients $A_{\pm, jk}, \dots, D_{\pm,jk}$ should be chosen such that the equations of motion are satisfied and the quadratic Hamiltonian takes the canonical form
\begin{equation}
H_{2,F} = \int d\sigma \, \mathcal H_{2,F} = \int d\sigma \big(P_L \cdot \partial_\tau \zeta_L+ P_R \cdot \partial_\tau \zeta_R - \mathcal L_{2,F} \big) = \int dp \sum_{j=1,2} \sum_{\mu=\pm} \omega_\mu b_{\mu,j}^\dagger b_{\mu,j}~,
\end{equation}
where the conjugate momenta are
\begin{equation}
P_{L,j} = i \zeta_{L,j}^\star~, \qquad P_{R,j} = i \zeta_{R,j}^\star~.
\end{equation}
Plugging the mode expansion \eqref{eq:oscillators-F} into the equations of motion \eqref{eq:eom-F} leads to the system of linear equations
\begin{align}
0 &= (\omega_\pm + p + \gamma_3) A_\pm - m C_\pm - \tilde{\mathbf{Z}}_+^\star D_\pm^\star~, \\
0 &= (\omega_\pm - p +\gamma_3) C_\pm - m \tilde{A}_\pm + \tilde{\mathbf{Z}}_+^\dagger B_\pm^\star~, \\
0 &= (\omega_\pm + p -\gamma_3) B_\pm + m D_\pm + \tilde{\mathbf{Z}}_+^\star C_\pm^\star~, \\
0 &= (\omega_\pm -p-\gamma_3) D_\pm + m B_\pm -\tilde{\mathbf{Z}}_+^\dagger A_\pm^\star~.
\end{align}
Requiring that these equations admit a non-trivial solution gives the dispersion relations of the fermionic excitations.
They depend on the parameters in the R-R fluxes, and read
\begin{equation}
\sqrt{\omega_\pm^2 + \hat{\gamma}_2^2} = \sqrt{p^2 + \hat{\gamma}_1^2} \pm \gamma_3~,
\end{equation}
with
\begin{equation}
\hat{\gamma}_1^2 = \ys \left( \frac{\zs}{\gamma_3^2}+1\right)- \frac{(\yz)^2}{\gamma_3^2}~, \qquad \hat{\gamma}_2^2 = \zs \left(\frac{\ys}{\gamma_3^2}-1\right)-\frac{(\yz)^2}{\gamma_3^2}~.
\end{equation}
Note that on the solution \eqref{eq:eqsol}, $\hat{\gamma}_1^2 = \gamma_1^2$ and $\hat{\gamma}_2^2 = \gamma_2^2$, as expected for a supersymmetric theory.
Conversely, requiring that the bosonic modes all have fermionic partners with the same dispersion relation, imposes the constraints \eqref{eq:sugraup} and \eqref{eq:susyup}, which were derived in the previous section from requiring supersymmetry in the pp-wave limit.

Then, solving the equations of motion gives the relations
\begin{equation}
A_+ = \frac{1}{d_+} V_+ B_+^\star~, \qquad C_+ = -\frac{1}{d_+}W_+ B_+^\star~, \qquad B_- = \frac{1}{d_-}V_- A_-^\star ~, \qquad D_- = \frac{1}{d_-}W_- A_-^\star~,
\end{equation}
\begin{equation}
D_+ = -\frac{m}{d_+} ( f_+ + \tilde{\mathbf{Z}}_+^\dagger \tilde{\mathbf{Z}}_+^t) B_+~, \qquad C_- = \frac{m}{d_-} (f_- + \tilde{\mathbf{Z}}_+^\dagger \tilde{\mathbf{Z}}_+^t) A_-~,
\end{equation}
with
\begin{align}
V_\pm &= - m (\omega_\pm - p \pm \gamma_3) \tilde{\mathbf{Z}}_+^\star - m (\omega_\pm -p \mp \gamma_3) \tilde{\mathbf{Z}}_+^\dagger ~, \qquad
W_\pm = m^2 \tilde{\mathbf{Z}}_+^\star + g_\pm \tilde{\mathbf{Z}}_+^\dagger ~,
\end{align}
and
\begin{equation}
\begin{aligned}
f_\pm &= (\omega_\pm + p \pm \gamma_3)(\omega_\pm -p \pm \gamma_3) -m^2~, \qquad
g_\pm = (\omega_\pm + p \pm \gamma_3)(\omega_\pm -p \mp \gamma_3) -||\mathbf{z}_+||^2~, \\
d_\pm &= (\omega_\pm - p \mp \gamma_3) f_\pm - (\omega_\pm - p \pm \gamma_3) ||\mathbf{z}_+||^2 = (\omega_\pm - p \pm \gamma_3) g_\pm - (\omega_\pm - p \mp \gamma_3) m^2 ~.
\end{aligned}
\end{equation}
The last step consists of fixing the as yet unconstrained coefficients of the matrices $A_-$ and $B_+$.
On the one hand, the Poisson brackets between the fields impose (the left-hand sides are $2 \times 2$ matrices)
\begin{align}
\label{eq:PBi}
0 &= \{\tilde{\zeta}_L, \tilde{\zeta}_L^t\} = A_+(p) B_+^t(p) + B_+(-p) A_+^t(-p) + A_-(p) B_-^t(p) + B_-(-p) A_-^t(-p)~, \\
0 &= \{\tilde{\zeta}_L, \tilde{\zeta}_R^t \} = A_+(p) D_+^t(p) + B_+(-p) C_+^t(-p) + A_-(p) D_-^t(p) + B_-(-p) C_-^t(-p)~, \\
0 &= \{\tilde{\zeta}_R, \tilde{\zeta}_R^t \} = C_+(p) D_+^t(p) + D_+(-p) C_+^t(-p) + C_-(p) D_-^t(p) + D_-(-p) C_-^t(-p)~, \\
0 &= \{\tilde{\zeta}_L, \tilde{\zeta}_R^\dagger\} = A_+(p) C_+^\dagger(p) + A_-(-p) C_-^\dagger(-p) + B_+(p) D_+^\dagger(p) + B_-(-p) D_-^\dagger(-p)~,\\
1 &= \{\tilde{\zeta}_L, \tilde{\zeta}_L^\dagger\} = A_+(p) A_+^\dagger(p) + A_-(-p) A_-^\dagger(-p) + B_+(p) B_+^\dagger(p) + B_-(-p) B_-^\dagger(-p)~, \\
1 &= \{\tilde{\zeta}_R, \tilde{\zeta}_R^\dagger\} = C_+(p) C_+^\dagger(p) + C_-(-p) C_-^\dagger(-p) + D_+(p) D_+^\dagger(p) + D_-(-p) D_-^\dagger(-p)~.
\label{eq:PBf}
\end{align}
These equations are linear equations for $\hat{A}=A_- A_-^\dagger$ and $\hat{B}=B_+ B_+^\dagger$.
On the other hand, requiring that the Hamiltonian takes the canonical form gives the equations
\begin{align}
\label{eq:norm1}
1 &= A_+^\dagger(p) A_+(p) + B_+^t(p) B_+^\star(p) + C_+^\dagger(p) C_+(p) + D_+^t(p) D_+^\star(p)~, \\
\label{eq:norm2}
1 &=A_-^\dagger(p) A_-(p) + B_-^t(p) B_-^\star(p) + C_-^\dagger(p) C_-(p) + D_-^t(p) D_-^\star(p)~, \\
\label{eq:norm3}
0 &= A_+^\dagger(p) A_-(p) +B_+^t(p) B_-^\star(p) + C_+^\dagger(p) C_-(p)+ D_+^t(p) D_-^\star(p)~, \\
\label{eq:norm4}
0 &= A_+^\dagger(p) B_-(-p) +B_+^t(p) A_-^\star(-p) + C_+^\dagger(p) D_-(-p)+ D_+^t(p) C_-^\star(-p)~.
\end{align}
The first two equations \eqref{eq:norm1} and \eqref{eq:norm2} are linear equations for $\check{A}=A_-^\dagger A_-$ and $\check{B}=B_+^\dagger B_+$ (note the opposite order with respect to $\hat A$ and $\hat B$).
The solution to \eqref{eq:PBi}--\eqref{eq:PBf} is unique, and so is the solution to \eqref{eq:norm1} and \eqref{eq:norm2}.
Namely,
\begin{equation}
\begin{aligned}
\label{eq:sol}
\hat{A}_{11} &= \hat{A}_{22} = \check{A}_{11} = \check{A}_{22} = \frac{m^2 + (\bar{\omega}-p)(\omega_+ - p + \gamma_3)}{4 \omega_+ \bar{\omega}}~, &\qquad
\hat{A}_{12} &=\hat{A}_{21}=\check{A}_{12}=\check{A}_{21}=0~, \\
\hat{B}_{11} &= \hat{B}_{22} = \check{B}_{11}=\check{B}_{22} =\frac{m^2 + (\bar{\omega}-p)(\omega_- - p - \gamma_3)}{4 \omega_- \bar{\omega}}~, &\quad
\hat{B}_{12}&=\hat{B}_{21}=\check{B}_{12}=\check{B}_{21}=0~.
\end{aligned}
\end{equation}
Therefore, we must have
\begin{equation}
\begin{aligned}
|A_{-,11}| &= |A_{-,22}|~, &\qquad |A_{-,12}| &= |A_{-,21}|~, &\qquad A_{-,12} A_{-,11}^\star &= - A_{-,21}^\star A_{-,22}~, \\
|B_{+,11}| &= |B_{+,22}|~, &\qquad |B_{+,12}| &= |B_{+,21}|~, &\qquad B_{+,12} B_{+,11}^\star &= - B_{+,21}^\star B_{+,22}~.
\end{aligned}
\end{equation}
There are two residual freedoms, acting as $A_- \rightarrow \mathcal U_- A_-$ and $B_+ \rightarrow \mathcal U_+ B_+$ where $\mathcal U_\pm$ are unitary matrices.
For simplicity we will take the matrices $A_-$ and $B_+$ to be diagonal, with equal real entries on the diagonal.

\paragraph{Rational limit.} As the expressions for the coefficients are a bit obscure, let us consider the rational limit (or any limit with $\mathbf{z}_+=0$).
In this case, the matrix $\mathbf{Z}_+=0$, and the unitary transformation \eqref{eq:rotation} diagonalises the equations of motion: $\tilde{\zeta}_{L,1}$ only couples to $\tilde{\zeta}_{R,1}$, and $\tilde{\zeta}_{L,2}$ only couples to $\tilde{\zeta}_{R,2}$.
The equations of motion then impose the relations
\begin{equation}
A_{+} = B_{-} = C_{+}= D_{-}=0~, \qquad C_{-} = (\bar{\omega}+p) A_{-}~, \qquad D_{+} = -(\bar{\omega} + p) B_{+} ~,
\end{equation}
while the canonical Poisson brackets (or equivalently requiring that the Hamiltonian takes the canonical form) lead to
\begin{equation}
A_- A_-^\dagger + C_- C_-^\dagger=1~,\qquad B_+ B_+^\dagger+ D_+ D_+^\dagger =1~.
\end{equation}
The above solution with diagonal $A_-$ and $B_+$ is then
\begin{equation}
\begin{aligned}
&A_{-,11} = A_{-,22} = B_{+,11} = B_{+,22} = \sqrt{\frac{\bar{\omega}-p}{2\bar{\omega}}}~, \\
&C_{-,11} = C_{-,22} = -D_{+,11} =-D_{+,22}= \sqrt{\frac{\bar{\omega}+p}{2\bar{\omega}}}~.
\end{aligned}
\end{equation}

%%%%%%%%%%%%%%%%%%%%%%%%%%%%%%%%%%%%%%%%%%%%%%%%%%%%%%%%%%%%%%%%%%%%%%%%%%%%%%%%
\section{Tree-level S-matrix}\label{sec:Smatrix}

Now that we have diagonalised the quadratic Hamiltonian and obtained the asymptotic states, we can study how these excitations interact.
The first interaction terms in the light-cone gauge-fixed theory appear at quartic order through $\mathcal L_4$.
This quartic Lagrangian, as well as its counterpart, the quartic Hamiltonian density $\mathcal H_4$, can be rewritten in terms of oscillators using the mode expansion worked out in the previous section.
It takes the schematic form
\begin{equation}
\begin{aligned}
\mathcal H_4 = \int dp_1 dp_2 dp_3 dp_4 \Big(& h_4^0 \, e^{i (\omega_1 + \omega_2 + \omega_3 + \omega_4) \tau } e^{-i (p_1+p_2+p_3+p_4)\sigma} \\
&+ h_3^1 \, e^{i (\omega_1 + \omega_2 + \omega_3 - \omega_4)\tau} e^{-i (p_1+p_2+p_3-p_4)\sigma} \\
&+ h_2^2 \, e^{i (\omega_1 + \omega_2 - \omega_3 - \omega_4)\tau} e^{-i (p_1+p_2-p_3-p_4)\sigma} + \hc \Big)~,
\end{aligned}
\end{equation}
where we have suppressed all quantum numbers for readability.
The quantities $h_{n_c}^{n_a}$ contains $n_c$ creation operators and $n_a$ annihilation operators.
These operators, as well as the energies $\omega$, are labelled by $j=1,2$ and $\mu=\pm$ and depend on the momenta (for instance, $\omega_1 = \omega_{j_1,\mu_1}(p_1)$ with $j_1=1,2$ and $\mu_1=\pm$).
The tree-level S-matrix is obtained through a double integration
\begin{equation}
\begin{aligned}
\mathbb{T} &= \int d\tau H_4 = \int d\tau d\sigma \, \mathcal H_4 & \\
&= \int dp_1 dp_2 dp_3 dp_4 \Big( h_4^0 \, \delta(p_1+p_2+p_3+p_4) \delta(\omega_1 + \omega_2 + \omega_3 + \omega_4) \\
&\phantom{= \int dp_1 dp_2 dp_3 dp_4 } \quad h_3^1 \, \delta(p_1+p_2+p_3-p_4) \delta(\omega_1 + \omega_2 + \omega_3 - \omega_4)\\
&\phantom{= \int dp_1 dp_2 dp_3 dp_4 } \quad h_2^2 \, \delta(p_1+p_2-p_3-p_4) \delta(\omega_1 + \omega_2 - \omega_3 - \omega_4) + \hc
\Big)~.
\end{aligned}
\end{equation}

\subsection{Elliptic tree-level S-matrix}

To compute the tree-level S-matrix, we only consider massive modes, and focus on a choice of parameters such that the full background, not only its pp-wave limit, admits eight 6d supersymmetries.
As discussed in \secref{sec:sugra-susy} and \appref{app:Gram}, this imposes that the vectors $\mathbf{y}_\pm$ and $\mathbf{z}_\pm$ lie in a 2-plane.
Henceforth, we will assume that this 2-plane is spanned by the $(1,4)$ directions, so that one may use the parametrisation \eqref{eq:explicitsolution}.
At this point, let us note that such a choice implies $\beta_{12}=\beta_{13}=0$ and $\beta=|\beta_{14}|$ for the quantities defined in~\eqref{eq:beta}.
The rotation matrix $\mathcal V$ given in eq.~\eqref{eq:rotation-V} is well-defined when $\beta_{12}^2 +\beta_{13}^2 \neq 0$ (assuming the non-trivial $\beta \neq 0$ case).
However, in the limit
\begin{equation}
(\mathbf{y}_+)_{2,3} \rightarrow \epsilon (\tilde{\mathbf{y}}_+)_{2,3}~, \qquad (\mathbf{z_+})_{2,3}\rightarrow \epsilon (\tilde{\mathbf{z}}_+)_{2,3}~, \qquad \beta_{12} \rightarrow \epsilon \tilde{\beta}_{12}~, \qquad \beta_{23} \rightarrow \epsilon \tilde{\beta}_{23}~, \qquad \epsilon \rightarrow 0~,
\end{equation}
the rotation matrix remains finite, with
\begin{equation}
\mathcal V \rightarrow \left\{ \begin{aligned}
&1 \qquad  & \beta_{14}<0 \\
&\frac{-i \tilde{\beta}_{13} \sigma_1 -i \tilde{\beta}_{12} \sigma_2 }{\sqrt{(\tilde{\beta}_{12})^2+(\tilde{\beta}_{13})^2}} \qquad & \beta_{14}>0
\end{aligned} \right.~, \qquad \qquad \tilde{\mathbf{Z}}_+ \rightarrow \frac{1}{m}(-\beta \sigma_1 + (\yz) \sigma_2)~.
\end{equation}
This shows that the rotation \eqref{eq:rotation-V} is still well-defined even when restricting to the $(1,4)$ plane.

We find that, solving the energy- and momentum-conservation $\delta$-functions, the only contribution to $\mathbb{T}$ comes from the terms with an equal number of creation and annihilation operators, and with either $(p_1=p_3, \mu_1=\mu_3, p_2=p_4,\mu_2=\mu_4)$ or $(p_1=p_4,\mu_1=\mu_4,p_2=p_3,\mu_2=\mu_3)$.
Moreover, the integration over the momenta $p_3$ and $p_4$ yields the Jacobian $J = \frac{1}{|\omega_1' - \omega_2'|}$.
The S-matrix encodes the scattering between an excitation with momentum $p_1$ and another with momentum $p_2$, created above the vacuum $\left|0\right>$.
For such a two-particle state we use the notation
\begin{equation}
\left| a^\dagger_{\mu,j} a^\dagger_{\nu,k} \right> = a^\dagger_{\mu,j}(p_1) a^\dagger_{\nu,k}(p_2) \left|0\right>~, \qquad p_1 > p_2~.
\end{equation}
The tree-level S-matrix acts on the incoming states as follows:
\unskip\footnote{With respect to the notation used in~\cite{Hoare:2023zti} we define
\begin{equation*}
l_1 = 2\mathcal A_{\pm \pm}^\text{there} + \mathcal B_{\pm \pm}^\text{there}~, \qquad l_2 = 2\mathcal A_{\pm \mp}^\text{there} + \mathcal B_{\pm \mp}^\text{there}~, \qquad l_3 = - 2\mathcal G^\text{there}~, \qquad c=-\mathcal D^\text{there}~.
\end{equation*}}
\paragraph{Boson-Boson}
\begin{equation} \label{eq:BB}
\begin{aligned}
\mathbb T \left|a_{\pm,1}^\dagger a_{\pm,1}^\dagger\right> &= (-l_1+c)\left|a_{\pm,1}^\dagger a_{\pm,1}^\dagger\right> - l_7\left|b_{\pm,1}^\dagger b_{\pm,2}^\dagger\right> +l_7^\star \left|b_{\pm,2}^\dagger b_{\pm,1}^\dagger\right> ~,\\
\mathbb T \left|a_{\pm,1}^\dagger a_{\mp,1}^\dagger\right> &= (-l_2 +c)\left|a_{\pm,1}^\dagger a_{\mp,1}^\dagger\right> - l_4\left|b_{\pm,1}^\dagger b_{\mp,1}^\dagger\right> -l_4^\star \left|b_{\pm,2}^\dagger b_{\mp,2}^\dagger\right> ~, \\
\mathbb T \left|a_{\pm,1}^\dagger a_{\pm,2}^\dagger\right> &= (-l_3+c)\left|a_{\pm,1}^\dagger a_{\pm,2}^\dagger\right> + l_5\left|b_{\pm,1}^\dagger b_{\pm,2}^\dagger\right> - l_5^\star \left|b_{\pm,2}^\dagger b_{\pm,1}^\dagger\right> ~,\\
\mathbb T \left|a_{\pm,1}^\dagger a_{\mp,2}^\dagger\right> &= (-l_3+c)\left|a_{\pm,1}^\dagger a_{\mp,2}^\dagger\right> -l_6\left|b_{\pm,1}^\dagger b_{\mp,1}^\dagger\right> -l_6^\star \left|b_{\pm,2}^\dagger b_{\mp,2}^\dagger\right> ~,\\
\vspace{0.3 cm} \\
\mathbb T \left|a_{\pm,2}^\dagger a_{\pm,2}^\dagger\right> &= (l_1+c)\left|a_{\pm,2}^\dagger a_{\pm,2}^\dagger\right> +l_7\left|b_{\pm,1}^\dagger b_{\pm,2}^\dagger\right> -l_7^\star \left|b_{\pm,2}^\dagger b_{\pm,1}^\dagger\right> ~,\\
\mathbb T \left|a_{\pm,2}^\dagger a_{\mp,2}^\dagger\right> &= (l_2+c)\left|a_{\pm,2}^\dagger a_{\mp,2}^\dagger\right> +l_4\left|b_{\pm,1}^\dagger b_{\mp,1}^\dagger\right> + l_4^\star\left|b_{\pm,2}^\dagger b_{\mp,2}^\dagger\right> ~,\\
\mathbb T \left|a_{\pm,2}^\dagger a_{\pm,1}^\dagger\right> &= (l_3+c)\left|a_{\pm,2}^\dagger a_{\pm,1}^\dagger\right> + l_5\left|b_{\pm,1}^\dagger b_{\pm,2}^\dagger\right> -l_5^\star \left|b_{\pm,2}^\dagger b_{\pm,1}^\dagger\right> ~,\\
\mathbb T \left|a_{\pm,2}^\dagger a_{\mp,1}^\dagger\right> &= (l_3+c)\left|a_{\pm,2}^\dagger a_{\mp,1}^\dagger\right> -l_6\left|b_{\pm,1}^\dagger b_{\mp,1}^\dagger\right> -l_6^\star \left|b_{\pm,2}^\dagger b_{\mp,2}^\dagger\right> ~,
\end{aligned}
\end{equation}
\paragraph{Fermion-Fermion}

\begin{equation} \label{eq:FF}
\begin{aligned}
\mathbb T \left|b_{\pm,1}^\dagger b_{\pm,1}^\dagger\right> &= c \left|b_{\pm,1}^\dagger b_{\pm,1}^\dagger\right>~, \qquad \mathbb T \left|b_{\pm,1}^\dagger b_{\mp,2}^\dagger\right> = c \left|b_{\pm,1}^\dagger b_{\mp,2}^\dagger\right>~, \\
\mathbb T \left|b_{\pm,2}^\dagger b_{\pm,2}^\dagger\right> &= c \left|b_{\pm,2}^\dagger b_{\pm,2}^\dagger\right>~, \qquad
\mathbb T \left|b_{\pm,2}^\dagger b_{\mp,1}^\dagger\right> = c \left|b_{\pm,2}^\dagger b_{\mp,1}^\dagger\right>~, \\
\vspace{0.3 cm} \\
\mathbb T \left|b_{\pm,1}^\dagger b_{\mp,1}^\dagger\right> &=c \left| b_{\pm,1}^\dagger b_{\mp,1}^\dagger \right>-l_4^\star \left(\left|a_{\pm,1}^\dagger a_{\mp,1}^\dagger\right> -\left|a_{\pm,2}^\dagger a_{\mp,2}^\dagger\right> \right)-l_6^\star \left(\left|a_{\pm,1}^\dagger a_{\mp,2}^\dagger\right> + \left|a_{\pm,2}^\dagger a_{\mp,1}^\dagger\right>\right) ~,\\
\mathbb T \left|b_{\pm,2}^\dagger b_{\mp,2}^\dagger\right> &= c \left| b_{\pm,2}^\dagger b_{\mp,2}^\dagger \right>-l_4 \left(\left|a_{\pm,1}^\dagger a_{\mp,1}^\dagger\right> - \left|a_{\pm,2}^\dagger a_{\mp,2}^\dagger\right> \right)-l_6 \left(\left|a_{\pm,1}^\dagger a_{\mp,2}^\dagger\right> + \left|a_{\pm,2}^\dagger a_{\mp,1}^\dagger\right> \right) ~,\\
\mathbb T \left|b_{\pm,1}^\dagger b_{\pm,2}^\dagger\right> &= c \left| b_{\pm,1}^\dagger b_{\pm,2}^\dagger \right>+ l_5^\star\left(\left|a_{\pm,1}^\dagger a_{\pm,2}^\dagger\right> +\left|a_{\pm,2}^\dagger a_{\pm,1}^\dagger\right> \right) -l_7^\star\left(\left|a_{\pm,1}^\dagger a_{\pm,1}^\dagger\right> -\left|a_{\pm,2}^\dagger a_{\pm,2}^\dagger\right>\right) ~,\\
\mathbb T \left|b_{\pm,2}^\dagger b_{\pm,1}^\dagger\right> &= c \left| b_{\pm,2}^\dagger b_{\pm,1}^\dagger \right> - l_5 \left( \left|a_{\pm,1}^\dagger a_{\pm,2}^\dagger\right> + \left|a_{\pm,2}^\dagger a_{\pm,1}^\dagger\right> \right)+l_7\left(\left|a_{\pm,1}^\dagger a_{\pm,1}^\dagger\right> - \left|a_{\pm,2}^\dagger a_{\pm,2}^\dagger\right>\right) ~,
\end{aligned}
\end{equation}

\paragraph{Boson-Fermion}
\begin{equation} \label{eq:BF}
\begin{aligned}
\mathbb T \left|a_{\pm,1}^\dagger b_{\pm,1}^\dagger\right> &= (-\frac{1}{2}(l_1+l_3)+c) \left|a_{\pm,1}^\dagger b_{\pm,1}^\dagger\right> + l_5 \left|b_{\pm,1}^\dagger a_{\pm,1}^\dagger\right> + l_7\left|b_{\pm,1}^\dagger a_{\pm,2}^\dagger\right> ~,\\
\mathbb T \left|a_{\pm,1}^\dagger b_{\mp,1}^\dagger\right> &= (-\frac{1}{2}(l_2+l_3) +c) \left|a_{\pm,1}^\dagger b_{\mp,1}^\dagger\right> + l_4^\star \left|b_{\pm,2}^\dagger a_{\mp,2}^\dagger\right> -l_6^\star\left|b_{\pm,2}^\dagger a_{\mp,1}^\dagger\right> ~,\\
\mathbb T \left|a_{\pm,1}^\dagger b_{\pm,2}^\dagger\right> &= (-\frac{1}{2}(l_1+l_3) +c) \left|a_{\pm,1}^\dagger b_{\pm,2}^\dagger\right> + l_5^\star \left|b_{\pm,2}^\dagger a_{\pm,1}^\dagger\right> +l_7^\star\left|b_{\pm,2}^\dagger a_{\pm,2}^\dagger\right> ~,\\
\mathbb T \left|a_{\pm,1}^\dagger b_{\mp,2}^\dagger\right> &= (-\frac{1}{2}(l_2+l_3)+c) \left|a_{\pm,1}^\dagger b_{\mp,2}^\dagger\right> - l_4 \left|b_{\pm,1}^\dagger a_{\mp,2}^\dagger\right>+ l_6\left|b_{\pm,1}^\dagger a_{\mp,1}^\dagger\right> ~,\\
\vspace{0.3 cm} \\
\mathbb T \left|a_{\pm,2}^\dagger b_{\pm,2}^\dagger\right> &= (+\frac{1}{2}(l_1+l_3) +c) \left|a_{\pm,2}^\dagger b_{\pm,2}^\dagger\right> - l_5^\star \left|b_{\pm,2}^\dagger a_{\pm,2}^\dagger\right> + l_7^\star\left|b_{\pm,2}^\dagger a_{\pm,1}^\dagger\right> ~,\\
\mathbb T \left|a_{\pm,2}^\dagger b_{\mp,2}^\dagger\right> &= (+\frac{1}{2}(l_2+l_3)+c) \left|a_{\pm,2}^\dagger b_{\mp,2}^\dagger\right> - l_4 \left|b_{\pm,1}^\dagger a_{\mp,1}^\dagger\right> -l_6\left|b_{\pm,1}^\dagger a_{\mp,2}^\dagger\right> ~,\\
\mathbb T \left|a_{\pm,2}^\dagger b_{\pm,1}^\dagger\right> &= (+\frac{1}{2}(l_1+l_3) +c) \left|a_{\pm,2}^\dagger b_{\pm,1}^\dagger\right> - l_5 \left|b_{\pm,1}^\dagger a_{\pm,2}^\dagger\right> +l_7\left|b_{\pm,1}^\dagger a_{\pm,1}^\dagger\right> ~,\\
\mathbb T \left|a_{\pm,2}^\dagger b_{\mp,1}^\dagger\right> &= (+\frac{1}{2}(l_2+l_3) +c) \left|a_{\pm,2}^\dagger b_{\mp,1}^\dagger\right> +l_4^\star \left|b_{\pm,2}^\dagger a_{\mp,1}^\dagger\right> + l_6^\star\left|b_{\pm,2}^\dagger a_{\mp,2}^\dagger\right> ~,
\end{aligned}
\end{equation}

\paragraph{Fermion-Boson}
\begin{equation} \label{eq:FB}
\begin{aligned}
\mathbb T \left|b_{\pm,1}^\dagger a_{\pm,1}^\dagger \right> &= (-\frac{1}{2}(l_1-l_3)+c) \left|b_{\pm,1}^\dagger a_{\pm,1}^\dagger \right> + l_5^\star \left|a_{\pm,1}^\dagger b_{\pm,1}^\dagger \right> + l_7^\star\left|a_{\pm,2}^\dagger b_{\pm,1}^\dagger \right> ~,\\
\mathbb T \left| b_{\mp,1}^\dagger a_{\pm,1}^\dagger\right> &= (-\frac{1}{2}(l_2-l_3) +c) \left| b_{\mp,1}^\dagger a_{\pm,1}^\dagger\right> - l_4^\star \left|a_{\mp,2}^\dagger b_{\pm,2}^\dagger \right> +l_6^\star\left|a_{\mp,1}^\dagger b_{\pm,2}^\dagger \right> ~,\\
\mathbb T \left|b_{\pm,2}^\dagger a_{\pm,1}^\dagger \right> &= (-\frac{1}{2}(l_1-l_3) +c) \left| b_{\pm,2}^\dagger a_{\pm,1}^\dagger\right> + l_5 \left| a_{\pm,1}^\dagger b_{\pm,2}^\dagger\right> +l_7\left| a_{\pm,2}^\dagger b_{\pm,2}^\dagger\right> ~,\\
\mathbb T \left| b_{\mp,2}^\dagger a_{\pm,1}^\dagger\right> &= (-\frac{1}{2}(l_2-l_3)+c) \left|b_{\mp,2}^\dagger a_{\pm,1}^\dagger \right> +l_4 \left|a_{\mp,2}^\dagger b_{\pm,1}^\dagger \right>- l_6\left|a_{\mp,1}^\dagger b_{\pm,1}^\dagger \right> ~,\\
\vspace{0.3 cm} \\
\mathbb T \left| b_{\pm,2}^\dagger a_{\pm,2}^\dagger\right> &= (+\frac{1}{2}(l_1-l_3) +c) \left| b_{\pm,2}^\dagger a_{\pm,2}^\dagger\right> - l_5 \left| a_{\pm,2}^\dagger b_{\pm,2}^\dagger\right> + l_7\left| a_{\pm,1}^\dagger b_{\pm,2}^\dagger\right> ~,\\
\mathbb T \left|b_{\mp,2}^\dagger a_{\pm,2}^\dagger \right> &= (+\frac{1}{2}(l_2-l_3)+c) \left| b_{\mp,2}^\dagger a_{\pm,2}^\dagger\right> + l_4 \left| a_{\mp,1}^\dagger b_{\pm,1}^\dagger\right> +l_6\left| a_{\mp,2}^\dagger b_{\pm,1}^\dagger\right> ~,\\
\mathbb T \left| b_{\pm,1}^\dagger a_{\pm,2}^\dagger\right> &= (+\frac{1}{2}(l_1-l_3) +c) \left| b_{\pm,1}^\dagger a_{\pm,2}^\dagger\right> - l_5^\star \left|a_{\pm,2}^\dagger b_{\pm,1}^\dagger \right> +l_7^\star\left| a_{\pm,1}^\dagger b_{\pm,1}^\dagger\right> ~,\\
\mathbb T \left| b_{\mp,1}^\dagger a_{\pm,2}^\dagger\right> &= (+\frac{1}{2}(l_2-l_3) +c) \left| b_{\mp,1}^\dagger a_{\pm,2}^\dagger\right> -l_4^\star \left| a_{\mp,1}^\dagger b_{\pm,2}^\dagger\right> - l_6^\star\left| a_{\mp,2}^\dagger b_{\pm,2}^\dagger\right> ~.
\end{aligned}
\end{equation}

All coefficients depend on the momenta $p_1$ and $p_2$, and carry two implicit indices $\mu_1,\mu_2 \in \{\pm\}$ that correspond to the quantum numbers of the incoming (and also outgoing) excitations.
For instance, $l_1 = l_{1,\pm \pm}(p_1,p_2)$ and $l_2 = l_{2,\pm \mp}(p_1,p_2)$.
Their explicit expressions are given by
\unskip\footnote{Note that we have used the explicit form of the dispersion relations~\eqref{eq:disp} to simplify these expressions.}
\begin{equation}
\begin{aligned}
l_{1} &= \frac{1}{2} \frac{p_1 \omega_{\mu_2}(p_2) + p_2 \omega_{\mu_1}(p_1)}{p_1^2-p_2^2} \Big(p_1^2 + p_2^2 + 2 (\bar{\omega}(p_1)-\mu_1 \gamma_1)(\bar{\omega}(p_2)-\mu_2 \gamma_1)\Big)~, \\
l_{2} &= \frac{1}{2} \frac{p_1 \omega_{\mu_2}(p_2) + p_2 \omega_{\mu_1}(p_1)}{p_1^2-p_2^2} \Big(p_1^2 + p_2^2 - 2 (\bar{\omega}(p_1)-\mu_1 \gamma_1)(\bar{\omega}(p_2)-\mu_2 \gamma_1)\Big)~, \\
l_{3} &= -\frac{1}{2} (p_1 \omega_{\mu_2}(p_2) + p_2 \omega_{\mu_1}(p_1))~, \\
c &= -\left(a-\frac{1}{2} \right)\left(p_1 \omega_{\mu_2}(p_2) - p_2 \omega_{\mu_1}(p_1)\right)~,
\end{aligned}
\end{equation}
and
\begin{equation}
\begin{aligned}
l_{4,\mu_1 \mu_2} &= \frac{1}{\sqrt{d_{\mu_1}(p_1)}\sqrt{d_{\mu_2}(p_2)}} \Big[ (\bar{\omega}_1 - p_1 -\mu_1 \gamma_1)(\bar{\omega}_2 - p_2 -\mu_2 \gamma_1)\times \\
&\quad\Big(l_4^{(0)} (f^{++}_{\mu_1 \mu_2} +f^{+-}_{\mu_1 \mu_2} +f^{-+}_{\mu_1 \mu_2} +f^{--}_{\mu_1 \mu_2}) + l_4^{(3)} (f^{++}_{\mu_1 \mu_2} + f^{--}_{\mu_1 \mu_2}) \\
&\quad + l_4^{(1)} (f^{++}_{\mu_1 \mu_2} - f^{--}_{\mu_1 \mu_2}) + l_4^{(2)} (f^{+-}_{\mu_1 \mu_2} - f^{-+}_{\mu_1 \mu_2}) \Big)\Big]~, \\
l_{5,\mu_1 \mu_2} &= \frac{1}{\sqrt{d_{\mu_1}(p_1)}\sqrt{d_{\mu_2}(p_2)}} \Big[(\bar{\omega}_1 - p_1 -\mu_1 \gamma_1)(\bar{\omega}_2 - p_2 -\mu_2 \gamma_1)\times \\
&\quad\Big(l_5^{(0)} (f^{++}_{\mu_1 \mu_2} +f^{+-}_{\mu_1 \mu_2} +f^{-+}_{\mu_1 \mu_2} +f^{--}_{\mu_1 \mu_2}) + l_5^{(3)} (f_{\mu_1 \mu_2}^{+-} + f_{\mu_1 \mu_2}^{-+}) \\
&\quad + l_5^{(1)} (f^{++}_{\mu_1 \mu_2} - f^{--}_{\mu_1 \mu_2}) + l_5^{(2)} (f^{+-}_{\mu_1 \mu_2} - f^{-+}_{\mu_1 \mu_2}) \Big) \Big]~, \\
l_{6,\mu_1 \mu_2} &= \frac{1}{\sqrt{d_{\mu_1}(p_1)}\sqrt{d_{\mu_2}(p_2)}} \Big[ (\bar{\omega}_1 - p_1 -\mu_1 \gamma_1)(\bar{\omega}_2 - p_2 -\mu_2 \gamma_1)\times \\
&\quad\Big(l_6^{(0)} (f^{++}_{\mu_1 \mu_2} +f^{+-}_{\mu_1 \mu_2} +f^{-+}_{\mu_1 \mu_2} +f^{--}_{\mu_1 \mu_2}) + l_6^{(3)} (f^{+-}_{\mu_1 \mu_2} + f^{-+}_{\mu_1 \mu_2}) \\
&\quad+ l_6^{(1)} (f^{++}_{\mu_1 \mu_2} - f^{--}_{\mu_1 \mu_2}) + l_6^{(2)} (f^{+-}_{\mu_1 \mu_2} - f^{-+}_{\mu_1 \mu_2}) \Big)\Big]~, \\
l_{7,\mu_1 \mu_2} &= \frac{1}{\sqrt{d_{\mu_1}(p_1)}\sqrt{d_{\mu_2}(p_2)}} \Big[ (\bar{\omega}_1 - p_1 -\mu_1 \gamma_1)(\bar{\omega}_2 - p_2 -\mu_2 \gamma_1)\times \\
&\quad\Big(l_7^{(0)} (f^{++}_{\mu_1 \mu_2} +f^{+-}_{\mu_1 \mu_2} +f^{-+}_{\mu_1 \mu_2} +f^{--}_{\mu_1 \mu_2}) + l_7^{(3)} (f^{++}_{\mu_1 \mu_2} + f^{--}_{\mu_1 \mu_2}) \\
&\quad + l_7^{(1)} (f^{++}_{\mu_1 \mu_2} - f^{--}_{\mu_1 \mu_2}) + l_7^{(2)} (f^{+-}_{\mu_1 \mu_2} - f^{-+}_{\mu_1 \mu_2}) \Big)\Big]~.
\end{aligned}
\end{equation}
In the above, we introduced the auxiliary quantities
\begin{equation}
f^{\nu_1 \nu_2}_{\mu_1 \mu_2} = \sqrt{\bar{\omega}_1-\nu_1\gamma_2+\mu_1 \gamma_3} \sqrt{\bar{\omega}_2-\nu_2\gamma_2+\mu_2 \gamma_3}~,
\end{equation}
as well as
\begin{equation}
\begin{aligned}
l_4^{(0)} &= +\frac{\gamma_3}{4(p_1+p_2)} \Big((\bar{\omega}_1 +\mu_1 \gamma_3)(\bar{\omega}_2 + \mu_2 \gamma_3) - \gamma_1^2 +\gamma_3^2 + p_1 p_2 \Big) + i \frac{\gamma_3^2}{4} \frac{\yz}{\beta}~, \\
l_4^{(1)} &= -\frac{1}{4} \gamma_2 \gamma_3~, \qquad l_4^{(2)} = -\frac{\gamma_2\gamma_3}{4(p_1+p_2)} \Big(\bar{\omega}_1 - \bar{\omega}_2\Big)~, \\
l_4^{(3)} &= \frac{\beta^2}{4 \gamma_1^2 \gamma_3 (p_1+p_2)} (\gamma_1^2- (\bar{\omega}_1+p_1)(\bar{\omega}_2+p_2))~,
\end{aligned}
\end{equation}
\begin{equation}
\begin{aligned}
l_5^{(0)} &= -\frac{\gamma_3}{4(p_1-p_2)} \Big((\bar{\omega}_1 +\mu_1 \gamma_3)(\bar{\omega}_2 + \mu_2 \gamma_3) + \gamma_1^2 -\gamma_3^2 + p_1 p_2 \Big) + i \frac{\gamma_3^2}{4} \frac{\yz}{\beta}~, \\
l_5^{(1)} &= -\frac{\gamma_2\gamma_3}{4(p_1-p_2)} \Big(\bar{\omega}_1 + \bar{\omega}_2\Big)~, \qquad l_5^{(2)} = -\frac{1}{4} \gamma_2 \gamma_3~, \\
l_5^{(3)} &= \frac{\beta^2}{4 \gamma_1^2 \gamma_3 (p_1-p_2)} (\gamma_1^2+ (\bar{\omega}_1+p_1)(\bar{\omega}_2+p_2))~,
\end{aligned}
\end{equation}
\begin{equation}
\begin{aligned}
l_6^{(0)} &= i \frac{\gamma_1 \gamma_2}{\beta} \frac{\gamma_3^2}{4}~, \qquad l^{(3)} = -i \frac{\beta \gamma_2}{4 \gamma_1} \frac{1}{p_1-p_2}(p_1-p_2 + \bar{\omega}_1-\bar{\omega}_2)~, \\
l_6^{(1)} &= i \frac{\beta}{4 \gamma_1} \frac{1}{p_1-p_2} (p_1 \bar{\omega}_2 - p_2 \bar{\omega}_1) + (\yz) \frac{\gamma_3}{4 \gamma_1} \frac{p_1-p_2 + \bar{\omega}_1-\bar{\omega}_2}{p_1-p_2}~, \\
l_6^{(2)} &= -i \frac{\beta}{4 \gamma_1} \frac{1}{p_1-p_2} (\gamma_1^2 - (\bar{\omega}_1+\mu_1\gamma_3)(\bar{\omega_2}+\mu_2 \gamma_3) + (p_1 - \mu_1 \gamma_3)(p_2-\mu_2 \gamma_3))~,
\end{aligned}
\end{equation}
\begin{equation}
\begin{aligned}
l_7^{(0)} &= i \frac{\gamma_1 \gamma_2}{\beta} \frac{\gamma_3^2}{4}~, \qquad l_7^{(3)} = -i \frac{\beta \gamma_2}{4 \gamma_1} \frac{1}{p_1+p_2}(p_1 + p_2 + \bar{\omega}_1+\bar{\omega}_2)~, \\
l_7^{(1)} &= -i \frac{\beta}{4 \gamma_1} \frac{1}{p_1+p_2} (\gamma_1^2 + (\bar{\omega}_1+\mu_1\gamma_3)(\bar{\omega_2}+\mu_2 \gamma_3) - (p_1 - \mu_1 \gamma_3)(p_2-\mu_2 \gamma_3))~,\\
l_7^{(2)} &= -i \frac{\beta}{4 \gamma_1} \frac{1}{p_1+p_2}(p_1 \bar{\omega}_2 - p_2 \bar{\omega}_1) +(\yz) \frac{\gamma_3}{4 \gamma_1} \frac{p_1+p_2 + \bar{\omega}_1+\bar{\omega}_2}{p_1+p_2}~.
\end{aligned}
\end{equation}
The processes governed by $l_{6,\pm \mp}$ and $l_{7,\pm \pm}$ are new with respect to the undeformed theory (this will be discussed in more detail in the following subsections when considering the rational (\secref{sec:Smatrix-rational}) and trigonometric (\secref{sec:Smatrix-trigonometric}) limits of the S-matrix).
The tree-level S-matrix is compatible with physical unitarity, $\mathbb{T}^\dagger = \mathbb{T}$, as well as braiding unitarity $\mathbb{T} + P_{12}^g\mathbb{T}P_{12}^g=0$.
\unskip\footnote{The graded permutation operator $P^g$ acts on a two-particle state as $P^g \left| c^\dagger_{\ind{A}} c^\dagger_{\ind{B}}\right> = (-1)^{\epsilon_{\ind{A}}\epsilon_{\ind{B}}} \left| c^\dagger_{\ind{B}}c^\dagger_{\ind{A}}\right>$, where $\epsilon_{\ind{A}}$ is $0$ if $A$ is a bosonic index, and $\epsilon_{\ind{A}}$ is $1$ if $A$ is a fermionic index.}
It is invariant under a $\alg{u}(1) \oplus \alg{u}(1)$ symmetry generated by $\mathbf{M}$ and $\mathbf{Q}$, under which the fields are charged as
\begin{equation}
\begin{aligned}
(\mathbf{M},\mathbf{Q}) \left|a^\dagger_{\pm,1}\right> &= (\pm 1,0)\left|a^\dagger_{\pm,1}\right>~, \qquad (\mathbf{M},\mathbf{Q}) \left|b^\dagger_{\pm,1}\right> = (\pm 1,\pm 1)\left|b^\dagger_{\pm,1}\right>~,\\
(\mathbf{M},\mathbf{Q}) \left|a^\dagger_{\pm,2}\right> &= (\pm 1,0)\left|a^\dagger_{\pm,2}\right>~, \qquad (\mathbf{M},\mathbf{Q}) \left|b^\dagger_{\pm,2}\right> = (\pm 1,\mp 1)\left|b^\dagger_{\pm,2}\right>~.
\end{aligned}
\end{equation}
There are no reflection processes under the symmetry generated by $\mathbf{M}$: an incoming two-particle state with momenta and charges $(p_1,\mu_1)$ and $(p_2,\mu_2)$ is always mapped to an outgoing state where the excitation with momentum $p_1$ has charge $\mu_1$ and the excitation with momentum $p_2$ has charge $\mu_2$.
The invariance under $\mathbf{Q}$ is manifest in the quadratic Lagrangian $\mathcal L_{2,F}$, where it is realised on the fields as in eq.~\eqref{eq:symL2F}.
On the other hand, the symmetry generated by $\mathbf{M}$ is only manifest in the oscillator basis diagonalising the quadratic Hamiltonian.

Remarkably, despite its complicated structure, the tree-level S-matrix satisfies the classical Yang-Baxter equation
\begin{equation}
[ \mathbb T_{23}, \mathbb{T}_{13}] + [ \mathbb T_{23}, \mathbb{T}_{12}] + [ \mathbb T_{13}, \mathbb{T}_{12}] =0~,
\end{equation}
where
\begin{equation}
\mathbb{T}_{12} = \mathbb{T} \otimes 1~, \qquad \mathbb{T}_{23} = 1 \otimes \mathbb{T}~, \qquad \mathbb{T}_{13} = P_{23}^g \mathbb{T}_{12} P_{23}^g~.
\end{equation}
We checked this equation numerically in Mathematica for several generic deformation parameters and momenta.
Note that since we work with the Green-Schwarz action expanded to quadratic order in the fermions, we cannot obtain the S-matrix elements for the four-fermion processes.
The diagonal terms in eq.~\eqref{eq:FF} have been added by hand so that the classical Yang-Baxter equation remains satisfied even when considering four-fermion interactions.
Non-diagonal four-fermion contributions are not necessary, and in fact they would break the $\mathbf{Q}$ symmetry of the S-matrix.
From this analysis, we conclude that the elliptic deformation of the $\AdS_3 \times \Sp^3 \times \To^4$ superstring is integrable at the classical level, at least in the massive sector.

\subsection{Factorisation}

The massive tree-level S-matrix $\mathbb{T}$ is a $64 \times 64$ matrix.
Given that it is invariant under $\mathbf{M}$ with no reflection processes, it decomposes into four blocks $\mathbb{T}_{\pm \pm}$, $\mathbb{T}_{\pm \mp}$ each of size $16 \times 16$.
An interesting question is then if these further factorise into $4 \times 4$ blocks of 6-vertex or 8-vertex type.
If this is the case, then one could write, for each $16 \times 16$ block,
\begin{equation}
\label{eq:Sfactorised}
\mathbb{T}_{\mu_1 \mu_2} = 1 \otimes \mathcal T^{(2)}_{\mu_1 \mu_2} + \mathcal T^{(1)}_{\mu_1 \mu_2} \otimes 1~,
\end{equation}
where $\mathcal T^{(1)}_{\mu_1 \mu_2}$ and $\mathcal T^{(2)}_{\mu_1 \mu_2}$ may a priori be different $4 \times 4$ matrices, acting on the tensor product of a couple $(\phi_{\mu_1}, \psi_{\mu_2})$, where $\left|\phi\right>$ is bosonic and $\left|\psi\right>$ is fermionic, as ($j=1,2$)
\begin{equation}
\begin{aligned}
\mathcal T_{\mu_1 \mu_2}^{(j)} \left|\phi_{\mu_1} \phi_{\mu_2} \right> &= r_{1,\mu_1 \mu_2}^{(j)} \left|\phi_{\mu_1} \phi_{\mu_2} \right> + r_{8,\mu_1 \mu_2}^{(j)} \left| \psi_{\mu_1} \psi_{\mu_2} \right>~, \\
\mathcal T_{\mu_1 \mu_2}^{(j)} \left|\phi_{\mu_1} \psi_{\mu_2} \right> &= r_{2,\mu_1 \mu_2}^{(j)} \left| \phi_{\mu_1} \psi_{\mu_2}\right> + r_{6,\mu_1 \mu_2}^{(j)} \left| \psi_{\mu_1} \phi_{\mu_2}\right>~, \\
\mathcal T_{\mu_1 \mu_2}^{(j)} \left|\psi_{\mu_1} \phi_{\mu_2} \right> &= r_{3,\mu_1 \mu_2}^{(j)} \left| \psi_{\mu_1} \phi_{\mu_2}\right> + r_{5,\mu_1 \mu_2}^{(j)} \left| \phi_{\mu_1} \psi_{\mu_2}\right>~, \\
\mathcal T_{\mu_1 \mu_2}^{(j)} \left|\psi_{\mu_1} \psi_{\mu_2} \right> &= r_{4,\mu_1 \mu_2}^{(j)} \left|\psi_{\mu_1} \psi_{\mu_2} \right> + r_{7,\mu_1 \mu_2}^{(j)} \left| \phi_{\mu_1} \phi_{\mu_2} \right>~.
\end{aligned}
\end{equation}
The diagonal elements of \eqref{eq:BB} to \eqref{eq:FB} are compatible with this factorised structure with $\mathcal T=\mathcal T^{(1)} = \mathcal T^{(2)}$, coefficients
\begin{equation}
\label{eq:r-diag}
\begin{aligned}
r_{1,\pm \pm} &= \frac{1}{2}(l_{1,\pm\pm}+c_{\pm \pm})~, &\qquad
r_{1,\pm \mp}&= \frac{1}{2}(l_{2,\pm\mp}+c_{\pm \mp})~, \\
r_{2,\pm \pm}&=\frac{1}{2}(l_{3,\pm \pm}+c_{\pm \pm})~, &\qquad
r_{2,\pm \mp}&=\frac{1}{2}(l_{3,\pm \mp}+c_{\pm \mp})~, \\
r_{3,\pm \pm}&= \frac{1}{2}(-l_{3,\pm \pm}+c_{\pm \pm})~, &\qquad
r_{3,\pm \mp}&= \frac{1}{2}(-l_{3,\pm \mp}+c_{\pm \mp})~, \\
r_{4,\pm\pm}&=\frac{1}{2}(-l_{1,\pm\pm}+c_{\pm \pm})~, &\qquad
r_{4,\pm\mp}&=\frac{1}{2}(-l_{2,\pm\mp}+c_{\pm \mp})~,
\end{aligned}
\end{equation}
and an identification of states
\begin{equation}
\label{eq:change-basis}
\begin{aligned}
&\left|a_{\pm,1}^\dagger\right> = \left|\psi_\pm \otimes \psi_\pm\right>~, \qquad \left|a_{\pm,2}^\dagger\right> = \left|\phi_\pm \otimes \phi_\pm\right>~, \qquad
&\begin{pmatrix}
\left|b_{\pm,1}^\dagger\right> \\ \left|b_{\pm,2}^\dagger\right>
\end{pmatrix} = \mathcal U_\pm
\begin{pmatrix}\left| \phi_\pm \otimes \psi_\pm\right> \\
\left|\psi_\pm \otimes \phi_\pm\right>
\end{pmatrix}~.
\end{aligned}
\end{equation}
Here $\mathcal U_\pm$ represent two unitary one-particle change of basis matrices between fermionic degrees of freedom.
The diagonal elements are blind to this change of basis, but it may be required for the full tree-level S-matrix to factorise.

Let us consider the generic case where $l_4 \neq 0 \neq l_5$ and $l_6 \neq 0 \neq l_7$, and an identification of states of the form \eqref{eq:change-basis}.
Then, \eqref{eq:Sfactorised} imposes the following relations,
\begin{equation}
\begin{aligned}
r_{5,\pm \pm}^{(1)} &= r_{6,\pm \pm}^{(2)} =- \frac{\mathcal U_{\pm,12}(p_2)}{\mathcal U_{\pm,12}(p_1)} l_{5,\pm\pm}~, &\qquad r_{6,\pm \pm}^{(1)} &= r_{5,\pm \pm}^{(2)} = -\frac{\mathcal U_{\pm,11}(p_2)}{\mathcal U_{\pm,11}(p_1)}l_{5,\pm\pm}~, \\
r_{7,\pm \pm}^{(1)} &= r_{8,\pm \pm}^{(2)} =\frac{\mathcal U_{\pm,11}(p_2)}{\mathcal U_{\pm,12}(p_1)} l_{7,\pm\pm}~, &\qquad r_{8, \pm \pm}^{(1)} &= r_{7,\pm \pm}^{(2)} = -\frac{\mathcal U_{\pm,12}(p_2)}{\mathcal U_{\pm,11}(p_1)}l_{7,\pm\pm} ~, \\
r_{5,\pm \mp}^{(1)} &= r_{6,\pm \mp}^{(2)} = -\frac{\mathcal U_{\mp,22}(p_2)}{\mathcal U_{\pm,12}(p_1)} l_{6,\pm \mp}~, &\qquad r_{6,\pm \mp}^{(1)} &= r_{5,\pm \mp}^{(2)} = -\frac{\mathcal U_{\mp,21}(p_2)}{\mathcal U_{\mp,11}(p_1)} l_{6,\pm \mp}~, \\
r_{7,\pm \mp}^{(1)} &= r_{8,\pm \mp} ^{(2)}= - \frac{\mathcal U_{\mp,21}(p_2)}{\mathcal U_{\pm,12}(p_1)} l_{4,\pm \mp}~, &\qquad r_{8,\pm \mp}^{(1)} &= r_{7, \pm \mp}^{(2)} = \frac{\mathcal U_{\mp,22}(p_2)}{\mathcal U_{\mp,11}(p_1)} l_{4,\pm \mp}~,
\end{aligned}
\end{equation}
where
\begin{equation}
\mathcal U_{\pm,21} = -\frac{1}{2 \mathcal U_{\pm, 12}}~, \qquad \mathcal U_{\pm, 22} = \frac{1}{2 \mathcal U_{\pm,11}}~,
\end{equation}
and the tree-level S-matrix elements must be related through
\begin{equation}
\begin{aligned}
l_{5,\pm \pm}^\star &= \frac{\mathcal U_{\pm,11}(p_2) \mathcal U_{\pm,12}(p_2)}{\mathcal U_{\pm,11}(p_1) \mathcal U_{\pm,12}(p_1) } l_{5,\pm \pm}~, \qquad l_{7,\pm \pm}^\star = -\frac{\mathcal U_{\pm,11}(p_2) \mathcal U_{\pm,12}(p_2)}{\mathcal U_{\pm,11}(p_1) \mathcal U_{\pm,12}(p_1) } l_{7,\pm \pm}~, \\
l_{4,\pm \mp}^\star &= \frac{\mathcal U_{\mp,22}(p_2) \mathcal U_{\mp,21}(p_2)}{\mathcal U_{\pm,11}(p_1) \mathcal U_{\pm,12}(p_1) } l_{4,\pm \mp} ~, \qquad l_{6,\pm \mp}^\star = -\frac{\mathcal U_{\mp,22}(p_2) \mathcal U_{\mp,21}(p_2)}{\mathcal U_{\pm,11}(p_1) \mathcal U_{\pm,12}(p_1) } l_{6,\pm \mp}~.
\end{aligned}
\end{equation}
In particular, such a factorisation is only possible if
\begin{equation}
\label{eq:fact-cond}
\frac{l_{5,\pm \pm}^\star}{l_{5,\pm \pm}} =-\frac{l_{7,\pm \pm}^\star}{l_{7,\pm \pm}}~, \qquad \frac{l_{6,\pm \mp}^\star}{l_{6,\pm \mp}} =-\frac{l_{8,\pm \mp}^\star}{l_{8,\pm \mp}}~.
\end{equation}
One can check that this requirement is not satisfied for the tree-level S-matrix for generic deformation parameters.
Therefore, for the elliptic deformation, it is not possible to write the tree-level S-matrix $\mathbb{T}$ in the factorised form~\eqref{eq:Sfactorised}.
This is consistent with the computation in~\cite{Hoare:2023zti}, where given the coefficients for $r_{1,2,3,4}$, it was not possible to find $r_{5,6,7,8}$ such that the 8-vertex S-matrix solves the classical Yang-Baxter equation.
Note that allowing for an additional one-particle change of basis for the bosons, in particular a different identification
\begin{equation}
\left|a_{+,1}^\dagger\right> = \left|\psi_+ \otimes \psi_+\right>~, \quad \left|a_{-,1}^\dagger\right> = \left|\phi_- \otimes \phi_-\right>~, \quad \left|a_{+,2}^\dagger\right> = \left|\phi_+ \otimes \phi_+\right>~, \quad \left|a_{-,2}^\dagger\right> = \left|\psi_- \otimes \psi_-\right>~,
\end{equation}
leads to the same inconsistency.
We did not explore the possibility of having more general change of bases that mix the bosonic and fermionic degrees of freedom.

\subsection{Rational limit}\label{sec:Smatrix-rational}

Let us start by considering the rational limit $\gamma_2=0$ and $\gamma_1=\gamma_3=1$, for which the dispersion relation becomes $\omega_\pm = \hat{\omega} \pm 1$ where $\hat{\omega} = \sqrt{p^2+1}$.
The off-diagonal matrix elements are given by (we use the $\text{R}$ superscript to specify that these expressions are only valid in the rational limit)
\begin{equation}
\begin{aligned}
l_4^\text{R}&= -\frac{p_1 p_2}{2(p_1 + p_2)} (\sqrt{\hat{\omega}_1+p_1}\sqrt{\hat{\omega}_2+p_2} - \sqrt{\hat{\omega}_1-p_1}\sqrt{\hat{\omega}_2-p_2})~, \\
l_5^\text{R}&= -\frac{p_1 p_2}{2(p_1 - p_2)} (\sqrt{\hat{\omega}_1+p_1}\sqrt{\hat{\omega}_2+p_2} + \sqrt{\hat{\omega}_1-p_1}\sqrt{\hat{\omega}_2-p_2})~, \\
l_6^\text{R} &= l_7^\text{R}=0~.
\end{aligned}
\end{equation}
These agree with the usual massive tree-level S-matrix results for the $\AdS_3 \times \Sp^3 \times \To^4$ superstring in the standard light-cone gauge as obtained in~\cite{Hoare:2013pma}.
For the diagonal matrix elements, instead of considering each coefficient separately, it is convenient to look at the linear combinations that enter the tree-level S-matrix.
One finds that the action of the tree-level S-matrix can again be brought into the form \eqref{eq:BB} to \eqref{eq:FB}, with
\unskip\footnote{We stress that our notation for the diagonal elements is such that, for instance,
\begin{equation*}
l_1^\text{R} \neq \left.l_1\right|_{\gamma_2 \rightarrow 0, \gamma_{1,3} \rightarrow 1} ~, \qquad (l_1^\text{R} + c^\text{R})\left|c^\dagger_{\pm}c^\dagger_{\pm}\right> = \left.(l_1 + c)\right|_{\gamma_2 \rightarrow 0, \gamma_{1,3} \rightarrow 1} \left|c^\dagger_{\pm}c^\dagger_{\pm}\right>~.
\end{equation*}
The coefficient $c^\text{R}$ is identified as acting diagonally on all the two-particle states.
Moreover, $c^\text{R}$ (or $\mathcal O$), should be understood as an operator acting on the two-particle state to its right-hand side.
}
\begin{equation}
\begin{aligned}
l_1^\text{R}&= \frac{(p_1+p_2)^2}{2(p_1 \hat{\omega}_2-p_2 \hat{\omega}_1)} = \frac{(p_1+p_2)(p_1 \hat{\omega}_2 + p_2 \hat{\omega}_1)}{2 (p_1-p_2)}~, \\
l_2^\text{R} &= \frac{(p_1-p_2)^2}{2(p_1 \hat{\omega}_2-p_2 \hat{\omega}_1)} = \frac{(p_1-p_2)(p_1 \hat{\omega}_2 + p_2 \hat{\omega}_1)}{2(p_1+p_2)}~, \\
l_3^\text{R} &= -\frac{p_1^2-p_2^2}{2(p_1 \hat{\omega}_2-p_2 \hat{\omega}_1)} = -\frac{1}{2} (p_1 \hat{\omega}_2 + p_2 \hat{\omega}_1)~, \\
c^\text{R} &=-\big(a-\frac{1}{2}\big) (p_1 \hat{\omega}_2- p_2 \hat{\omega}_1)- \mathcal O~,
\end{aligned}
\end{equation}
where
\begin{equation}
\label{eq:OJT}
\mathcal O = a (p_1 \mathbf{L}_2 - p_2 \mathbf{L}_1) + (a-1) (p_1 \mathbf{J}_2 - p_2 \mathbf{J}_1)~.
\end{equation}
The operators $\mathbf{L}$ and $\mathbf{J}$ are defined below, with the index indicating on which excitation the operator acts.
On top of the $\alg{u}(1)$ generated by $\mathbf Q$, the tree-level S-matrix is invariant under an additional $\alg{u}(1) \oplus \alg{u}(1)$ symmetry generated by $\mathbf{L}$ and $\mathbf{J}$, under which the one-particle states have charges
\begin{equation}
\begin{aligned}
&(\mathbf{L},\mathbf{J})\left|a_{\pm,1}^\dagger \right> =(\pm 1, 0)\left|a_{\pm,1}^\dagger \right>~, \qquad (\mathbf{L},\mathbf{J}) \left|b_{\pm,1}^\dagger \right> =(\pm \frac{1}{2}, \pm \frac{1}{2})\left|b_{\pm,1}^\dagger \right>~, \\
&(\mathbf{L},\mathbf{J}) \left|a_{\pm, 2}^\dagger \right> = (0, \pm1) \left|a_{\pm,2}^\dagger \right>~,\qquad (\mathbf{L},\mathbf{J}) \left|b_{\pm,2}^\dagger \right> = (\pm \frac{1}{2}, \pm \frac{1}{2})\left|b_{\pm,2}^\dagger \right>~.
\end{aligned}
\end{equation}
Note that in the elliptic deformation only $\mathbf{Q}$ and the linear combination $\mathbf{M}=\mathbf{L}+\mathbf{J}$ survive.
The $\AdS_3$ excitations $a_{\pm,1}^\dagger$ are only charged under $\mathbf{L}$, while the $\Sp^3$ excitations $a_{\pm,2}^\dagger$ are only charged under $\mathbf{J}$.
The fermionic excitations on the other hand are charged under both operators, with half-integer quantum numbers.
Given that the fermions $b^\dagger_{+,1}$ and $b^\dagger_{+,2}$ (respectively $b^\dagger_{-,1}$ and $b^\dagger_{-,2}$) have the same charges under both $\mathbf{L}$ and $\mathbf{J}$, a one-particle change of basis between these excitations does not spoil this special structure.
The coefficients $l_1^\text{R}, l_2^\text{R}, l_3^\text{R}$ also agree with the known results, but the diagonal tree-level S-matrix elements are shifted by the $JT$ operator $\mathcal O$, as expected from the different choice of light-cone gauge~\cite{Borsato:2023oru}.

The S-matrix factorises with $\mathcal T^{(1)} = \mathcal T^{(2)}$, states identified as in eq.~\eqref{eq:change-basis} with $\mathcal U_\pm =1$, and factorised S-matrix elements
\begin{equation}
\begin{aligned}
r_{1,\pm \pm} &= \frac{1}{2}(l_1^\text{R}+c^\text{R})~, &\qquad r_{8,\pm \pm} &=0~, &\qquad r_{1,\pm \mp} &= \frac{1}{2}(l_2^\text{R}+c^\text{R})~,&\qquad r_{8,\pm \mp} &= l_4^\text{R}~,\\
r_{2, \pm \pm} &= \frac{1}{2}(l_3^\text{R}+c^\text{R})~, &\qquad r_{6, \pm \pm} &= -l_5^\text{R} ~, &\qquad r_{2, \pm \mp} &= \frac{1}{2}(l_3^\text{R}+c^\text{R})~, &\qquad r_{6,\pm \mp} &= 0~, \\
r_{3, \pm \pm} &= \frac{1}{2}(-l_3^\text{R}+c^\text{R})~, &\qquad r_{5, \pm \pm} &= -l_5^\text{R} ~, &\qquad r_{3, \pm \mp} &= \frac{1}{2}(-l_3^\text{R}+c^\text{R})~, &\qquad r_{5,\pm \mp} &= 0~, \\
r_{4, \pm \pm} &= \frac{1}{2}(- l_1^\text{R}+c^\text{R})~, &\qquad r_{7,\pm \pm} &=0~, &\qquad r_{4, \pm \mp} &= \frac{1}{2}(- l_2^\text{R}+c^\text{R})~,&\qquad r_{7,\pm \mp} &= l_4^\text{R}~.
\end{aligned}
\end{equation}
The factorised states have charges
\begin{equation}
(\mathbf{L},\mathbf{J})\left|\psi_\pm \right> =(\pm \frac{1}{2}, 0)\left|\psi_\pm \right>~, \qquad (\mathbf{L},\mathbf{J})\left|\phi_\pm \right> =(0, \pm \frac{1}{2})\left|\phi_\pm\right>~,
\end{equation}
while the charge $\mathbf{Q}$ does not factorise.
The S-matrix $\mathcal T$ again satisfies the classical Yang-Baxter equation.
An exact S-matrix, satisfying the quantum Yang-Baxter equation, is known to all-loop order~\cite{Borsato:2013qpa}.

\subsection{Time-like trigonometric limit}\label{sec:Smatrix-trigonometric}

Let us now consider the trigonometric limit with $\gamma_1=1+\kappa^2$, $\gamma_2 = 0$ and $\gamma_3 = 1-\kappa^2$.
The excitations then have dispersion relations $\omega_\pm = \sqrt{p^2+(1+\kappa^2)^2} \pm (1-\kappa^2)$, which can be written as
\begin{equation}
\omega_\pm = \check{\omega}_\pm \pm 1~, \qquad \check{\omega}_\pm = \sqrt{p^2+(1+\kappa^2)^2} \mp \kappa^2~.
\end{equation}
The motivation behind this is that $\check{\omega}_\pm$ corresponds to the dispersion relation of the massive excitations when computed in the standard light-cone gauge (upon swapping the definition of $\omega_+$ and $\omega_-$)~\cite{Hoare:2014oua}, which makes it easier to compare our results to the existing literature.
The tree-level S-matrix can again be brought into the form \eqref{eq:BB} to \eqref{eq:FB}, with coefficients
\begin{equation}
\begin{aligned}
l_{1,\pm \pm}^\text{T} &= \frac{(p_1+p_2)(p_1 \check{\omega}_{\pm}(p_2) + p_2 \check{\omega}_{\pm}(p_1))\mp 4 \kappa^2(\check{\omega}_{\pm}(p_1)\mp 1)(\check{\omega}_{\pm}(p_2)\mp 1)}{2(p_1-p_2)}~, \\
l_{2,\pm \mp}^\text{T} &= \frac{(p_1-p_2)(p_1 \check{\omega}_{\mp}(p_2) + p_2 \check{\omega}_{\pm}(p_1))\mp 4 \kappa^2(\check{\omega}_{\pm}(p_1)\mp 1)(\check{\omega}_{\mp}(p_2)\pm 1)}{2(p_1+p_2)}~, \\
l_{3,\mu_1 \mu_2}^\text{T} &= -\frac{1}{2} (p_1 \check{\omega}_{\mu_2}(p_2) + p_2 \check{\omega}_{\mu_1}(p_1))~, \qquad
c^\text{T} =-\big(a-\frac{1}{2}\big) (p_1 \check{\omega}_2- p_2 \check{\omega}_1)- \mathcal O~,
\end{aligned}
\end{equation}
where $\mathcal O$ is the same as in \eqref{eq:OJT} and again originates from the different choice in the light-cone gauge-fixing.
The coefficients $l_{1,2,3}^\text{T}$ are in agreement with the tree-level S-matrix elements as computed from the unilateral inhomogeneous Yang-Baxter deformation in the standard light-cone gauge~\cite{Seibold:2021lju}.
The coefficients $l_{4,5,6,7}$ depend on the R-R fluxes in the theory.
As discussed in \secref{sec:sugra-susy}, there are two branches of supersymmetric backgrounds, each preserving 8 supersymmetries.

The first supersymmetric trigonometric branch is characterised by
\begin{equation}
\psi=\pi~, \qquad \beta = 2 \kappa \big|(1-\kappa^2) \sin \frac{\phi}{2} \big|~, \qquad \yz = 2 \kappa^2 \sin \phi~, \qquad \gamma_2=0~.
\end{equation}
Letting $\sigma = \operatorname{sgn} \big( (1-\kappa^2) \sin \frac{\phi}{2}\big)$, one can then check that
\begin{equation}
\begin{aligned}
l_4^{\text{T}_1} = \varphi_{\mu_1}^\sigma(p_1) \varphi_{\mu_2}^\sigma(p_2) \mathcal C_{\mu_1 \mu_2}~, \qquad l_5^{\text{T}_1} = \varphi_{\mu_1}^\sigma(p_1) \varphi^{-\sigma}_{\mu_2}(p_2) \mathcal H_{\mu_1 \mu_2}~, \qquad l_6^{\text{T}_1} = l_7^{\text{T}_1} =0~,
\end{aligned}
\end{equation}
where
\begin{align}
\label{eq:Smatrix-C}
\mathcal C_{\mu_1 \mu_2} = -\frac{1+\kappa^2}{p_1+p_2} \sqrt{\check{\omega}_{\mu_1}(p_1)^2-1} \sqrt{\check{\omega}_{\mu_2}(p_2)^2-1} \sinh \Big(\frac{1}{2} \text{arcsinh} \frac{p_1}{1+\kappa^2} + \frac{1}{2} \text{arcsinh} \frac{p_2}{1+\kappa^2} \Big)~, \\
\mathcal H_{\mu_1 \mu_2} = -\frac{1+\kappa^2}{p_1-p_2} \sqrt{\check{\omega}_{\mu_1}(p_1)^2-1} \sqrt{\check{\omega}_{\mu_2}(p_2)^2-1} \cosh \Big(\frac{1}{2} \text{arcsinh} \frac{p_1}{1+\kappa^2} + \frac{1}{2} \text{arcsinh} \frac{p_2}{1+\kappa^2} \Big) ~,
\label{eq:Smatrix-H}
\end{align}
and the phases are given by
\begin{equation}
\begin{aligned}
\varphi_\mu(p) &= \frac{1}{\sqrt{1+\kappa^2}} \frac{\sqrt{\bar{\omega}-p}}{\sqrt{(\bar{\omega}-p)(\bar{\omega}+\mu (1-\kappa^2))-2 \kappa^2 \sin^2 \frac{\phi}{2}}} \times \\
&\Big( \kappa \sqrt{1- \frac{1}{1+\kappa^2} \sin^2\frac{\phi}{2}} \sqrt{\bar{\omega} - \mu (1+\kappa^2)} + i \frac{p (\mu \cos \frac{\phi}{2} + i \frac{\kappa}{ 1+\kappa^2} \sin^2 \frac{\phi}{2} )}{\sqrt{1- \frac{1}{1+\kappa^2} \sin^2\frac{\phi}{2}} \sqrt{\bar{\omega} - \mu (1+\kappa^2)}} \Big)~,
\end{aligned}
\end{equation}
with $|\varphi_\mu(p)|=1$.
Note that $\mathcal H_{\mu_1 \mu_2}$ and $\mathcal C_{\mu_1 \mu_2}$ do not depend on the angle $\phi$; all the dependence is absorbed into the phases $\varphi_\mu(p)$.
Consequently, the $\phi \neq 0$ case is related to the $\phi=0$ case by a diagonal one-particle change of basis.
We recall that for $\phi=0$ we have $\mathbf{z}_\pm=0$, the supergravity background corresponds to the unilateral inhomogeneous Yang-Baxter deformation, and the phases simplify to
\begin{equation}
\left.\varphi_\mu(p) \right|_{\phi=0} = \frac{\sqrt{p+i \kappa (1-\mu \check{\omega}_\mu)}}{\sqrt{p-i \kappa (1-\mu \check{\omega}_\mu)}}~.
\end{equation}
These $\phi=0$ results are in agreement with the tree-level S-matrix computed in~\cite{Seibold:2021lju} for the Yang-Baxter deformation.
Because $l_6^{\text{T}_1}=l_7^{\text{T}_1}=0$, the tree-level S-matrix is invariant under the three $\alg{u}(1)$s generated by $\mathbf{Q}$, $\mathbf{L}$ and $\mathbf{J}$.
The tree-level S-matrix can again be written in the factorised form \eqref{eq:Sfactorised} with $\mathcal T^{(1)} = \mathcal T^{(2)}$, diagonal elements \eqref{eq:r-diag}, and
\begin{equation}
\begin{aligned}
& r_{5,\pm \pm}=r_{6,\pm\pm} = - \mathcal H~, \qquad r_{7,\pm \mp}=r_{8 \pm \mp} = \mathcal C~, \qquad \mathcal U_\pm = \begin{pmatrix} \varphi_\pm^\star & 0 \\ 0 & \varphi_\pm\end{pmatrix}~.
\end{aligned}
\end{equation}
The factorised S-matrix satisfies the classical Yang-Baxter equation.
Note that for this case, an exact S-matrix satisfying the quantum Yang-Baxter equation and reproducing the above tree-level results is known to all-loop order~\cite{Hoare:2014oua} (modulo dressing factors).
It can be bootstrapped using the $q$-deformed symmetries.

The second supersymmetric trigonometric branch is characterised by
\begin{equation}
\phi=0~, \qquad \beta=0~, \qquad \yz=0~, \qquad \gamma_2=0~,
\end{equation}
with finite ratio
\begin{equation}
\frac{\yz}{\beta} = \frac{2 \kappa}{1-\kappa^2} \sin \frac{\psi}{2} ~, \qquad \frac{\gamma_1 \gamma_2}{\beta}= \frac{2 \kappa}{1-\kappa^2} \cos \frac{\psi}{2} ~.
\end{equation}
In this case, one can check that
\begin{equation}
\begin{aligned}
l_{4}^{\text{T}_2} &= \text{Re} \left[ \varphi_{\mu_1}(p_1) \varphi_{\mu_2}(p_2)\right]_{\phi=0} \mathcal C_{\mu_1 \mu_2} +i \sin \frac{\psi}{2}\, \text{Im} \left[\varphi_{\mu_1}(p_1) \varphi_{\mu_2}(p_2)\right]_{\phi=0} \mathcal C_{\mu_1 \mu_2} ~, \\
l_{5}^{\text{T}_2} &= \text{Re} \left[ \varphi_{\mu_1}(p_1) \varphi^\star_{\mu_2}(p_2) \right]_{\phi=0} \mathcal H_{\mu_1 \mu_2} +i \sin \frac{\psi}{2}\, \text{Im} \left[\varphi_{\mu_1}(p_1) \varphi^\star_{\mu_2}(p_2) \right]_{\phi=0} \mathcal H_{\mu_1 \mu_2} ~, \\
l_{6}^{\text{T}_2} &= i \cos \frac{\psi}{2}\, \text{Im} \left[ \varphi_{\mu_1}(p_1) \varphi_{\mu_2}(p_2)\right]_{\phi=0} \mathcal C_{\mu_1 \mu_2} ~, \\
l_{7}^{\text{T}_2} &= i \cos \frac{\psi}{2}\, \text{Im} \left[ \varphi_{\mu_1}(p_1) \varphi^\star_{\mu_2}(p_2)\right]_{\phi=0} \mathcal H_{\mu_1 \mu_2} ~.
\end{aligned}
\end{equation}
The factorisation condition \eqref{eq:fact-cond} is only satisfied for $\psi=0$ and $\psi =\pi$.
The latter is a special case of the first trigonometric branch, with $l_6^{\text{T}_2}=l_7^{\text{T}_2}=0$.
Let us therefore focus on the former case.
Then it is possible to factorise the S-matrix with $\mathcal{T}^{(1)} = \mathcal T^{(2)}$ and
\begin{equation}
\begin{aligned}
&r_{5,\pm \pm} = r_{6,\pm \pm} = - \text{Re} \left[ \varphi_{\pm}(p_1) \varphi^\star_{\pm}(p_2) \right]_{\phi=0}\mathcal H_{\pm \pm}~,\\
&r_{7,\pm \pm} = r_{8,\pm \pm} = \mp \text{Im} \left[ \varphi_{\pm}(p_1) \varphi^\star_{\pm}(p_2) \right]_{\phi=0}\mathcal H_{\pm \pm}~, \\
&r_{7,\pm \mp} = r_{8,\pm \mp} = \text{Re} \left[ \varphi_{\pm}(p_1) \varphi^\star_{\mp}(p_2) \right]_{\phi=0}\mathcal C_{\pm \mp}~,\\
&r_{5,\pm \mp} = r_{6,\pm \mp} = \pm \text{Im} \left[ \varphi_{\pm}(p_1) \varphi^\star_{\mp}(p_2) \right]_{\phi=0}\mathcal C_{\pm \mp}~.
\end{aligned}
\end{equation}
The required change of basis matrices are
\begin{equation}
\mathcal U_\pm = \frac{1}{\sqrt{2}} \begin{pmatrix} 1 & \mp i \\ \mp i & 1 \end{pmatrix}~.
\end{equation}
The factorised S-matrix $\mathcal T$ again solves the classical Yang-Baxter equation.

\subsection{Space-like trigonometric and limit to the \texorpdfstring{$\AdS_2 \times \Sp^2$}{AdS3 × S3} S-matrix}

Two other interesting limits, already discussed in \secref{sec:sugra-susy}, are the space-like trigonometric limits corresponding to taking $\gamma_3 = \gamma_+$ or $\gamma_3 =\gamma_-$.
In terms of the original deformation parameters, these two limits correspond respectively to
\begin{equation}
\gamma_3 = \gamma_+~, \qquad \alpha_2 = \alpha_3~, \qquad \gamma_1 = \frac{\sqrt{\text{T}}}{\sqrt{\alpha_1}}~, \qquad \gamma_2 = \sqrt{\text{T}} \left(\frac{\sqrt{\alpha_1}}{\alpha_2} - \frac{1}{\sqrt{\alpha_1}}\right)~, \qquad \gamma_3 = \sqrt{\text{T}} \frac{\sqrt{\alpha_1}}{\alpha_2}~,
\end{equation}
and
\begin{equation}
\gamma_3 = \gamma_-~, \qquad \alpha_2 = \alpha_1~, \qquad \gamma_1 = \frac{\sqrt{\text{T}}}{\sqrt{\alpha_3}}~, \qquad \gamma_2 = \sqrt{\text{T}} \left(\frac{1}{\sqrt{\alpha_3}} - \frac{\sqrt{\alpha_3}}{\alpha_2}\right)~, \qquad \gamma_3 = \sqrt{\text{T}} \frac{\sqrt{\alpha_3}}{\alpha_2}~.
\end{equation}
The dispersion relations become
\begin{equation}
\omega_\pm^2 = p^2 + 2 \gamma_3 (\gamma_1 \pm \sqrt{p^2 + \gamma_1^2})~.
\end{equation}
From this it follows that $\omega_-(p=0)=0$ so that this excitation becomes gapless.
Additionally taking the limit $\alpha_1 \rightarrow 0$ for the first case, and $\alpha_3 \rightarrow 0$ for the second case,
\unskip\footnote{This corresponds to
\begin{equation*}
\gamma_1 \rightarrow + \infty~, \qquad \gamma_2 \rightarrow -c \infty~, \qquad \gamma_3 \rightarrow 0~, \qquad \gamma_1 \gamma_3 = \frac{1}{\alpha_2}~,
\end{equation*}
with $c=+1$ in the first case (when considering $\gamma_3 =\gamma_+$) and $c=-1$ in the second case (when considering $\gamma_3=\gamma_-$).
}
one finds the relativistic dispersion relations
\begin{equation}
\omega_+^2 = p^2 + \frac{4}{\alpha_2}~, \qquad \omega_-^2 = p^2~.
\end{equation}
One of the excitations is massive with mass $m = 2/\sqrt{\alpha_2}$, while the other becomes massless.
The S-matrix elements between the massive excitations remain finite in this limit, with
\begin{equation}
\begin{aligned}
l_{1,++} &= \frac{1}{2} \frac{p_1^2+p_2^2}{p_1^2-p_2^2}(p_1 \omega_2+p_2 \omega_1) ~, \qquad l_{3,++} = -\frac{1}{2} (p_1 \omega_2 + p_2 \omega_1)~, \\
l_{5,++} &= - \frac{1}{4} \frac{p_1 p_2}{p_1-p_2} (\sqrt{\omega_1+p_1}\sqrt{\omega_2+p_2} + \sqrt{\omega_1-p_1}\sqrt{\omega_2-p_2})~, \\
l_{7,++} &= -i c \frac{1}{4} \frac{p_1 p_2}{p_1+p_2} (\sqrt{\omega_1+p_1}\sqrt{\omega_2+p_2} - \sqrt{\omega_1-p_1}\sqrt{\omega_2-p_2})~,
\end{aligned}
\end{equation}
where $c=1$ for $\gamma_3 = \gamma_+$ and $c=-1$ for $\gamma_3=\gamma_-$.
Since $l_{5,++} \in \mathbb{R}$ and $l_{7,++} \in i \mathbb{R}$, the condition \eqref{eq:fact-cond} is satisfied.
The $(++)$ sector of the tree-level S-matrix factorises with $\mathcal T_{++}^{(1)} = \mathcal T_{++}^{(2)}$, and the coefficients and change of basis matrix given by
\begin{equation}
r_{5,++} = r_{6,++} = - l_{5,++}~, \qquad r_{7,++} = r_{8,++} = -i l_{7,++}~, \qquad \mathcal U_+ = \frac{1}{\sqrt{2}} \begin{pmatrix} 1 & -i \\ -i & 1\end{pmatrix}~.
\end{equation}
The factorised S-matrix $\mathcal T_{++}$ satisfies the classical Yang-Baxter equation and reproduces the tree-level S-matrix elements in the massive sector of the $\AdS_2 \times \Sp^2 \times \To^6$ superstring in the usual BMN light-cone gauge~\cite{Abbott:2013kka}, for which an all-loop S-matrix is known~\cite{Hoare:2014kma}.
Similarly, we can compute the limit of the mixed-mass S-matrix elements in the $(+-)$ and $(-+)$ sectors.
The limit is again finite, with coefficients
\begin{equation}
\begin{aligned}
l_{2,+-} &= -\frac{1}{2} (p_1 |p_2| + p_2 \omega_1)~, &\qquad l_{3, +-} &= -\frac{1}{2}(p_1 |p_2| + p_2 \omega_1)~, \\
l_{2,-+} &= +\frac{1}{2} (p_1 \omega_2 + p_2 |p_1|) ~, &\qquad l_{3, -+} &= -\frac{1}{2}(p_1 \omega_2 + p_2 |p_1|)~,
\end{aligned}
\end{equation}
and
\begin{equation}
\begin{aligned}
l_{4,+-} &= - \operatorname{sgn}(p_2) \frac{p_1}{2 \sqrt{2}} \sqrt{\omega_1 |p_2| + p_1 p_2}~, &\qquad l_{6,+-} &=i \frac{p_1}{2 \sqrt{2}} \sqrt{\omega_1 |p_2| + p_1 p_2}~, \\
l_{4,-+} &=- \operatorname{sgn}(p_1) \frac{p_2}{2 \sqrt{2}} \sqrt{\omega_2 |p_1| + p_2 p_1}~, &\qquad l_{6,-+} &= i \frac{p_2}{2 \sqrt{2}} \sqrt{\omega_2 |p_1| + p_2 p_1}~.
\end{aligned}
\end{equation}
Finally, some of the S-matrix elements in the $(--)$ sector diverge when $\operatorname{sgn}(p_1)=\operatorname{sgn}(p_2)$.
This corresponds to collinear scattering, which is not physical for massless excitations.
For the head-on collision, with $p_1>0$ and $p_2<0$, the S-matrix elements remain finite, with
\begin{equation}
l_{1,--} = 0~, \qquad l_{3,--}=0~, \qquad l_{5,--} = \frac{m}{4} \sqrt{|p_1 p_2|} ~, \qquad l_{7,--} = -i \frac{m}{4} \sqrt{|p_1 p_2|} ~.
\end{equation}
All these results are in agreement with the perturbative calculations of the tree-level S-matrix for strings on $\AdS_2 \times \Sp^2 \times \To^6$~\cite{Abbott:2013kka}.
To conclude, the S-matrix with $\gamma_3=\gamma_+$ or $\gamma_3=\gamma_-$ interpolates between the $\AdS_3 \times \Sp^3$ S-matrix in the alternative light-cone gauge-fixing considered in this paper and the $\AdS_2 \times \Sp^2$ S-matrix in the standard light-cone gauge.

%%%%%%%%%%%%%%%%%%%%%%%%%%%%%%%%%%%%%%%%%%%%%%%%%%%%%%%%%%%%%%%%%%%%%%%%%%%%%%%%
\section{Conclusions}\label{sec:conclusions}

We have shown that the elliptic $\AdS_3 \times \Sp^3 \times \To^4$ type IIB supergravity background~\eqref{eq:ellipticmetricAdS3xS3,-,eq:fluxbasis} on the supersymmetric locus~\eqref{eq:susylocus} preserves 8 supersymmetries, that is half of the undeformed background.
Together with the bosonic symmetries that are not broken by the deformation these supersymmetries form a single copy of the $\alg{psu}(1,1|2)$ superalgebra.
From the elliptic deformation, it is possible to take the time-like and space-like trigonometric limits, the latter of which interpolates between the undeformed $\AdS_3 \times \Sp^3 \times \To^4$ and $\AdS_2 \times \Sp^2 \times \To^6$ backgrounds with 8 supersymmetries.

Analysing the Green-Schwarz action for this elliptic background in uniform light-cone gauge, our second result is that the massive sector of the tree-level two-particle S-matrix satisfies the classical Yang-Baxter equation up to quadratic order in fermions.
Moreover, imposing integrability and symmetries we conjecture compatible processes quartic in fermions.
It is notable that the structure of the tree-level S-matrix is fundamentally different to the known rational and time-like trigonometric limits (with $\phi = 0$).
In particular, there are new non-vanishing amplitudes that are reminiscent of an 8-vertex structure.
As a result, the four $16 \times 16$ blocks of the S-matrix do not generically factorise into a tensor product of $4 \times 4$ matrices, hence this deformation falls outside the classification of~\cite{deLeeuw:2021ufg}.

Together, these two results provide strong evidence that the worldsheet theory of free strings propagating in the elliptic $\AdS_3 \times \Sp^3 \times \To^4$ background on the supersymmetric locus is classically integrable.

\medskip

One of the curious features of the elliptic $\AdS_3\times\Sp^3\times\To^4$ background is the dependence of the R-R fluxes, up to $\grp{O}(4)_{\mathrm{T-d}}$ transformations, on a single free angle $\phi$.
Constructing a (deformed) semi-symmetric space sigma model description of the worldsheet theory and a Lax connection is a key step to confirming classical integrability.
We would expect such a description to depend on the angle $\phi$ in addition to the deformation parameters $\alpha_1, \alpha_2, \alpha_3$.

There are different possible approaches to this question, two of which we will briefly outline here.
In a suitable kappa-symmetry gauge, the Green-Schwarz action can be written as the semi-symmetric space sigma model for the permutation supercoset~\cite{Babichenko:2009dk}
\begin{equation}
\frac{\grp{PSU}(1,1|2) \times \grp{PSU}(1,1|2)}{\grp{SU}(1,1) \times \grp{SU}(2)} ~,
\end{equation}
coupled via the Virasoro constraints to four free compact bosons.
One approach would be to construct an elliptic deformation of this semi-symmetric space sigma model for permutation supercosets in conformal gauge.
Rational and trigonometric integrable deformations of such symmetric sigma models have been extensively studied, however, much less is known about their elliptic deformations.
Using the machinery of 4d Chern-Simons \cite{Costello:2019tri}, which can be used to systematically construct classically integrable field theories and their Lax connections, an elliptic deformation of the $\grp{SL}(N)$ PCM has been determined~\cite{Lacroix:2023qlz}.
Whether it is possible to generalise this to symmetric and semi-symmetric spaces remains an open problem.
It may be fruitful to instead consider a hybrid formulation of the superstring based on the $\grp{PSU}(1,1|2)$ PCM~\cite{Berkovits:1999im} and deform this theory.

A second approach would be to start from the Lax connection of the bosonic theory~\cite{Cherednik:1981df} and use supersymmetry to build a Lax connection for the superstring worldsheet theory to leading order in fermions.
For example, it may be possible to follow the strategy of~\cite{Wulff:2014kja}, however, one apparent challenge with this is that the Lax connection of the bosonic theory is not naturally written in terms of the $\grp{SL}(2;\Real)_L \times \grp{SU}(2)_L$ conserved current.
Nevertheless, this would provide useful data for the construction of a deformed symmetric sigma model.

\medskip

In parallel to confirming the classical integrability, it is also important to investigate the quantum integrability of the elliptic worldsheet theory.
As a starting point, this could entail constructing the exact massive dispersion relation and S-matrix in uniform light-cone gauge, compatible with the tree-level results computed here.
This exact S-matrix should satisfy the quantum Yang-Baxter equation.
To do this, it would be helpful to better understand the expected symmetries of the light-cone gauge-fixed theory.
On the one hand, we have found that the deformed tree-level S-matrix has a $\grp{U}(1)^2$ symmetry.
However, only one of these, the one that acts on the fermions and originates from the $\grp{O}(4)_{\mathrm{T-d}}$ T-duality group, is a symmetry of the light-cone gauge-fixed Lagrangian.
Curiously, the other appears after we rewrite the Lagrangian in terms of oscillators and it is not clear if we should expect it to survive to higher orders in fields or $\hbar$.

On the other hand, the superisometry group of the elliptic background is $\grp{PSU}(1,1|2)$.
In our analysis we have defined light-cone coordinates $X^\pm$ associated to two directions $\Lambda_\pm$ in the corresponding superalgebra $\alg{psu}(1,1|2)$.
In addition to the conserved worldsheet energy and momentum, we also expect the worldsheet S-matrix to be invariant under the 4 supersymmetries that commute with $\Lambda_+$.
Since these supercharges will not commute with $\Lambda_-$ their action on two-particle states will become dynamical and depend non-trivially on the worldsheet momentum~\cite{Arutyunov:2006ak,Beisert:2005tm}.
Determining this action will help to constrain the S-matrix.

To completely fix the structure of the S-matrix up to dressing phases in the time-like trigonometric case~\cite{Hoare:2014oua}, the action of the $q$-deformed symmetries is required.
Here, we expect that the right-acting symmetry of the undeformed, or rational, limit should not be truly broken, but rather elliptically-deformed.
Understanding this putative elliptic deformation of $\alg{psu}(1,1|2)$ may then allow the structure of the exact massive dispersion relation and S-matrix to be determined.
If this is possible it would have important implications for understanding the undeformed $\AdS_2 \times \Sp^2 \times \To^6$ superstring as a limit of the space-like trigonometric case, where, thus far, it is not possible to fix the massive dispersion relation and S-matrix using symmetries alone~\cite{Hoare:2014kma}.
Moreover, since half of the modes become massless, this limit could also provide insights into massless $\AdS_2 \times \Sp^2 \times \To^6$ scattering~\cite{Fontanella:2017rvu,DeLeeuw:2020ahx}.

\medskip

We conclude by commenting on a number of possible generalisations.
The first is to allow non-vanishing NS-NS flux.
Doing so while satisfying the supergravity equations is straightforward.
Following \cite{Hoare:2022asa} we can simply parametrise $H_3 = F_{3,5} = \mathbf{x}^{(1)}_5 f_3^{(1)} + \mathbf{x}^{(2)}_5 f_3^{(2)} + \mathbf{x}^{(3)}_5 f_3^{(3)} + \mathbf{x}^{(4)}_5 f_3^{(4)}$.
The 4-vectors $\mathbf{x}^{(1,2,3,4)}$ now become 5-vectors and we can generate mixed-flux backgrounds by acting with the $\grp{SO}(5)$ U-duality group on the elliptic supergravity background constructed in this paper.
\unskip\footnote{As observed in \cite{Hoare:2022asa}, since away from the rational limit the vectors $\mathbf{x}^{(1,2,3,4)}$ span at least a 2-plane, it is not possible to construct a deformed background supported only by NS-NS flux.}
While in the special case of the time-like trigonometric deformation with $\phi = 0$ it was shown in \cite{Hoare:2022asa} that this procedure preserves 8 supercharges and classical integrability, it is not clear if this will be true more generally due to the S-duality transformation involved.
It would also be interesting to construct an elliptic $\AdS_3 \times \Sp^3\times \Sp^3 \times \Sp^1$ superstring, again preserving classical integrability and 8 supercharges now forming a copy of $\alg{d}(2,1;\alpha)$.

Finally, it is also possible to embed the current-current deformation of the $\grp{SL}(2;\Real) \times \grp{SU}(2)$ WZW sigma model~\cite{Sfetsos:2013wia} in both a classically integrable sigma model based on a semi-symmetric space~\cite{Hoare:2022vnw} and in type IIB supergravity \cite{Itsios:2023kma,Sfetsos:2014cea} with both constructions preserving 8 supersymmetries.
These worldsheet theories are expected to be related to the trigonometric deformations through Poisson-Lie duality with respect to the $q$-deformed $\alg{psu}(1,1|2)$ and its bosonic subalgebra respectively~\cite{Hoare:2018ebg}.
Constructing an exact S-matrix in uniform-light cone gauge for these theories and understanding the relation to the S-matrices of the trigonometric deformations is an open question.

%%%%%%%%%%%%%%%%%%%%%%%%%%%%%%%%%%%%%%%%%%%%%%%%%%%%%%%%%%%%%%%%%%%%%%%%%%%%%%%%
\section*{Acknowledgments}

We would like to thank R.~Borsato, S.~Driezen, S.~Frolov, A.~Retore, A.~Sfondrini and A.~Torrielli for interesting discussions and A.~Retore for collaboration on the precursor to this paper~\cite{Hoare:2023zti}.
The work of BH was supported by a UKRI Future Leaders Fellowship (grant number MR/T018909/1).
FS is supported by the Swiss National Science Foundation (SNSF) through the Ambizione grant ``Effective strings, membranes, and their solvability'' (project number 223544), and also received funding from the Deutsche Forschungsgemeinschaft (DFG, German Research Foundation) under the Collaborative Research Center 1624 ``Higher structures, moduli spaces and integrability'', project number 506632645.

%%%%%%%%%%%%%%%%%%%%%%%%%%%%%%%%%%%%%%%%%%%%%%%%%%%%%%%%%%%%%%%%%%%%%%%%%%%%%%%%
\appendix

%%%%%%%%%%%%%%%%%%%%%%%%%%%%%%%%%%%%%%%%%%%%%%%%%%%%%%%%%%%%%%%%%%%%%%%%%%%%%%%%
\section{Dirac matrix conventions and spinors}\label{app:gammamatrix}

We use the following basis for the 10d Dirac matrices
\unskip\footnote{This choice is the same as in \cite{Seibold:2020ywq}, up to the identification
\begin{equation*}
\Gamma_3 = \Gamma_1^{\mathrm{there}} ~, \qquad \Gamma_1 = \Gamma_2^{\mathrm{there}} \qquad \Gamma_2 = \Gamma_3^{\mathrm{there}} ~,
\end{equation*}
which takes into account that our tangent-space light-cone directions are $0,3$, while they are $0,1$ in \cite{Seibold:2020ywq}.}
\begin{equation}
\begin{aligned}
\Gamma_0 &= \sigma_1 \otimes \gamma_0 \otimes 1_4~, &\qquad \Gamma_3 &= - \sigma_2 \otimes 1_4 \otimes \gamma_5~, \\
\Gamma_1 &= \sigma_1 \otimes \gamma_1 \otimes 1_4~, &\qquad \Gamma_4 &= \sigma_1 \otimes \gamma_3 \otimes 1_4~, \\
\Gamma_2 &= \sigma_1 \otimes \gamma_2 \otimes 1_4~, &\qquad \Gamma_5 &= \sigma_1 \otimes \gamma_4 \otimes 1_4~, \\
\Gamma_6 &= - \sigma_2 \otimes 1_4 \otimes \gamma_1~, &\qquad \Gamma_8 &= - \sigma_2 \otimes 1_4 \otimes \gamma_3~, \\
\Gamma_7 &= -\sigma_2 \otimes 1_4 \otimes \gamma_2~, &\qquad \Gamma_9 &= - \sigma_2 \otimes 1_4 \otimes \gamma_4~,
\end{aligned}
\end{equation}
with
\begin{equation}
\gamma_0 = i \sigma_3 \otimes 1_2 = i \gamma_5~, \qquad \gamma_1 = \sigma_2 \otimes \sigma_2~, \qquad \gamma_2 = -\sigma_2 \otimes \sigma_1~, \qquad \gamma_3 = \sigma_1 \otimes 1_2~, \qquad \gamma_4 = \sigma_2 \otimes \sigma_3~.
\end{equation}
Note that
\begin{equation}
\Gamma_6 \Gamma_7 \Gamma_8 \Gamma_9 = - 1_2 \otimes 1_2 \otimes 1_2 \otimes \sigma_3 \otimes 1_2~.
\end{equation}
The associated conjugation matrix is
\begin{equation}\label{eq:conjugationmatrix}
\mathcal C = i \sigma_2 \otimes K \otimes K~, \qquad K = -i 1_2 \otimes \sigma_2~,
\end{equation}
satisfying
\begin{equation}
\mathcal C^t + \mathcal C =0~, \qquad \mathcal C^t \mathcal C = 1_{32}~, \qquad \Gamma_a^t + \mathcal C \Gamma_a \mathcal C^{-1} =0~.
\end{equation}
The Majorana condition then reads
\begin{equation}
\bar{\theta} = \theta^\dagger \Gamma^0 = \theta^t \mathcal C~.
\end{equation}
We also define
\begin{equation}
\Gamma_{11} = \Gamma_0 \Gamma_1 \Gamma_2 \dots \Gamma_9 = \sigma_3 \otimes 1_{16}~.
\end{equation}
With our convention for the self-duality of the 5-form, the Weyl condition on the fermions reads
\begin{equation}
(1-\Gamma_{11}) \theta = 0~.
\end{equation}
The contraction of the 3-form and 5-form fluxes with the Dirac matrices gives
\begin{equation}
\begin{aligned}
\slashed{F}_3 &= 2 \mathrm{T}^{-\frac12} (\mathbf{x}^{(1)}_4\Gamma^{024} + \mathbf{x}^{(2)}_4 \Gamma^{045} + \mathbf{x}^{(3)}_4\Gamma^{015} + \mathbf{x}^{(4)}_4\Gamma^{012} ) (1_{32} + \Gamma^{11} \Gamma^{6789})~, \\
\slashed{F}_5 &= 2 \mathrm{T}^{-\frac12}( \Gamma^{024} J_1 + \Gamma^{045} J_2 + \Gamma^{015} J_3 + \Gamma^{012} J_4 ) (1_{32}+\Gamma^{6789})(1_{32} +\Gamma^{11} \Gamma^{6789}) ~, \\
J_i &= \mathbf{x}^{(i)}_1 \Gamma^{67} + \mathbf{x}^{(i)}_2 \Gamma^{68} + \mathbf{x}^{(i)}_3 \Gamma^{69} ~.
\end{aligned}
\end{equation}
Let us also note the following conjugation properties
\begin{equation}
(\slashed{F}_1)^t \mathcal C = - \mathcal C \slashed{F}_1~, \qquad (\slashed{F}_3)^t \mathcal C = + \mathcal C \slashed{F}_3~, \qquad (\slashed{F}_5)^t \mathcal C = - \mathcal C \slashed{F}_5~, \qquad \mathcal S^t (1_2 \otimes \mathcal C) = (1_2 \otimes \mathcal C)\mathcal S~.
\end{equation}

\medskip

In the analysis of supersymmetries in \secref{sec:sugra-susy} the equations for the vectors $\mathbf{x}^{(1)}$, $\mathbf{x}^{(2)}$, $\mathbf{x}^{(3)}$ and $\mathbf{x}^{(4)}$ are invariant under the $\grp{O}(4)_{\mathrm{T-d}}$ T-duality group.
The T-duality group also has an action on spinors valued in $\alg{s}_6$, where here we are interested in the action of the connected component $\grp{SO}(4)_{\mathrm{T-d}}$.
Letting $\epsilon = (\epsilon^1 ,\epsilon^2) \in \alg{s}_6$ where $\epsilon^I$ are two 10d Majorana-Weyl spinors satisfying~\eqref{eq:chiral} and~\eqref{eq:6dspin}, this action can be determined from the condition
\begin{equation}
\mathcal{R}^{-1} \widetilde\Omega_\ind{M} \mathcal{R} \epsilon = \Omega_\ind{M} \epsilon ~,
\end{equation}
where $\Omega_\ind{M}$ is defined in eq.~\eqref{eq:gravitino3} in terms of the R-R fluxes and $\widetilde\Omega_\ind{M}$ takes the same form, but now defined in terms of the $\grp{SO}(4)_{\mathrm{T-d}}$-transformed fluxes.
Parametrising the action of $\grp{SO}(4)_{\mathrm{T-d}}$ on the vectors as
\begin{equation}
\begin{pmatrix} \mathbf{x}^{(i)}_4 + i \mathbf{x}^{(i)}_1 & \mathbf{x}^{(i)}_2 + i \mathbf{x}^{(i)}_3 \\
-\mathbf{x}^{(i)}_2 + i \mathbf{x}^{(i)}_3 & \mathbf{x}^{(i)}_4 - i \mathbf{x}^{(i)}_1 \end{pmatrix} \to \mathscr{N}_L
\begin{pmatrix} \mathbf{x}^{(i)}_4 + i \mathbf{x}^{(i)}_1 & \mathbf{x}^{(i)}_2 + i \mathbf{x}^{(i)}_3 \\
-\mathbf{x}^{(i)}_2 + i \mathbf{x}^{(i)}_3 & \mathbf{x}^{(i)}_4 - i \mathbf{x}^{(i)}_1 \end{pmatrix}
\mathscr{N}_R ~, \qquad \mathscr{N}_L, \mathscr{N}_R \in \grp{SU}(2) ~,
\end{equation}
we find that
\begin{equation}\begin{split}
\mathcal{R} &
= \varrho_+ \otimes \varrho_+ \otimes 1_2 \otimes 1_2 \otimes (\varrho_- \otimes \mathscr{N}_L + \varrho_+ \otimes 1_2)
+ \varrho_+ \otimes \varrho_- \otimes 1_2 \otimes 1_2 \otimes 1_2 \otimes 1_2
\\ & \quad +
\varrho_- \otimes \varrho_+ \otimes 1_2 \otimes 1_2 \otimes (\varrho_- \otimes \mathscr{N}_R^{-1} + \varrho_+ \otimes 1_2)
+ \varrho_- \otimes \varrho_- \otimes 1_2 \otimes 1_2 \otimes 1_2 \otimes 1_2 ~,
\end{split}\end{equation}
where we have defined $\varrho_\pm = \frac12(1_2 \pm \sigma_3)$.
Therefore, if $\epsilon \in \alg{s}_6$ satisfies the gravitino Killing spinor equation~\eqref{eq:gravitino3} for some choice of R-R fluxes, $\mathcal{R} \epsilon$ will satisfy the Killing spinor equations for the $\grp{SO}(4)_{\mathrm{T-d}}$-transformed fluxes.

\medskip

For the computation of the light-cone gauge-fixed theory in \secref{sec:lcgf} it is necessary to impose a kappa-symmetry gauge on the fermions.
Imposing the gauge $\Glc^+ \theta^I = 0$, we can parametrise the doublet of 10d Majorana-Weyl spinors as
\begin{equation}
\theta^1 = \frac{1}{\sqrt{2}}
\begin{pmatrix}
1 \\
0
\end{pmatrix}
\otimes
\begin{pmatrix}
0 \\ 0 \\ \zeta_{R,1} \\ \zeta_{R,2} \\ 0 \\ 0 \\ i \zeta_{R,2}^\star \\ -i \zeta_{R,1}^\star \\ i \zeta_{R,4}^\star \\ -i \zeta_{R,3}^\star \\ 0 \\ 0 \\ \zeta_{R,3} \\ \zeta_{R,4}\\ 0 \\ 0
\end{pmatrix}
~, \qquad
\theta^2 = \frac{1}{\sqrt{2}}\begin{pmatrix}
1 \\
0
\end{pmatrix}
\otimes
\begin{pmatrix}
0 \\ 0 \\ \zeta_{L,1} \\ \zeta_{L,2} \\ 0 \\ 0 \\ i \zeta_{L,2}^\star \\ -i \zeta_{L,1}^\star \\ i \zeta_{L,4}^\star \\ -i \zeta_{L,3}^\star \\ 0 \\ 0 \\ \zeta_{L,3} \\ \zeta_{L,4}\\ 0 \\ 0
\end{pmatrix}~.
\end{equation}

%%%%%%%%%%%%%%%%%%%%%%%%%%%%%%%%%%%%%%%%%%%%%%%%%%%%%%%%%%%%%%%%%%%%%%%%%%%%%%%%
\section{The supersymmetric locus}\label{app:Gram}

The R-R fluxes are parametrised by four vectors $\mathbf{y}_\pm, \mathbf{z}_\pm \in \mathbb{R}^4$.
Their Gram matrix is given by
\begin{equation}
\begin{aligned}
G &= \begin{pmatrix}
||\mathbf{y}_+||^2 & \mathbf{y}_+ \cdot \mathbf{z}_+ & \mathbf{y}_+ \cdot \mathbf{y}_- &\mathbf{y}_+ \cdot\mathbf{z}_- \\
\mathbf{y}_+ \cdot \mathbf{z}_+ & ||\mathbf{z}_+||^2 & \mathbf{z}_+ \cdot \mathbf{y}_- & \mathbf{z}_+ \cdot \mathbf{z}_- \\
\mathbf{y}_+ \cdot \mathbf{y}_- & \mathbf{z}_+ \cdot \mathbf{y}_- & ||\mathbf{y}_-||^2 & \mathbf{y}_- \cdot \mathbf{z}_- \\
\mathbf{y}_+ \cdot \mathbf{z}_- & \mathbf{z}_+ \cdot \mathbf{z}_- & \mathbf{y}_- \cdot \mathbf{z}_- & ||\mathbf{z}_-||^2
\end{pmatrix} \\
&= \begin{pmatrix}
||\mathbf{y}_+||^2 & \mathbf{y}_+ \cdot \mathbf{z}_+ & -\gamma_1 \gamma_3 & 0 \\
\mathbf{y}_+ \cdot \mathbf{z}_+ & \gamma_1^2-\gamma_2^2-||\mathbf{y}_+||^2 & 0 & -\gamma_2 \gamma_3 \\
-\gamma_1 \gamma_3 & 0 & ||\mathbf{y}_-||^2 & \mathbf{y}_- \cdot \mathbf{z}_- \\
0 & -\gamma_2 \gamma_3 & \mathbf{y}_- \cdot \mathbf{z}_- & \gamma_1^2-\gamma_2^2-||\mathbf{y}_-||^2
\end{pmatrix}~,
\end{aligned}
\end{equation}
where in the second equality we used the constraints from supergravity.
For generic values of the parameters $\rank G=4$.
It reduces to $\rank G=2$ if and only if
\begin{align}
\label{eq:G1}
&||\mathbf{y}_+||^2(\gamma_1^2-\gamma_2^2+\gamma_3^2-||\mathbf{y}_+||^2)-\gamma_1^2 \gamma_3^2 - (\mathbf{y}_+ \cdot \mathbf{z}_+)^2=0~, \\
\label{eq:G2}
&(\mathbf{y}_+ \cdot \mathbf{z}_+)(||\mathbf{y}_-||^2-\gamma_1^2) -\gamma_1 \gamma_2 (\mathbf{y}_- \cdot \mathbf{z}_-) =0~, \\
\label{eq:G3}
&(\mathbf{y}_- \cdot \mathbf{z}_-)(||\mathbf{y}_+||^2-\gamma_1^2) -\gamma_1 \gamma_2 (\mathbf{y}_+ \cdot \mathbf{z}_+) =0~.
\end{align}
This can be checked by requiring that all minors of $G$ vanish.
The first equation \eqref{eq:G1} is equivalent to \eqref{eq:susyup} on the supergravity constraint \eqref{eq:sugraup}.
Solving \eqref{eq:G2} and \eqref{eq:G3} for $||\mathbf{y}_+||^2$ and $(\mathbf{y}_+ \cdot \mathbf{z}_+)$, and plugging into \eqref{eq:G1} then gives \eqref{eq:susyupt}, on the supergravity constraint \eqref{eq:sugraupt}.
As discussed in the main text, these two equations can be solved parametrically in terms of angles $\phi \in (-\pi,\pi]$ and $\psi \in (-\pi, \pi]$, with the parametrisation \eqref{eq:eqsol} and \eqref{eq:eqsolt}.
The equations \eqref{eq:G2} and \eqref{eq:G3} then impose the following relations between the angles:
\begin{equation}
\begin{aligned}
\sin \psi &= \frac{ 2 \gamma_1 \gamma_2 \sin \phi}{\gamma_1^2 + \gamma_2^2 - \gamma_3^2 - \sqrt{(\gamma_+^2-\gamma_3^2)(\gamma_-^2-\gamma_3^2)} \cos \phi}~,\\
\cos \psi &= \frac{(\gamma_1^2 + \gamma_2^2 - \gamma_3^2) \cos \phi - \sqrt{(\gamma_+^2-\gamma_3^2)(\gamma_-^2-\gamma_3^2)} }{\gamma_1^2 + \gamma_2^2 - \gamma_3^2 - \sqrt{(\gamma_+^2-\gamma_3^2)(\gamma_-^2-\gamma_3^2)} \cos \phi}~.
\end{aligned}
\end{equation}
This is equivalent to the relation between angles~\eqref{eq:anglegamma}.
We therefore conclude that the background preserves 8 supersymmetries if and only if $\rank G=2$, meaning that the vectors parametrising the R-R fluxes span a 2-plane.

To obtain this result we assumed $||\mathbf{y}_+||^2 \neq \gamma_1^2$ and $||\mathbf{y}_-||^2 \neq \gamma_1^2$.
This is no longer true for the rational limit.
The rank of the Gram matrix further reduces to $\rank G=1$ in the rational limit $\mathbf{z}_+ = \mathbf{z}_-=0$, $||\mathbf{y}_+||^2 = ||\mathbf{y}_-||^2 = \gamma_1^2$ and $\gamma_1=\gamma_3$, $\gamma_2=0$.
In this case, all the vectors parametrising the R-R flux lie on the same line.

%%%%%%%%%%%%%%%%%%%%%%%%%%%%%%%%%%%%%%%%%%%%%%%%%%%%%%%%%%%%%%%%%%%%%%%%%%%%%%%%
\section{Killing vectors}\label{app:c1}

Since it is constructed out of left-invariant Maurer-Cartan forms, the elliptic metric \eqref{eq:ellipticmetricAdS3xS3} possesses an $\grp{SL}(2; \Real)_L \times \grp{SU}(2)_L$ symmetry.
The associated Killing vectors are
\begin{equation}
\begin{aligned}
v_1 &=\sin(2 T) \tanh(2 U) \partial_T - \cos(2 T) \partial_U - \sin(2 T)\sech(2 U) \partial_V~, \\
v_2 &= \cos(2 T) \tanh(2 U) \partial_T +\sin(2 T) \partial_U- \cos(2 T)\sech(2 U)\partial_V~, \\
v_3 &= \partial_T~,
\end{aligned}
\end{equation}
and
\begin{equation}
\begin{aligned}
v_4 &= \sin(2 \Phi) \tan(2 X) \partial_\Phi +\cos(2 \Phi) \partial_X + \sin(2 \Phi) \sec(2 X) \partial_Y~, \\
v_5 &= \cos(2 \Phi) \tan(2 X) \partial_\Phi - \sin(2 \Phi) \partial_X + \cos(2 \Phi) \sec(2 X) \partial_Y~, \\
v_6 &= \partial_\Phi~.
\end{aligned}
\end{equation}
They satisfy the Killing vector equations
\begin{equation}
\nabla_\ind{M} v_{j, \ind{N}} + \nabla_\ind{N} v_{j,\ind{M} }= 0 ~, \qquad v_j = v_j^\ind{M} \partial_\ind{M}~, \qquad j=1,\dots 6~,
\end{equation}
as well as the $\alg{sl}(2; \Real) \oplus \alg{su}(2)$ algebra
\begin{equation}
\begin{aligned}
&\com{v_1}{v_2} = -2 v_3~, &\qquad &\com{v_3}{v_1}=+2 v_2~, \qquad \com{v_3}{v_2}=-2 v_1 \\
& \com{v_4}{v_5} = + 2 v_6~, &\qquad &\com{v_6}{v_4} = +2 v_5~, \qquad \com{v_6}{v_5} = - 2 v_4~.
\end{aligned}
\end{equation}
The Killing vectors transform under the left-acting $\grp{SL}(2; \Real)_L \times \grp{SU}(2)_L$ symmetry, as can readily be seen from the fact that the components are not invariant under, e.g.,~translations by $T$ or $\Phi$.
They are however singlets under the right-acting symmetry (which only survives in the rational case, and contains in particular the Killing vectors $\partial_V$ and $\partial_Y$).
The Killing vectors $v_3 = \partial_T$ and $v_6 = \partial_\Phi$ generate translations in the $T$ and $\Phi$ directions, which are the ones used to impose uniform light-cone gauge.

%%%%%%%%%%%%%%%%%%%%%%%%%%%%%%%%%%%%%%%%%%%%%%%%%%%%%%%%%%%%%%%%%%%%%%%%%%%%%%%%
\section{Killing spinors}\label{app:c2}

After imposing the chirality condition \eqref{eq:chiral} and the 6d condition \eqref{eq:6dspin}, the spinor doublet $\epsilon \in \alg{s}$ spans a 16-dimensional subspace within the total 64-dimensional space.
In our basis of Dirac matrices, the non-vanishing positions are given by 3, 4, 7, 8, 11, 12, 15, 16, 35, 36, 39, 40, 43, 44, 47, 48.
We will write these 6d spinors as $\varepsilon_{a \alpha A}$ with indices $a=\pm$, $\alpha=\pm$ and $A=\pm$.
A basis of complex (not necessarily Majorana-Weyl) 6d Killing spinors satisfying the constraint
\begin{equation}
F_\ind{MN} \vert_{\mathfrak{s}_6} \varepsilon_{a \alpha A}=0~, \qquad \forall M,N \in \{T,U,V,\Phi,X,Y,\Psi^6, \Psi^7, \Psi^8, \Psi^9\}~,
\end{equation}
is given by
\begin{equation}
\begin{aligned}
\varepsilon_{+++} &= i \begin{pmatrix} Y_- \\ 1 \end{pmatrix} \otimes T_{11} + \Sigma_{13}^{+-} \begin{pmatrix} 1 \\ Y_- \end{pmatrix} \otimes T_{21}~, \\
\varepsilon_{++-} &= (-i) \Sigma_{13}^{-+} \begin{pmatrix} Y_+ \\ 1 \end{pmatrix} \otimes T_{12} + (-1) \begin{pmatrix} 1 \\ Y_+ \end{pmatrix} \otimes T_{22}~, \\
\varepsilon_{+-+} &= - \begin{pmatrix} X_{+-} - X_{-+} \Sigma_{12}^{+-} \\ X_{++} \Sigma_{13}^{+-} +X_{--} \Sigma_{23}^{--} \end{pmatrix} \otimes T_{11} + i \begin{pmatrix} X_{++} + X_{--} \Sigma_{12}^{++} \\ X_{+-} \Sigma_{13}^{+-} - X_{-+} \Sigma_{23}^{+-} \end{pmatrix} \otimes T_{21}~, \\
\varepsilon_{+--} &= \begin{pmatrix} X_{+-} \Sigma_{13}^{-+} + X_{-+} \Sigma_{13}^{-+} \\ X_{++} -X_{--} \Sigma_{12}^{--} \end{pmatrix} \otimes T_{12} - i \begin{pmatrix} X_{++} \Sigma_{13}^{-+} - X_{--} \Sigma_{23}^{++} \\ X_{+-} +X_{-+} \Sigma_{12}^{-+}\end{pmatrix} \otimes T_{22}~, \\
\varepsilon_{-++} &= i \begin{pmatrix} X_{+-} -X_{-+} \Sigma_{12}^{+-} \\ X_{++} \Sigma_{13}^{+-} + X_{--} \Sigma_{23}^{--} \end{pmatrix} \otimes T_{11} + (-1)\begin{pmatrix} X_{++} +X_{--} \Sigma_{12}^{++} \\ X_{+-} \Sigma_{13}^{+-} - X_{-+} \Sigma_{23}^{+-} \end{pmatrix} \otimes T_{21}~, \\
\varepsilon_{-+-} &= (-i) \begin{pmatrix} X_{+-} \Sigma_{13}^{-+} + X_{-+} \Sigma_{13}^{-+} \\ X_{++}-X_{--} \Sigma_{12}^{--} \end{pmatrix} \otimes T_{12} + \begin{pmatrix} X_{++} \Sigma_{13}^{-+} - X_{++} \Sigma_{23}^{++} \\ X_{+-} + X_{-+} \Sigma_{12}^{-+}\end{pmatrix} \otimes T_{22}~, \\
\varepsilon_{--+} &= (-1) \begin{pmatrix} Y_- \\ 1 \end{pmatrix} \otimes T_{11} + (-i \Sigma_{13}^{+-}) \begin{pmatrix} 1 \\ Y_- \end{pmatrix} \otimes T_{21}~, \\
\varepsilon_{---} &= \Sigma_{13}^{-+} \begin{pmatrix} Y_+ \\ 1 \end{pmatrix} \otimes T_{12} + i \begin{pmatrix} 1\\ Y_+ \end{pmatrix} \otimes T_{22}~.
\end{aligned}
\end{equation}
To simplify the expressions, we have introduced the auxiliary quantities
\begin{equation}
\begin{aligned}
T_{11} &= \sqrt{B Y_+} \begin{pmatrix} 1 \\ 0\end{pmatrix}
\otimes \begin{pmatrix} i \\ 1 \end{pmatrix} \otimes \begin{pmatrix} 1 \\ 0\end{pmatrix}~, &\qquad T_{21} &= \sqrt{B Y_+} \begin{pmatrix} 1 \\ 0\end{pmatrix}
\otimes \begin{pmatrix} 1 \\ i \end{pmatrix} \otimes \begin{pmatrix} 1 \\ 0\end{pmatrix}~, \\
T_{12} &= \sqrt{B Y_-} \operatorname{sgn}(s)\begin{pmatrix} 1 \\ 0\end{pmatrix}
\otimes \begin{pmatrix} i \\ 1 \end{pmatrix} \otimes \begin{pmatrix} 0 \\ 1\end{pmatrix}~, &\qquad T_{22} &= \sqrt{B Y_-} \operatorname{sgn}(s)\begin{pmatrix} 1 \\ 0\end{pmatrix}
\otimes \begin{pmatrix} 1 \\ i \end{pmatrix} \otimes \begin{pmatrix} 0 \\ 1\end{pmatrix}~,
\end{aligned}
\end{equation}
\begin{equation}
Y_\pm = \frac{s \sqrt{1-t^2} \Sigma_1 \pm \sqrt{1-s^2} \Sigma_2}{\sqrt{s^2-t^2}\Sigma_3}~, \qquad Y_+ Y_- = 1~,
\end{equation}
\begin{equation}
B = \frac{(s^2-t^2) \sqrt{\mathrm{T}}}{8 s (1+s) (1-t^2) \sqrt{\gamma_3^2}} \frac{\Sigma_3}{\Sigma_1}~, \qquad X_{\mu_1 \mu_2} = \frac{s}{\sqrt{2}} \frac{\Sigma_1}{\Sigma_3} \sqrt{\frac{(1+\mu_1 s)(1+\mu_2 t)}{s(s +\mu_1 \mu_2 t)}}~,
\end{equation}
as well as
\begin{equation}
\begin{aligned}
\Sigma_{12}^{\mu_1 \mu_2} &= \frac{\left(1+i \mu_1 \tan \frac{\phi}{2}\right) \left(t+i \mu_2 s \tan \frac{\phi}{2}\right)}{\Sigma_1 \Sigma_2}~, &\qquad \Sigma_{12}^{++} \Sigma_{12}^{--}&=1~, &\qquad \Sigma_{12}^{+-} \Sigma_{12}^{-+} &= 1~,\\
\Sigma_{13}^{\mu_1 \mu_2} &= \frac{\left(1+i \mu_1 \tan \frac{\phi}{2}\right) \left(1+i \mu_2 s \tan \frac{\phi}{2}\right)}{\Sigma_1 \Sigma_3 \, \operatorname{sgn}(s)}~, &\qquad \Sigma_{13}^{++} \Sigma_{13}^{--}&=1~, &\qquad \Sigma_{13}^{+-} \Sigma_{13}^{-+} &=1~, \\
\Sigma_{23}^{\mu_1 \mu_2} &= \frac{\left(t+i \mu_1 s \tan \frac{\phi}{2}\right)\left(1+i \mu_2 s \tan \frac{\phi}{2}\right)}{\Sigma_2 \Sigma_3 \, \operatorname{sgn}(s)}~, &\qquad \Sigma_{23}^{++} \Sigma_{23}^{--} &=1~, &\qquad \Sigma_{23}^{+-} \Sigma_{23}^{-+} &=1~,
\end{aligned}
\end{equation}
where $\Sigma_{1,2,3}$ are defined in eq.~\eqref{eq:sigma1234}.
One can check that for the region of parameter space defined in eq.~\eqref{eq:realityst} we have $B Y_+ > 0$ and $B Y_- > 0$.

Defining the constant (independent of space-time coordinates $\Psi^\ind{M}$) matrices
\begin{equation}
\begin{aligned}
\mathbf{L}_3 &= \sqrt{\alpha_2} \Omega_0~, &\qquad \mathbf{L}_\pm &= \frac{\sqrt{\alpha_1} \Omega_1 \pm i \sqrt{\alpha_3} \Omega_2}{2}~, \\
\mathbf{J}_3 &= \sqrt{\alpha_2} \Omega_3 ~, &\qquad \mathbf{J}_\pm &= \frac{\sqrt{\alpha_1} \Omega_4 \pm i \sqrt{\alpha_3} \Omega_5}{2}~,
\end{aligned}
\end{equation}
one can check that the basis of spinors defined above are eigenvectors of $\mathbf{L}_\pm$ and $\mathbf{J}_\pm$, with
\begin{equation}
\begin{aligned}
\mathbf{L}_3 \varepsilon_{\pm \alpha A} &= \pm i \varepsilon_{\pm \alpha A}~, \qquad
\mathbf{J}_3\varepsilon_{a \pm A} &= \pm i \varepsilon_{a \pm A}~.
\end{aligned}
\end{equation}
Moreover, in this basis, $\mathbf{L}_\pm$ and $\mathbf{J}_\pm$ act as raising and lowering operators, with
\begin{equation}
\mathbf{L}_\pm \varepsilon_{\pm\alpha A} = 0~, \qquad \mathbf{L}_\pm \varepsilon_{\mp \alpha A} = \mp i \varepsilon_{\pm \alpha A}~, \qquad
\mathbf{J}_\pm \varepsilon_{a \pm A} = 0~, \qquad \mathbf{J}_\pm \varepsilon_{a \mp A} = i \varepsilon_{a \pm A}~.
\end{equation}
The most general Killing spinor solving the dilatino and gravitino equations is a linear combination of the above constant basis spinors,
\begin{equation}
\epsilon = \sum_{a = \pm} \sum_{\alpha = \pm} \sum_{A=\pm} \varepsilon_{a \alpha A} f_{a \alpha A}(\Psi) = \vec{\varepsilon} \cdot \vec{f}~.
\end{equation}
The gravitino equations then translate to the following equations on the vector of functions $\vec{f}(\Psi)$
\begin{equation}
\begin{aligned}
\partial_V \vec{f} &= (- \sigma_2 \otimes 1_2 \otimes 1_2) \vec{f} ~, \\
\partial_U \vec{f}&= (\sinh(2 V) i \sigma_3 + \cosh(2 V) \sigma_1) \vec{f}~, \\
\partial_T \vec{f} &= \left(\cosh(2 U) \cosh(2 V) i \sigma_3 + \cosh(2 U) \sinh(2 V) \sigma_1 + \sinh(2 U) \sigma_2 \right) \vec{f}~, \\
\partial_Y \vec{f} &= i (1_2 \otimes \sigma_1 \otimes 1_2) \vec{f}, \\
\partial_X \vec{f} &= i(-\sin(2 Y) \sigma_3 + \cos(2 Y) \sigma_2) \vec{f}, \\
\partial_\Phi \vec{f} &= i \left(\cos(2 X) \cos(2 Y) \sigma_3 + \cos(2 X) \sin(2 Y) \sigma_2 - \sin(2 X) \sigma_1\right) \vec{f}.
\end{aligned}
\end{equation}
These are solved by
\begin{equation}
\vec{f} = (f_A \otimes f_S \otimes 1_2 ) \vec{f}_0~, \qquad f_A = e^{-\sigma_2 V} e^{\sigma_1 U} e^{i \sigma_3 T} ~, \qquad f_S = e^{i \sigma_1 Y} e^{i \sigma_2 X} e^{i \sigma_3 \Phi}~,
\end{equation}
with $\vec{f}_0$ a constant 8-dimensional vector.
A basis of 6d spinors solving the dilatino and gravitino equations is then given by
\begin{equation}
\epsilon_{a \alpha A} = \sum_{b=\pm} \sum_{\beta = \pm} \sum_{B = \pm}(f_A \otimes f_S \otimes 1_2)_{b \beta B, a \alpha A} \varepsilon_{b \beta B} = \left( \vec{\varepsilon} \cdot (f_A \otimes f_S \otimes 1_2) \right)_{a \alpha A}~.
\end{equation}

\begin{bibtex}[\jobname]

@article{Abbott:2013kka,
author = "Abbott, Michael C. and Murugan, Jeff and Sundin, Per and Wulff, Linus",
title = "{Scattering in AdS(2)/CFT(1) and the BES Phase}",
eprint = "1308.1370",
archivePrefix = "arXiv",
primaryClass = "hep-th",
reportNumber = "MIFPA-13-25, QGASLAB-13-07",
doi = "10.1007/JHEP10(2013)066",
journal = "JHEP",
volume = "10",
pages = "066",
year = "2013"
}

@article{Arutyunov:2004yx,
author = "Arutyunov, Gleb and Frolov, Sergey",
title = "{Integrable Hamiltonian for classical strings on AdS(5) x S**5}",
eprint = "hep-th/0411089",
archivePrefix = "arXiv",
reportNumber = "AEI-2004-105",
doi = "10.1088/1126-6708/2005/02/059",
journal = "JHEP",
volume = "02",
pages = "059",
year = "2005"
}

@article{Arutyunov:2005hd,
author = "Arutyunov, Gleb and Frolov, Sergey",
title = "{Uniform light-cone gauge for strings in $AdS_5 \times S^5$: Solving SU(1|1) sector}",
eprint = "hep-th/0510208",
archivePrefix = "arXiv",
reportNumber = "ITP-UU-05-47, SPIN-05-32, AEI-2005-160",
doi = "10.1088/1126-6708/2006/01/055",
journal = "JHEP",
volume = "01",
pages = "055",
year = "2006"
}

@article{Arutyunov:2006ak,
author = "Arutyunov, Gleb and Frolov, Sergey and Plefka, Jan and Zamaklar, Marija",
title = "{The Off-shell Symmetry Algebra of the Light-cone $AdS_5 \times S^5$ Superstring}",
eprint = "hep-th/0609157",
archivePrefix = "arXiv",
reportNumber = "AEI-2006-071, HU-EP-06-31, ITP-UU-06-39, SPIN-06-33, TCDMATH-06-13",
doi = "10.1088/1751-8113/40/13/018",
journal = "J. Phys. A",
volume = "40",
pages = "3583--3606",
year = "2007"
}

@article{Arutyunov:2014jfa,
author = "Arutyunov, Gleb and van Tongeren, Stijn J.",
title = "{Double Wick rotating Green-Schwarz strings}",
eprint = "1412.5137",
archivePrefix = "arXiv",
primaryClass = "hep-th",
doi = "10.1007/JHEP05(2015)027",
journal = "JHEP",
volume = "05",
pages = "027",
year = "2015"
}

@article{Babichenko:2009dk,
author = "Babichenko, A. and Stefanski, Jr., B. and Zarembo, K.",
title = "{Integrability and the AdS(3)/CFT(2) correspondence}",
eprint = "0912.1723",
archivePrefix = "arXiv",
primaryClass = "hep-th",
reportNumber = "ITEP-TH-59-09, LPTENS-09-36, UUITP-25-09",
doi = "10.1007/JHEP03(2010)058",
journal = "JHEP",
volume = "03",
pages = "058",
year = "2010"
}

@article{Beisert:2005tm,
author = "Beisert, Niklas",
title = "{The SU(2|2) dynamic S-matrix}",
eprint = "hep-th/0511082",
archivePrefix = "arXiv",
reportNumber = "PUTP-2181, NSF-KITP-05-92",
doi = "10.4310/ATMP.2008.v12.n5.a1",
journal = "Adv. Theor. Math. Phys.",
volume = "12",
pages = "945--979",
year = "2008"
}

@article{Bena:2003wd,
author = "Bena, Iosif and Polchinski, Joseph and Roiban, Radu",
title = "{Hidden Symmetries of the AdS$_5 \times S^5$ Superstring}",
eprint = "hep-th/0305116",
archivePrefix = "arXiv",
reportNumber = "NSF-KITP-03-34, UCLA-03-TEP-14",
doi = "10.1103/PhysRevD.69.046002",
journal = "Phys. Rev. D",
volume = "69",
pages = "046002",
year = "2004"
}

@article{Berkovits:1999im,
author = "Berkovits, Nathan and Vafa, Cumrun and Witten, Edward",
title = "{Conformal field theory of AdS background with Ramond-Ramond flux}",
eprint = "hep-th/9902098",
archivePrefix = "arXiv",
reportNumber = "IFT-P-012-99, HUTP-99-A004, IASSNS-HEP-99-5",
doi = "10.1088/1126-6708/1999/03/018",
journal = "JHEP",
volume = "03",
pages = "018",
year = "1999"
}

@article{Berkovits:1999zq,
author = "Berkovits, N. and Bershadsky, M. and Hauer, T. and Zhukov, S. and Zwiebach, B.",
title = "{Superstring theory on $AdS_2 \times S^2$ as a coset supermanifold}",
eprint = "hep-th/9907200",
archivePrefix = "arXiv",
reportNumber = "IFT-P-060-99, HUTP-99-A044, MIT-CTP-2878, CTP-MIT-2878",
doi = "10.1016/S0550-3213(99)00683-5",
journal = "Nucl. Phys. B",
volume = "567",
pages = "61--86",
year = "2000"
}

@article{Borsato:2013qpa,
author = "Borsato, Riccardo and Ohlsson Sax, Olof and Sfondrini, Alessandro and Stefa\'nski, Bogdan and Torrielli, Alessandro",
title = "{The all-loop integrable spin-chain for strings on AdS$_3 \times S^3 \times T^4$: the massive sector}",
eprint = "1303.5995",
archivePrefix = "arXiv",
primaryClass = "hep-th",
reportNumber = "SPIN-13-05, DMUS-MP-13-08, ITP-UU-13-08",
doi = "10.1007/JHEP08(2013)043",
journal = "JHEP",
volume = "08",
pages = "043",
year = "2013"
}

@article{Borsato:2014exa,
author = "Borsato, Riccardo and Ohlsson Sax, Olof and Sfondrini, Alessandro and Stefanski, Bogdan",
title = "{Towards the All-Loop Worldsheet S Matrix for $AdS_3\times S^3\times T^4$}",
eprint = "1403.4543",
archivePrefix = "arXiv",
primaryClass = "hep-th",
reportNumber = "IMPERIAL-TP-OOS-2014-01, HU-MATHEMATIK-2014-05, HU-EP-14-12, SPIN-14-11, ITP-UU-14-10",
doi = "10.1103/PhysRevLett.113.131601",
journal = "Phys. Rev. Lett.",
volume = "113",
number = "13",
pages = "131601",
year = "2014"
}

@article{Borsato:2023oru,
author = "Borsato, Riccardo and Driezen, Sibylle and Hoare, Ben and Retore, Ana L. and Seibold, Fiona K.",
title = "{Inequivalent light-cone gauge-fixings of strings on AdSn{\texttimes}Sn backgrounds}",
eprint = "2312.17056",
archivePrefix = "arXiv",
primaryClass = "hep-th",
doi = "10.1103/PhysRevD.109.106023",
journal = "Phys. Rev. D",
volume = "109",
number = "10",
pages = "106023",
year = "2024"
}

@article{Borsato:2024sru,
author = "Borsato, Riccardo and Driezen, Sibylle",
title = "{Particle production in a light-cone gauge fixed Jordanian deformation of AdS5\texttimes{}S5}",
eprint = "2412.08411",
archivePrefix = "arXiv",
primaryClass = "hep-th",
doi = "10.1103/PhysRevD.111.086010",
journal = "Phys. Rev. D",
volume = "111",
number = "8",
pages = "086010",
year = "2025"
}

@article{Cagnazzo:2012se,
author = "Cagnazzo, A. and Zarembo, K.",
title = "{B-field in AdS(3)/CFT(2) Correspondence and Integrability}",
eprint = "1209.4049",
archivePrefix = "arXiv",
primaryClass = "hep-th",
reportNumber = "NORDITA-2012-67, UUITP-24-12",
doi = "10.1007/JHEP11(2012)133",
journal = "JHEP",
volume = "11",
pages = "133",
year = "2012",
note = "Erratum: \texttt{JHEP 04, 003 (2013)}"
}

@article{Cherednik:1981df,
author = "Cherednik, I. V.",
title = "{Relativistically Invariant Quasiclassical Limits of Integrable Two-dimensional Quantum Models}",
doi = "10.1007/BF01086395",
journal = "Theor. Math. Phys.",
volume = "47",
pages = "422--425",
year = "1981"
}

@article{Costello:2019tri,
author = "Costello, Kevin and Yamazaki, Masahito",
title = "{Gauge Theory And Integrability, III}",
eprint = "1908.02289",
archivePrefix = "arXiv",
primaryClass = "hep-th",
reportNumber = "IPMU19-0110",
month = "8",
year = "2019"
}

@article{Cvetic:2002hi,
author = "Cvetic, Mirjam and Lu, Hong and Pope, C. N.",
title = "{Penrose limits, PP waves and deformed M2 branes}",
eprint = "hep-th/0203082",
archivePrefix = "arXiv",
reportNumber = "CTP-TAMU-04-02, UPR-982-T, MCTP-02-15",
doi = "10.1103/PhysRevD.69.046003",
journal = "Phys. Rev. D",
volume = "69",
pages = "046003",
year = "2004"
}

@article{Cvetic:2002nh,
author = "Cvetic, Mirjam and Lu, H. and Pope, C. N. and Stelle, K. S.",
title = "{Linearly realised world sheet supersymmetry in pp wave background}",
eprint = "hep-th/0209193",
archivePrefix = "arXiv",
reportNumber = "RUNHETC-2002-34, MIFP-02-01, IMPERIAL-TP-01-02-27, UPR-1013-T",
doi = "10.1016/S0550-3213(03)00263-3",
journal = "Nucl. Phys. B",
volume = "662",
pages = "89--119",
year = "2003"
}

@article{Delduc:2013fga,
author = "Delduc, Francois and Magro, Marc and Vicedo, Benoit",
title = "{On classical $q$-deformations of integrable sigma-models}",
eprint = "1308.3581",
archivePrefix = "arXiv",
primaryClass = "hep-th",
doi = "10.1007/JHEP11(2013)192",
journal = "JHEP",
volume = "11",
pages = "192",
year = "2013"
}

@article{Delduc:2013qra,
author = "Delduc, Francois and Magro, Marc and Vicedo, Benoit",
title = "{An integrable deformation of the $AdS_5 \times S^5$ superstring action}",
eprint = "1309.5850",
archivePrefix = "arXiv",
primaryClass = "hep-th",
doi = "10.1103/PhysRevLett.112.051601",
journal = "Phys. Rev. Lett.",
volume = "112",
number = "5",
pages = "051601",
year = "2014"
}

@article{Delduc:2014kha,
author = "Delduc, Francois and Magro, Marc and Vicedo, Benoit",
title = "{Derivation of the action and symmetries of the $q$-deformed $AdS_{5} \times S^{5}$ superstring}",
eprint = "1406.6286",
archivePrefix = "arXiv",
primaryClass = "hep-th",
doi = "10.1007/JHEP10(2014)132",
journal = "JHEP",
volume = "10",
pages = "132",
year = "2014"
}

@article{DeLeeuw:2020ahx,
author = "de Leeuw, Marius and Paletta, Chiara and Pribytok, Anton and Retore, Ana L. and Torrielli, Alessandro",
title = "{Free Fermions, vertex Hamiltonians, and lower-dimensional AdS/CFT}",
eprint = "2011.08217",
archivePrefix = "arXiv",
primaryClass = "hep-th",
reportNumber = "DMUS-MP-20/09; TCDMATH-20-14, DMUS-MP-20/09",
doi = "10.1007/JHEP02(2021)191",
journal = "JHEP",
volume = "02",
pages = "191",
year = "2021"
}

@article{deLeeuw:2021ufg,
author = "de Leeuw, Marius and Pribytok, Anton and Retore, Ana L. and Ryan, Paul",
title = "{Integrable deformations of AdS/CFT}",
eprint = "2109.00017",
archivePrefix = "arXiv",
primaryClass = "hep-th",
doi = "10.1007/JHEP05(2022)012",
journal = "JHEP",
volume = "05",
pages = "012",
year = "2022"
}

@article{Demulder:2023bux,
author = "Demulder, Saskia and Driezen, Sibylle and Knighton, Bob and Oling, Gerben and Retore, Ana L. and Seibold, Fiona K. and Sfondrini, Alessandro and Yan, Ziqi",
title = "{Exact approaches on the string worldsheet}",
eprint = "2312.12930",
archivePrefix = "arXiv",
primaryClass = "hep-th",
reportNumber = "NORDITA 2023-083",
doi = "10.1088/1751-8121/ad72be",
journal = "J. Phys. A",
volume = "57",
number = "42",
pages = "423001",
year = "2024"
}

@article{Figueroa-OFarrill:2004qhu,
author = "Figueroa-O'Farrill, Jose M. and Meessen, Patrick and Philip, Simon",
title = "{Supersymmetry and homogeneity of M-theory backgrounds}",
eprint = "hep-th/0409170",
archivePrefix = "arXiv",
reportNumber = "EMPG-04-09, CERN-PH-TH-2004-145",
doi = "10.1088/0264-9381/22/1/014",
journal = "Class. Quant. Grav.",
volume = "22",
pages = "207--226",
year = "2005"
}

@article{Fontanella:2017rvu,
author = "Fontanella, Andrea and Torrielli, Alessandro",
title = "{Massless $AdS_2$ scattering and Bethe ansatz}",
eprint = "1706.02634",
archivePrefix = "arXiv",
primaryClass = "hep-th",
reportNumber = "DMUS-MP-17-05",
doi = "10.1007/JHEP09(2017)075",
journal = "JHEP",
volume = "09",
pages = "075",
year = "2017"
}

@article{Frolov:2019xzi,
author = "Frolov, Sergey",
title = "{$T{\overline T}$, $\widetilde JJ$, $JT$ and $\widetilde JT$ deformations}",
eprint = "1907.12117",
archivePrefix = "arXiv",
primaryClass = "hep-th",
doi = "10.1088/1751-8121/ab581b",
journal = "J. Phys. A",
volume = "53",
number = "2",
pages = "025401",
year = "2020"
}

@article{Fukushima:2020kta,
author = "Fukushima, Osamu and Sakamoto, Jun-ichi and Yoshida, Kentaroh",
title = "{Comments on $\eta$-deformed principal chiral model from 4D Chern-Simons theory}",
eprint = "2003.07309",
archivePrefix = "arXiv",
primaryClass = "hep-th",
reportNumber = "KUNS-2802",
doi = "10.1016/j.nuclphysb.2020.115080",
journal = "Nucl. Phys. B",
volume = "957",
pages = "115080",
year = "2020"
}

@article{Hamidi:2025sgg,
author = "Hamidi, Rashad and Hoare, Ben",
title = "{Twists of trigonometric sigma models}",
eprint = "2504.18492",
archivePrefix = "arXiv",
primaryClass = "hep-th",
doi = "10.1007/JHEP08(2025)090",
journal = "JHEP",
volume = "08",
pages = "090",
year = "2025"
}

@article{Hoare:2013pma,
author = "Hoare, B. and Tseytlin, A. A.",
title = "{On string theory on $AdS_3 \times S^3 \times T^4$ with mixed 3-form flux: tree-level S-matrix}",
eprint = "1303.1037",
archivePrefix = "arXiv",
primaryClass = "hep-th",
reportNumber = "IMPERIAL-TP-AT-2013-01, HU-EP-13-10",
doi = "10.1016/j.nuclphysb.2013.05.005",
journal = "Nucl. Phys. B",
volume = "873",
pages = "682--727",
year = "2013"
}

@article{Hoare:2014kma,
author = "Hoare, Ben and Pittelli, Antonio and Torrielli, Alessandro",
title = "{Integrable S-matrices, massive and massless modes and the $AdS_2 \times S^2$ superstring}",
eprint = "1407.0303",
archivePrefix = "arXiv",
primaryClass = "hep-th",
reportNumber = "HU-EP-14-28, DMUS--MP--14-05",
doi = "10.1007/JHEP11(2014)051",
journal = "JHEP",
volume = "11",
pages = "051",
year = "2014"
}

@article{Hoare:2014oua,
author = "Hoare, Ben",
title = "{Towards a two-parameter q-deformation of AdS$_3 \times S^3 \times M^4$ superstrings}",
eprint = "1411.1266",
archivePrefix = "arXiv",
primaryClass = "hep-th",
reportNumber = "HU-EP-14-44",
doi = "10.1016/j.nuclphysb.2014.12.012",
journal = "Nucl. Phys. B",
volume = "891",
pages = "259--295",
year = "2015"
}

@article{Hoare:2018ebg,
author = "Hoare, Ben and Seibold, Fiona K.",
title = "{Poisson-Lie duals of the $\eta$-deformed $\mathrm{AdS}_2 \times \mathrm{S}^2 \times \mathrm{T}^6$ superstring}",
eprint = "1807.04608",
archivePrefix = "arXiv",
primaryClass = "hep-th",
doi = "10.1007/JHEP08(2018)107",
journal = "JHEP",
volume = "08",
pages = "107",
year = "2018"
}

@article{Hoare:2021dix,
author = "Hoare, Ben",
title = "{Integrable deformations of sigma models}",
eprint = "2109.14284",
archivePrefix = "arXiv",
primaryClass = "hep-th",
doi = "10.1088/1751-8121/ac4a1e",
journal = "J. Phys. A",
volume = "55",
number = "9",
pages = "093001",
year = "2022"
}

@article{Hoare:2022asa,
author = "Hoare, Ben and Seibold, Fiona K. and Tseytlin, Arkady A.",
title = "{Integrable supersymmetric deformations of AdS$_{3}$\texttimes{} S$^{3}$\texttimes{} T$^{4}$}",
eprint = "2206.12347",
archivePrefix = "arXiv",
primaryClass = "hep-th",
reportNumber = "Imperial-TP-AT-2022-02",
doi = "10.1007/JHEP09(2022)018",
journal = "JHEP",
volume = "09",
pages = "018",
year = "2022"
}

@article{Hoare:2022vnw,
author = "Hoare, Ben and Levine, Nat and Seibold, Fiona K.",
title = "{Bi-\ensuremath{\eta} and bi-\ensuremath{\lambda} deformations of \ensuremath{\Integer}$_{4}$ permutation supercosets}",
eprint = "2212.08625",
archivePrefix = "arXiv",
primaryClass = "hep-th",
reportNumber = "Imperial-TP-FS-2022-03",
doi = "10.1007/JHEP04(2023)024",
journal = "JHEP",
volume = "04",
pages = "024",
year = "2023"
}

@article{Hoare:2023zti,
author = "Hoare, Ben and Retore, Ana L. and Seibold, Fiona K.",
title = "{Elliptic deformations of the~$\mathsf{AdS}_3 \times \mathsf{S}^3 \times \mathsf{T}^4$ string}",
eprint = "2312.14031",
archivePrefix = "arXiv",
primaryClass = "hep-th",
reportNumber = "Imperial-TP-FS-2023-02",
doi = "10.1007/JHEP04(2024)042",
journal = "JHEP",
volume = "04",
pages = "042",
year = "2024"
}

@article{Itsios:2023kma,
author = "Itsios, Georgios and Sfetsos, Konstantinos and Siampos, Konstantinos",
title = "{Supersymmetric backgrounds from {\ensuremath{\lambda}}-deformations}",
eprint = "2310.17700",
archivePrefix = "arXiv",
primaryClass = "hep-th",
reportNumber = "HU-EP-23/57",
doi = "10.1007/JHEP01(2024)084",
journal = "JHEP",
volume = "01",
pages = "084",
year = "2024"
}

@article{Klimcik:2002zj,
author = "Klim\v{c}\'{i}k, Ctirad",
title = "{Yang-Baxter sigma models and dS/AdS T duality}",
eprint = "hep-th/0210095",
archivePrefix = "arXiv",
reportNumber = "IML-02-XY",
doi = "10.1088/1126-6708/2002/12/051",
journal = "JHEP",
volume = "12",
pages = "051",
year = "2002"
}

@article{Klimcik:2008eq,
author = "Klim\v{c}\'{i}k, Ctirad",
title = "{On integrability of the Yang-Baxter sigma-model}",
eprint = "0802.3518",
archivePrefix = "arXiv",
primaryClass = "hep-th",
doi = "10.1063/1.3116242",
journal = "J. Math. Phys.",
volume = "50",
pages = "043508",
year = "2009"
}

@article{Lacroix:2023qlz,
author = "Lacroix, Sylvain and Wallberg, Anders",
title = "{An elliptic integrable deformation of the Principal Chiral Model}",
eprint = "2311.09301",
archivePrefix = "arXiv",
primaryClass = "hep-th",
reportNumber = "CERN-TH-2023-205",
doi = "10.1007/JHEP05(2024)006",
journal = "JHEP",
volume = "05",
pages = "006",
year = "2024"
}

@article{Lloyd:2014bsa,
author = "Lloyd, Thomas and Ohlsson Sax, Olof and Sfondrini, Alessandro and Stefa\'nski, Jr., Bogdan",
title = "{The complete worldsheet S matrix of superstrings on AdS$_3 \times$ S$^3 \times$ T$^4$ with mixed three-form flux}",
eprint = "1410.0866",
archivePrefix = "arXiv",
primaryClass = "hep-th",
reportNumber = "IMPERIAL-TP-OOS-2014-04, HU-MATHEMATIK-2014-21, HU-EP-14-34",
doi = "10.1016/j.nuclphysb.2014.12.019",
journal = "Nucl. Phys. B",
volume = "891",
pages = "570--612",
year = "2015"
}

@article{Maldacena:1997re,
author = "Maldacena, Juan Martin",
title = "{The Large $N$ limit of superconformal field theories and supergravity}",
eprint = "hep-th/9711200",
archivePrefix = "arXiv",
reportNumber = "HUTP-97-A097, HUTP-98-A097",
doi = "10.4310/ATMP.1998.v2.n2.a1",
journal = "Adv. Theor. Math. Phys.",
volume = "2",
pages = "231--252",
year = "1998"
}

@article{Metsaev:1998it,
author = "Metsaev, R. R. and Tseytlin, Arkady A.",
title = "{Type IIB superstring action in $AdS_5 \times S^5$ background}",
eprint = "hep-th/9805028",
archivePrefix = "arXiv",
reportNumber = "FIAN-TD-98-21, IMPERIAL-TP-97-98-44, NSF-ITP-98-055",
doi = "10.1016/S0550-3213(98)00570-7",
journal = "Nucl. Phys. B",
volume = "533",
pages = "109--126",
year = "1998"
}

@article{Rughoonauth:2012qd,
author = "Rughoonauth, Nitin and Sundin, Per and Wulff, Linus",
title = "{Near BMN dynamics of the $AdS(3) \times S(3) \times S(3) \times S(1)$ superstring}",
eprint = "1204.4742",
archivePrefix = "arXiv",
primaryClass = "hep-th",
reportNumber = "MIFPA-12-17",
doi = "10.1007/JHEP07(2012)159",
journal = "JHEP",
volume = "07",
pages = "159",
year = "2012"
}

@article{Seibold:2019dvf,
author = "Seibold, Fiona K.",
title = "{Two-parameter integrable deformations of the $AdS_3 \times S^3 \times T^4$ superstring}",
eprint = "1907.05430",
archivePrefix = "arXiv",
primaryClass = "hep-th",
doi = "10.1007/JHEP10(2019)049",
journal = "JHEP",
volume = "10",
pages = "049",
year = "2019"
}

@article{Seibold:2020ywq,
author = "Seibold, Fiona K. and Van Tongeren, Stijn J. and Zimmermann, Yannik",
title = "{The twisted story of worldsheet scattering in $\eta$-deformed $AdS_5 \times S^5$}",
eprint = "2007.09136",
archivePrefix = "arXiv",
primaryClass = "hep-th",
doi = "10.1007/JHEP12(2020)043",
journal = "JHEP",
volume = "12",
pages = "043",
year = "2020"
}

@article{Seibold:2021lju,
author = "Seibold, Fiona K. and van Tongeren, Stijn J. and Zimmermann, Yannik",
title = "{On quantum deformations of AdS$_{3}$ \texttimes{} S$^{3}$ \texttimes{} T$^{4}$ and mirror duality}",
eprint = "2107.02564",
archivePrefix = "arXiv",
primaryClass = "hep-th",
reportNumber = "Imperial-TP-FS-2021-01, HU-EP-21/19",
doi = "10.1007/JHEP09(2021)110",
journal = "JHEP",
volume = "09",
pages = "110",
year = "2021"
}

@article{Sfetsos:2013wia,
author = "Sfetsos, Konstadinos",
title = "{Integrable interpolations: From exact CFTs to non-Abelian T-duals}",
eprint = "1312.4560",
archivePrefix = "arXiv",
primaryClass = "hep-th",
reportNumber = "DMUS-MP-13-23, DMUS--MP--13-23",
doi = "10.1016/j.nuclphysb.2014.01.004",
journal = "Nucl. Phys. B",
volume = "880",
pages = "225--246",
year = "2014"
}

@article{Sfetsos:2014cea,
author = "Sfetsos, Konstantinos and Thompson, Daniel C.",
title = "{Spacetimes for $\lambda$-deformations}",
eprint = "1410.1886",
archivePrefix = "arXiv",
primaryClass = "hep-th",
doi = "10.1007/JHEP12(2014)164",
journal = "JHEP",
volume = "12",
pages = "164",
year = "2014"
}

@article{Sorokin:2011rr,
author = "Sorokin, Dmitri and Tseytlin, Arkady and Wulff, Linus and Zarembo, Konstantin",
title = "{Superstrings in $AdS(2) \times S(2) \times T(6)$}",
eprint = "1104.1793",
archivePrefix = "arXiv",
primaryClass = "hep-th",
reportNumber = "MIFPA-11-11, NORDITA-2011-30, IMPERIAL-TP-AT-2011-2",
doi = "10.1088/1751-8113/44/27/275401",
journal = "J. Phys. A",
volume = "44",
pages = "275401",
year = "2011"
}

@article{Sundin:2013ypa,
author = "Sundin, Per and Wulff, Linus",
title = "{Worldsheet scattering in $AdS(3)/CFT(2)$}",
eprint = "1302.5349",
archivePrefix = "arXiv",
primaryClass = "hep-th",
reportNumber = "MIFPA-13-08",
doi = "10.1007/JHEP07(2013)007",
journal = "JHEP",
volume = "07",
pages = "007",
year = "2013"
}

@article{Wulff:2014kja,
author = "Wulff, Linus",
title = "{Superisometries and integrability of superstrings}",
eprint = "1402.3122",
archivePrefix = "arXiv",
primaryClass = "hep-th",
reportNumber = "IMPERIAL-TP-LW-2014-01",
doi = "10.1007/JHEP05(2014)115",
journal = "JHEP",
volume = "05",
pages = "115",
year = "2014"
}

@article{Zamolodchikov:1978xm,
author = "Zamolodchikov, Alexander B. and Zamolodchikov, Alexei B.",
editor = "Khalatnikov, I. M. and Mineev, V. P.",
title = "{Factorized s Matrices in Two-Dimensions as the Exact Solutions of Certain Relativistic Quantum Field Models}",
reportNumber = "ITEP-35-1978",
doi = "10.1016/0003-4916(79)90391-9",
journal = "Annals Phys.",
volume = "120",
pages = "253--291",
year = "1979"
}

@article{Fateev:1996ea,
author = "Fateev, V. A.",
title = "{The sigma model (dual) representation for a two-parameter family of integrable quantum field theories}",
doi = "10.1016/0550-3213(96)00256-8",
journal = "Nucl. Phys. B",
volume = "473",
pages = "509--538",
year = "1996"
}

\end{bibtex}

\bibliographystyle{nb}
\bibliography{\jobname}

\end{document}